\documentclass[12pt]{article}

\usepackage{preamble}
\usepackage{xr-hyper}
\usepackage{hyperref}
\usepackage{makecell}

\makeatletter
\newcommand*{\addFileDependency}[1]{
\typeout{(#1)}
\@addtofilelist{#1}
\IfFileExists{#1}{}{\typeout{No file #1.}}
}\makeatother

\def\boxit#1{\vbox{\hrule\hbox{\vrule\kern6pt
          \vbox{\kern6pt#1\kern6pt}\kern6pt\vrule}\hrule}}

\def\bUpsilon{{\boldsymbol{\Upsilon}}}

\begin{document}

\thispagestyle{empty}
\baselineskip=28pt

\begin{center}
{\LARGE{\bf Establishing the Parallels and Differences Between Right-Censored and Missing Covariates}}
\end{center}

\baselineskip=12pt

\vskip 2mm
\begin{center}
Jesus E. Vazquez$^{1*}$, Marissa C. Ashner$^{2}$, Yanyuan Ma$^3$, Karen Marder$^4$, and Tanya P. Garcia$^1$
\\
$^1$Department of Biostatistics, Gillings School of Global Public Health, University of North Carolina at Chapel Hill, Chapel Hill, NC, 27516\\
$^2$Department of Biostatistics and Bioinformatics, Duke University, Durham, NC, 27708\\
$^3$Department of Statistics, Penn State University, State College, PA 16802\\
$^4$Department of Neurology, Columbia University Medical Center, New York, NY 10032\\
$^*$ Correspondence to:  jesusvaz@ad.unc.edu;  X \@biostat\_vazquez\end{center}
%

\begin{center}
{\Large{\bf Abstract}}
\end{center}
\baselineskip=12pt


While right-censored time-to-event outcomes have been studied for decades, handling time-to-event  covariates, also known as right-censored covariates, is now of growing interest.  So far, the literature has treated right-censored covariates as distinct from missing covariates, overlooking the potential applicability of estimators to both scenarios. We bridge this gap by establishing connections between right-censored and missing covariates under various assumptions about censoring and missingness, allowing us to identify parallels and differences to determine when estimators can be used in both contexts. These connections reveal adaptations to five estimators for right-censored covariates in the unexplored area of informative covariate right-censoring and to formulate a new estimator for this setting, where the event time depends on the censoring time. We establish the asymptotic properties of the six estimators, evaluate their  robustness under incorrect distributional assumptions, and establish their comparative efficiency.  We conducted a simulation study to confirm our theoretical results, and then applied all estimators to a Huntington disease observational study to analyze cognitive impairments as a function of time to clinical diagnosis.

\baselineskip=12pt
\par\vfill\noindent
\underline{\bf Some Key Words}:
censored covariate; missing data; not at random; informative;
model misspecification; robustness; statistical efficiency
\par\medskip\noindent
\underline{\bf Short title}: Parallels between right-censored covariates and missing covariates
\clearpage\pagebreak\newpage

\pagenumbering{arabic}
\newlength{\gnat}
\setlength{\gnat}{22pt}
\baselineskip=\gnat

\clearpage\newpage

\section{Introduction}
\label{sec:introduction}

%
Efforts are now underway to design clinical trials for 
Huntington disease aimed at slowing or stopping the disease before irreversible damage occurs \parencite{tabrizi2022biological}. A key aspect of that design is modeling how disease symptoms worsen over time---the symptom trajectory---in the years leading up to a diagnosis because with knowledge of that symptom trajectory, researchers can identify when and how to test experimental therapies. Yet modeling the symptom trajectory is not easy because Huntington disease 
progresses slowly over decades, so studies that track symptoms often end before a diagnosis can be made. This makes time to diagnosis \emph{right-censored} (i.e., a patient will reach the criteria for a diagnosis sometime after {, or “to the right of”,} the last study visit, but exactly when is unknown), leaving researchers with the challenge of trying to model the symptom trajectory without full information about when a diagnosis occurs.

This challenge creates a unique statistical problem of modeling the symptom trajectory as a function of a \emph{right-censored covariate}, time to diagnosis. Tackling this problem by adopting estimators from the missing data literature has long been the go-to strategy \parencite{lotspeich2024making}, as the true covariate value---whether right-censored or missing---is sometimes observed and sometimes not. Right-censoring and missingness are well known to be different, yet many estimators in the right-censored covariate literature seem to share similar statistical properties as estimators used in the missing covariate literature \parencite{ Bartlettetal2014,Bernhardtetal2015,Ahnetal2018, Atemetal2019,MatsouakaAtem2020}. These similarities blur the distinctions between the right-censored covariate problem and missing covariate problem. When we blindly apply estimators designed for one problem to the other, sometimes they behave similarly, and at other times they do not, even though the assumptions made in both problems are similar.

That flip-flopping, where the estimators sometimes behave similarly and other times do not, led us to ask: How do the subtle differences between right censoring and missingness affect the estimators and their statistical properties? Neglecting this question leaves us vulnerable to using the incorrect estimator due to an unacknowledged blind spot. Not using the correct estimator  may result in bias and inefficiency in estimating the parameters indexing symptom trajectory models, posing two major challenges when striving to design a well-powered clinical trial that tests an experimental therapy.

We tackle this blind spot head-on. We start by outlining the right-censored and missing covariate problems, highlighting the parallels and distinctions between different types of covariate missingness (i.e., missing at random, missing not at random) and covariate right-censoring (i.e., noninformative, informative) (Sections \ref{sec:method-framework} and \ref{sec:dependence}). These insights help us identify when estimators for one problem can be applied to the other, when they cannot, and the driving reasons behind these differences (Sections \ref{sec:estimators} and \ref{sec:dependence}). We provide practical guidance on implementing the estimators, complete with R software and documentation for reproducibility (Section \ref{sec:implement}). We provide the first-ever theoretical and empirical justifications for when each estimator remains consistent despite distribution misspecifications and how they compare in statistical efficiency (Sections \ref{sec:summary} and \ref{sec:simulations_paper}). By applying the estimators to Huntington disease data
we show how the estimators perform in practice and how their different properties impact conclusions made about the progression patterns of cognitive symptom trajectories in the years leading up to a diagnosis (Section \ref{sec:real_data}). All these findings are distilled into practical guidance (Section \ref{sec:discussion}), empowering researchers to choose the right estimator without making blind assumptions.

\section{Framework for right-censoring and missingness}
\label{sec:method-framework}

\subsection{Notation and assumptions}
\label{sec:notation}

We consider the regression model 
\be
\label{eqn:equation_paper2}
{Y} = m(X, \bZ; \bbeta) + {\epsilon}, \ \ \ \epsilon\sim\Normal(0,\sigma^2),
\ee
which models the relationship between an outcome $Y$ and covariates $X$ and $\bZ$.  That relationship is determined by a mean function $m(\cdot)$, which is assumed to be known up to a  parameter $\bbeta$ that is of length $p$. We assume throughout that the values for $\bZ$ are always observed, but those for $X$ are not.

We consider two distinct problems in which $X$ is not always observed. The first, known as the right-censored covariate problem, is when $X$ is right-censored. Instead of observing $X$, we observe $W=\min(X,C)$ and $\Delta=I(X\leq C)$, where $C$ is a random censoring variable.  In our Huntington disease problem, $W$ represents the  minimum value between the time to diagnosis $X$ and the random time to study exit $C$, and $\Delta$ is the censoring indicator that is 1 if  $X$ is observed and 0 if not. The second, known as the missing covariate problem, is when $X$ is missing. In this problem, we observe a missingness indicator $R$ which is 1 if $X$ is observed and 0 if $X$ is missing.  The observed data when $X$ is right-censored are $\bO^{\rm cens}=(Y,W,\Delta,\bZ)$, and when $X$ is missing are $\bO^{\rm miss}=(Y,RX,R,\bZ)$. The first  goal is to derive consistent estimators for $\btheta\equiv (\bbeta^{\trans},\sigma)^{\trans}$ in both problems, and the second goal is to highlight how those estimators differ across the two problems.

Toward these goals, we first consider simple, yet common assumptions made in the statistical literature about the right-censored and missing covariate problems. Let $\independent$  represent statistical independence. When $X$ is potentially right-censored, we assume that $C$ and $Y$ are independent given $X$ and $\bZ$ (i.e.,  $C\independent Y |X,\bZ$) and that $X$ and $C$ are independent given $\bZ$ (i.e., $X\independent C|\bZ$).  The last assumption is known as noninformative covariate censoring and occurs when the event rate (e.g, clinical diagnosis) is the same for individuals who were observed or censored  conditional on the set of fully observed covariates \parencite{lee1998simple}. For example, overall, the event rate may not be the same among those who were observed or censored, but equal event rates are obtained conditional on ethnicity (i.e., $\bZ$). 

We will make assumptions in the missing covariate problem that parallel those used in the right-censored covariate problem so that we may pinpoint if and how the two problems differ. The analog to these assumptions when $X$ is missing are: $R$ and $Y$ are independent given $X$ and $\bZ$ (i.e., $R\independent Y|X,\bZ$), and  $X$ and $R$ are independent given $\bZ$ (i.e., $X\independent R|\bZ$).  The latter assumption is known as the missing at random assumption, implying that the missingness in $X$ is independent of $X$ given the observed data \parencite{little2019statistical}. The  noninformative covariate right-censoring and missing at random assumptions are analogous in that they assume $X$ is independent of $C$ or $R$. For that reason, it will help to refer to these assumptions as independent censoring and independent missingness because, in both cases, we assume conditional independence.

Throughout, we will use superscripts ``cens'' and ``miss'' to differentiate between right-censoring and missingness, respectively. We  will also use ``ind'' to denote  independence, which will refer to independent censoring (i.e., $X\independent C|\bZ$) for the right-censored covariate problem and to independent missingness (i.e., $X\independent R|\bZ$) for the missing covariate problem. To minimize notation we will let $X \independent (C,R)$ indicate independent right-censoring or missingness. We will denote different (conditional) densities $f$ with subscripts to differentiate between densities. For example,  $f_{Y|X,\bZ}(y,w,\bz;\btheta)$  denotes the conditional  density function of $Y$ given $(X,\bZ)$ evaluated at $Y=y$, $X=w$, and $\bZ=\bz$.  Similarly, $E_{X|Y,\bZ}(\cdot)$ will be used to denote the conditional expectation with respect to the density $f_{X|Y,\bZ}$. For a vector or matrix $\bM$,  we define $\bM^{\otimes2}\equiv\bM\bM\trans$.

\subsection{The distinction between right-censoring and missingness}
\label{sec:distinction}

At first glance, a right-censored covariate seems no different from a missing covariate since, in both cases, we do not always observe the true covariate value. Yet, right-censoring is different from missingness in one major way.  With missingness, we have no
information about the true covariate value. With right-censoring, we have partial information: if an individual had been in a study for five  years and the study ended before a diagnosis was made, we know their time to diagnosis is at least five years.  
That partial information, though seemingly small, impacts how we handle the unobserved $X$.  One way we see that impact is when computing the probability that $X$ is observed. That probability changes depending on whether we are handling a right-censored or missing covariate.  When $X$ is subject to right-censoring, the  probability $X$ is observed is 
\bse
\pi_{X,\bZ}^{\rm cens, ind}(x,\bz) &\equiv& \pr(\Delta=1|X=x,\bZ=\bz)=\int_{x<c}f_{C|\bZ}(c,\bz)dc. 
\ese
When $X$ is subject to missingness, the probability that $X$ is observed is
\bse
  \pi_{X,\bZ}^{\rm miss, ind}(x,\bz)&\equiv&\pr(R=1|X=x,\bZ=\bz) = \pr(R=1|\bZ=\bz) = \pi_{\bZ}^{\rm miss, ind}(\bz).
\ese

Justification of these calculations is given in Sections \ref{sec:thm1-proof-part1} and \ref{sec:thm1-proof-part2}. Comparing the above, we see that in the missing covariate problem, the probability that $X$ is observed is only a function of $\bZ$. In the right-censored covariate problem, however, that same probability is \emph{not} just a function of $\bZ$.  In that problem,  having $X$ observed means $C$ is larger than $X$, and that information is reflected by the domain of the integral bound (i.e., $x < c$). 
If we were to blindly use the probability from the missing covariate problem in the right-censored covariate problem, we would incorrectly assume that we could replace $\pi_{\bZ}^{\rm miss,ind}(\bz)\equiv \pr(R=1|\bZ=\bz)$ with  $\pi_{\bZ}^{\rm cens,ind}(\bz)\equiv\pr(\Delta=1|\bZ=\bz)$.  Yet, in the right-censored covariate problem, the probability $X$ is observed is  $\pi_{X,\bZ}^{\rm cens,ind}(x,\bz)$, which is not equal to $\pi_{\bZ}^{\rm cens,ind}(\bz)$. These differences affect the estimators used for covariate right-censoring and missingness, as shown next.

\section{Comparison of estimators for right-censoring and missingness
}
\label{sec:estimators}

%
The estimators we construct are solutions to so-called $m$-estimating equations $\sumi \bPhi_{\rm est}(\bO_i; \btheta) = \bzero$, where the estimator $\wh\btheta$ solves these equations \parencite{Tsiatis2006}. We show in Table \ref{tab:estimators} when and how these estimators differ between  the right-censored and missing covariate problems.

\subsection{Complete case (CC) estimator}
\label{sec:cc}

%
The simplest way to estimate the parameters $\btheta$ when $X$ is  right-censored or missing is using the CC estimator. This estimator discards all observations where some data are unobserved---right-censored or missing---and estimates the parameters with the remaining data.  The estimator for $\btheta$ based on only the complete cases is the solution to the estimating equations of the CC estimator in Table \ref{tab:estimators}. Apart from minor notational differences---using $\Delta$ or $R$ to denote when $X$ is observed---these estimating equations are the same for the right-censored covariate problem and the missing covariate problem. Thus, the  CC estimator requires no adjustment when adapting it from the missing data literature. 

\subsection{Inverse probability weighting (IPW) estimator}
\label{sec:ipw}

%
The IPW estimator is a modification of the CC estimator. It assigns weights to the data to reflect what the full data sample would be if all data were observed, then estimates the parameters. The weights used are the inverse of the probability that $X$ is observed and that probability depends on whether we have a right-censored covariate or a missing covariate.  In the right-censored covariate problem, the weights are $1/\pi_{X,\bZ}^{\rm cens, ind}(x,\bz)$, whereas in the missing covariate problem, the weights are  $1/\pi_{\bZ}^{\rm miss, ind}(\bz)$ (Table \ref{tab:estimators}; IPW estimator).  


\begin{table}[!t]
\caption{Estimating equations used in the right-censored covariate problem when censoring is noninformative (i.e., independent censoring) and in the missing covariate problem when missingness is at random (i.e., independent missingness).}
\small
\scalebox{0.89}{
 \begin{tabular}{>{\raggedright}p{1.05\linewidth}c}
\hline
\makecell{\uline{Complete case (CC) estimator}}\\
\makecell{ \footnotesize $\sumi \bPhi_{\rm CC}^{\rm cens,ind}(\bO_i^{\rm cens}; \btheta)=\sumi \delta_i   \bS_{\btheta}^F(y_i,w_i,\bz_i)=\0$}  &\\  %
\makecell{ \footnotesize $\sumi \bPhi_{\rm CC}^{\rm miss,ind}(\bO_i^{\rm miss}; \btheta)=\sumi r_i   \bS_{\btheta}^F(y_i,x_i,\bz_i)=\0$}  & \\[1em] 

\makecell{\uline{Inverse probability weighting (IPW) estimator}}\\
\makecell{ \footnotesize $\sumi \bPhi_{\rm IPW}^{\rm cens, ind}(\bO_i^{\rm cens}; \btheta) = \sumi \delta_i \bS_{\btheta}^F (y_i, w_i, \bz_i)/ \highlight{\pi_{X,\bZ}^{\rm cens, ind}(w_i,\bz_i)} = \bzero
$} &\\
\makecell{ \footnotesize $\sumi \bPhi_{\rm IPW}^{\rm miss, ind}(\bO_i^{\rm miss}; \btheta) = \sumi r_i \bS_{\btheta}^F (y_i, x_i, \bz_i)/\pi_{\bZ}^{\rm miss, ind}(\bz_i) = \bzero$}& \\[1em]
\makecell{\uline{Maximum likelihood estimator (MLE)}}\\
\makecell{ \footnotesize $\sumi \bPhi_{\rm MLE}^{\rm cens, ind}(\bO_i^{\rm cens};\btheta) \equiv \sum_{i=1}^n  \biggr[ \delta_i\bS_\btheta^F(y_i,w_i,\bz_i) + (1- \delta_i) \frac{\partial}{\partial \btheta^T}  \log \biggr\{ \highlight{\int_{w_i < x}} f_{Y|X,\bZ} (y_i, x, \bz_i; \btheta) f_{X|\bZ} (x,\bz_i) dx \biggr\}   \biggr]=\0$}& \\
\makecell{ \footnotesize $\sumi \bPhi_{\rm MLE}^{\rm miss, ind}(\bO_i^{\rm miss};\btheta) \equiv \sum_{i=1}^n  \biggr[ r_i\bS_\btheta^F(y_i,x_i,\bz_i)  + (1- r_i) \frac{\partial}{\partial \btheta^T}  \log \biggr\{ \int f_{Y|X,\bZ} (y_i, x, \bz_i; \btheta) f_{X|\bZ} (x,\bz_i) dx \biggr\}   \biggr]=\0$} &\\[1em]

\makecell{\uline{Augmented complete case (ACC) estimator}}\\
 \makecell{ \footnotesize $\sumi \bPhi_{\rm ACC}^{\rm cens,ind}(\bO_i^{\rm cens}; \btheta) \equiv \sum_{i=1}^n \left[ \delta_i \bS_{\btheta}^F(y,w_i,\bz_i)+ \{\delta_i- \highlight{\pi_{Y,\bZ}^{\rm cens,ind}(y_i, \bz_i)} \} \highlight{\bPsi_{\rm ACC, eff}^{\rm cens,ind}(y_i,\bz_i; \btheta)}  \right] = \bzero$} &\\
 \makecell{ \footnotesize $\sumi \bPhi_{\rm ACC}^{\rm miss, ind}(\bO_i^{\rm miss};\btheta)\equiv \sum_{i=1}^n \left[ r_i \bS_{\btheta}^F(y,x_i,\bz_i) + \{r_i- \pi_{\bZ}^{\rm miss,ind}( \bz_i) \} \bPsi_{\rm ACC,eff}^{\rm miss,ind}(y_i,\bz_i; \btheta)  \right] = \bzero$} &\\[1em]

\makecell{\uline{Modified augmented complete case  (MACC) estimator}}\\
\makecell{ \footnotesize $ 
\sumi \bPhi_{\rm MACC}^{\rm cens,ind}(\bO_i^{\rm cens};\btheta) \equiv \sum_{i=1}^n \left[ \delta_i \bS_{\btheta}^F(y,w_i,\bz_i) +  \{1- \delta_i/\highlight{\pi_{X,\bZ}^{\rm cens,ind}(w_i, \bz_i)}\} \highlight{\bPsi_{\rm MACC,eff}^{\rm cens,ind}(y_i,\bz_i; \btheta)}  \right] = \bzero$} &\\
\makecell{ \footnotesize $\sumi \bPhi_{\rm MACC}^{\rm miss,ind}(\bO_i^{\rm miss};\btheta) \equiv \sum_{i=1}^n \left[ r_i \bS_{\btheta}^F(y,x_i,\bz_i) +\{1- r_i/ \pi_{\bZ}^{\rm miss,ind}(\bz_i)\} \bPsi_{\rm MACC,eff}^{\rm miss,ind}(y_i,\bz_i; \btheta)  \right] = \bzero$} &\\ [1em]
\makecell{\uline{Augmented inverse probability weighting (AIPW) estimator}} \\ [-0.8em]
 \makecell{ \footnotesize $\sumi\bPhi_{\rm AIPW}^{\rm cens,ind}(\bO_i^{\rm cens};\btheta)\equiv\sum_{i=1}^n \left[ \delta_i\bS_\btheta^F(y_i,w_i,\bz_i)/\highlight{\pi_{X,\bZ}^{\rm cens,ind}(w_i,\bz_i)} + \{1- \delta_i/\highlight{\pi_{X,\bZ}^{\rm cens,ind}(w_i, \bz_i)}\} \highlight{\bPsi_{\rm AIPW,eff}^{\rm cens,ind}(y_i,\bz_i; \btheta)} \right] = \bzero$} &\\
\makecell{ \footnotesize $\sumi\bPhi_{\rm AIPW}^{\rm miss,ind}(\bO_i^{\rm miss};\btheta)\equiv\sum_{i=1}^n \left[ \delta_i\bS_\btheta^F(y_i,x_i,\bz_i)/\pi_{\bZ}^{\rm miss,ind}(\bz_i) + \{1- r_i/ \pi_{\bZ}^{\rm miss,ind}( \bz_i)\} \bPsi_{\rm AIPW,eff}^{\rm miss,ind}(y_i,\bz_i; \btheta) \right] = \bzero
$} &\\ [1em]
  \hline
\multicolumn{2}{p{\linewidth}}{
{\footnotesize \textbf{Note:} cens: right-censoring; miss: missingness; ind: independent censoring (i.e., $X\independent C|\bZ$) for the right-censored covariate problem and independent missingness (i.e., $X\independent R|\bZ$) for the missing covariate problem. All estimating equations use the score function $\bS_{\btheta}^F(y,x,\bz)=\partial \log f_{Y|X,\bZ}(y,x,\bz;\btheta)/\partial\btheta$. Differences from the missing covariate estimating equations are highlighted in $\highlight{\rm gray}$. Specific terms are defined below.
}} \\ 
\multicolumn{2}{p{\linewidth}}{
{\footnotesize 
$\pi_{X,\bZ}^{\rm cens, ind}(x,\bz)\equiv \pr(\Delta=1|X=x,\bZ=\bz)=\int_{x<c}f_{C|\bZ}(c,\bz)dc$, 
$\pi_{\bZ}^{\rm miss, ind}(\bz)\equiv \pr(R=1|\bZ=\bz)$;  
}}\\ 
%
\multicolumn{2}{p{\linewidth}}{
{\footnotesize 
$\pi_{Y,\bZ}^{\rm cens, ind}(y,\bz)\equiv \pr(\Delta=1|Y=y,\bZ=\bz)$, 
$\pi_{Y,\bZ}^{\rm miss, ind}(y,\bz)\equiv \pr(R=1|Y=y,\bZ=\bz)$;  
}}\\ 

\multicolumn{2}{p{\linewidth}}{
{\footnotesize 
$\bPsi_{\rm ACC,eff}^{\rm cens,ind}(y, \bz
; \btheta)= -E_{X|Y,\bZ, \Delta=1} \{ \bS_{\btheta}^F(y,X,\bz)\}$, 
$\bPsi_{\rm ACC, eff}^{\rm  miss,ind}(y, \bz; \btheta)= 
 -E_{X|Y,\bZ} \{ \bS_{\btheta}^F(y,X,\bz)\}$;  
}}\\ 
\multicolumn{2}{p{1.05\linewidth}}{
{\footnotesize 
$\bPsi_{\rm MACC,eff}^{\rm cens,ind}(y, \bz
; \btheta)= \frac{E_{X|Y,\bZ}[\{\pi_{X, \bZ}^{\rm cens,ind}(X,\bz)-1\} \bS^F_{\btheta} (y,X,\bz)]}{ E_{X|Y,\bZ}\{1-1/\pi_{X,\bZ}^{\rm cens,ind}(X,\bz) \}}$,
$\bPsi_{\rm MACC,eff}^{\rm miss,ind}(y, \bz
; \btheta)= E_{X|Y,\bZ} \{ \bS_{\btheta}^F(y,X,\bz)\} \pi_{\bZ}^{\rm miss,ind}(\bz) $;  
}}\\ 

\multicolumn{2}{p{1.05\linewidth}}{
{\footnotesize 
$\bPsi_{\rm AIPW,eff}^{\rm cens,ind}(y, \bz
; \btheta) = \frac{E_{X|Y,\bZ}[\{1-1/\pi_{X,\bZ}^{\rm cens,ind}(X,\bz)\} \bS^F_{\btheta} (y,X,\bz)]}{ E_{X|Y,\bZ}\{1-1/\pi_{X,\bZ}^{\rm cens,ind}(X,\bz) \} }$,
$\bPsi_{\rm AIPW,eff}^{\rm miss,ind}(y, \bz
; \btheta) = E_{X|Y,\bZ} \{ \bS_{\btheta}^F(y,X,\bz)\}$.  
}}\\
%
\hline
  \end{tabular}
  }
 \label{tab:estimators}
  \vspace{-1em}
  \end{table}

\subsection{Maximum likelihood estimator (MLE)}
\label{sec:likelihood}

%
The MLE maximizes the likelihood of the data to estimate the parameters $\btheta$. The resulting estimator is the solution to the MLE estimating equations shown in Table \ref{tab:estimators}. These equations contain two parts: the first is based on the likelihood of the data when $X$ is observed and the second  is based on the likelihood of the data when $X$ is not observed.  That second part involves integrating over the possible range of values for $X$, and that range differs for the right-censored covariate problem and for the missing covariate problem. When $X$ is right-censored, we have the partial information that $C<X$, so the integration is computed over this range. When $X$ is missing, we do not have similar partial information, so the integral is computed over all possible values for $X$. These differences mean that the estimating equations for the MLE cannot be interchanged between the right-censored covariate problem and the missing covariate problem.

\subsection{Augmented estimators}
\label{sec:augmention}

A challenge with the CC and IPW estimators is their inefficiency, as they discard any observations where some data are unobserved. Augmented estimators address this challenge by creating augmented versions of the CC and IPW estimators to increase their efficiency. The augmented versions are created by comparing the variance of the ``original" estimators (i.e., CC and IPW) with the variance of their augmented version (i.e., the augmented estimator).

The variances of two estimators can be compared by analyzing the variances of their influence functions since there is a one-to-one correspondence between estimators and influence functions \parencite{rotnitzky1997analysis, Tsiatis2006}. Let $\bUpsilon$ and $\bUpsilon_{\rm AUG}$ denote the influence functions for an original estimator and the augmented estimator. Suppose that
\be
\label{eqn:efficiency-condition}
\0 &=& {\rm cov} (\bUpsilon_{\rm AUG} - \bUpsilon, \bUpsilon_{\rm AUG}) = \var(\bUpsilon_{\rm AUG}) - {\rm cov} (\bUpsilon_{\rm AUG}, \bUpsilon).
\ee
This condition, we call the ``efficiency condition," ensures
that the 
augmented estimator is less variable (i.e., 
   $\var(\bUpsilon_{\rm AUG}) <  \var(\bUpsilon)$) since, when it holds, we have
\bse
\var(\bUpsilon)-\var(\bUpsilon_{\rm AUG}) = \var(\bUpsilon_{\rm AUG})+\var(\bUpsilon)-2\cov(\bUpsilon_{\rm AUG},\bUpsilon)
= \var(\bUpsilon_{\rm AUG}-\bUpsilon) > 0.
\ese
%
When the efficiency condition holds, the augmented estimator will be more efficient than the original estimator. We thus use the efficiency condition to create augmented versions of the CC and IPW estimators; derivations of these augmented estimators can be found in the proof of Theorem \ref{thm:all-robust-consistent} (Sections \ref{sec:thm1-proof-part1-efficiency} and \ref{sec:thm1-proof-part2-efficiency}). 

\subsubsection{Augmented CC (ACC) estimator}
\label{sec:ACC}

%
The ACC estimator, initially proposed by \textcite{Bartlettetal2014} for $X$ missing not at random, augments the CC estimator to increase efficiency. We now adapt the ACC estimator to account for independent covariate right-censoring and missingness. The estimating equations have two parts: the first is the same part used in the CC estimator and the second is the augmented part designed to satisfy the efficiency condition in equation \eqref{eqn:efficiency-condition}. This augmented part has two components: the first is a function of the probability that $X$ is observed and the second is a vector of length equal to that of $\btheta$ and a function of the complete data ($Y,\bZ$).  

The ACC estimating equations for the right-censored covariate problem and the missing covariate problem look very similar (Table \ref{tab:estimators}; ACC estimator). Yet, interchanging $(RX,R)$ with $(W,\Delta)$ in the estimating equation for the missing covariate problem will lead to an inconsistent estimator. For the right-censored covariate problem, consistency and the efficiency condition in equation \eqref{eqn:efficiency-condition} is satisfied when the augmented part uses the probability $\pi_{Y,\bZ}^{\rm cens}\equiv \pr(\Delta=1 | Y=y,\bZ=\bz)$ and augmented component $\bPsi_{\rm ACC, eff}^{\rm cens}$, as defined in Table \ref{tab:estimators}.  The probability and augmented component differ from those used in the missing covariate problem, where the probability is $\pi_{Y,\bZ}^{\rm miss}(y,\bz)\equiv \pr(R=1| Y=y,\bZ=\bz)$, which simplifies to $\pi_{\bZ}^{\rm miss, ind}$ because $R\independent Y|(X,\bZ)$, and the augmented component is $\bPsi_{\rm ACC, eff}^{\rm miss, ind}$.  

We prove in Theorem \ref{thm:all-robust-consistent} that the ACC estimator for the right-censored covariate problem produces a consistent estimator only when the probability $\pi_{Y,\bZ}^{\rm cens}$ is correctly specified, which is equivalent to requiring the  distribution $f_{\Delta|Y,\bZ}$ to be correctly specified. Correctly specifying $f_{\Delta|Y,\bZ}$ may initially seem like a deterrence, but because the data $(Y,\Delta,\bZ)$ are always observed, practical strategies exist for estimating $f_{\Delta|Y,\bZ}$ accurately; see Section \ref{sec:implement}.

\subsubsection{Modified ACC (MACC) estimator}
\label{sec:MACC}

%
The ACC estimator was originally designed based on the probability $\pi_{Y,\bZ}$ to take advantage of the fully observed data $(Y,\Delta,\bZ)$. Given the additional information obtained in the right-censored covariate problem, an analyst may be interested in using the probability $\pi_{X,\bZ}$. While this choice of probability leads to a consistent estimator, it also requires a different formulation of the ACC estimator, which we call the modified ACC (MACC) estimator. The different formulation of the augmented component is required to ensure higher efficiency than that of the CC estimator when using the probability $\pi_{X,\bZ}$. The MACC estimator may be preferable to the ACC estimator if there is a higher chance of correctly specifying the probability  $\pi_{X,\bZ}$ than correctly specifying the probability $\pi_{Y,\bZ}$. Even if there is a higher chance of specifying $\pi_{X,\bZ}$ correctly, correctly estimating this probability is more challenging  since we do not always observe $(X,\bZ)$. Still, in Section \ref{sec:implement}, we discuss practical strategies for appropriately estimating $\pi_{X,\bZ}^{\rm cens,ind}$. 

The albeit more challenging version only complicates the estimation process when $X$ is right-censored, not when $X$ is missing. When $X$ is missing, the ACC and MACC estimators are equivalent. If $\bPsi_{\rm ACC,eff}^{\rm miss,ind}(y, \bz; \btheta)$ and $\bPsi_{\rm MACC,eff}^{\rm miss,ind}(y, \bz; \btheta)$ are plugged into their respective estimating equations, the estimating equations corresponding to the MACC and the ACC estimators are the same (Table \ref{tab:estimators}). This result does not hold for the right-censored covariate problem since plugging $\bPsi_{\rm ACC,eff}^{\rm cens,ind}(y, \bz; \btheta)$ and $\bPsi_{\rm MACC,eff}^{\rm cens,ind}(y, \bz; \btheta)$ into their respective estimating equations will lead to different estimating equations.

\subsubsection{Augmented IPW (AIPW) estimator}
\label{sec:AIPW}

%
Another augmented estimator  typically used in the missing covariate problem is the AIPW estimator, which augments the IPW estimator to increase efficiency \parencite{rotnitzky1997analysis, SeamanVansteelandt2018}.  To the best of our knowledge, only one AIPW estimator has been developed for censored covariates; specifically, it is for a survival outcome with a time-varying covariate that is subject to interval-censoring  \parencite{Ahnetal2018}. We consider a non-survival outcome and the more general setting in which the covariate is randomly right-censored and not time-dependent. 

The AIPW estimator contains two parts in its estimating equations (Table \ref{tab:estimators}; AIPW estimator): the first part is the part used in the IPW estimator and the second is the augmented part designed to satisfy the efficiency condition in equation \eqref{eqn:efficiency-condition}. Similar to the ACC and MACC estimators, the AIPW estimator designed for the missing covariate problem cannot be applied to the right-censored covariate problem. While the estimating equations for the AIPW estimator look similar in both problems,  the probabilities $\pi_{X,\bZ}^{\rm cens, ind}$ and $\pi_{\bZ}^{\rm miss, ind}$ are different, as well as their augmented components. 

\subsubsection{Guaranteeing statistical efficiency and reducing computation cost}
\label{sec:efficiency-computational-augmented-estimators}

%
The efficiency condition in equation \eqref{eqn:efficiency-condition} was used to derive the augmented estimators---ACC, MACC, and AIPW---and guarantee their improved efficiency. However, that derivation assumed that the probability (i.e., $\pi_{X,\bZ}$ or $\pi_{Y,\bZ}$) and the augmented component (i.e., $\bPsi(y,\bz;\btheta)$) were correctly specified. If either are incorrectly specified, improved efficiency is not guaranteed. To guarantee the efficiency condition in equation \eqref{eqn:efficiency-condition} even when the probability or the augmented component is misspecified, we update the augmented component $\bPsi_{\rm updated}(y,\bz;\btheta)$ by pre-multiplying the augmented component (as defined in Table \ref{tab:estimators}) by a fixed matrix $\bLambda$, i.e., $\bPsi_{\rm updated}(y,\bz;\btheta) = \bLambda \bPsi(y, \bz; \btheta)$. 

We use the efficiency condition in equation \eqref{eqn:efficiency-condition} to derive the form of $\bLambda$. Instead of solving for the form of $\bPsi (y,\bz; \btheta)$ that satisfies the efficiency condition, we solve for $\bLambda$. This $\bLambda$ is a function of the probability (e.g., $\pi_{Y,\bZ}$) and the 
augmented component $\bPsi (y,\bz; \btheta)$; therefore, for any specification of the probability and 
the augmented component, we obtain a corresponding $\bLambda$ matrix that satisfies the efficiency condition in equation (\ref{eqn:efficiency-condition}). This strategy leads to an updated augmented component $\bPsi_{\rm updated}(y,\bz;\btheta)$ that modifies the estimating equations for the ACC, MACC, and AIPW estimators; see Table \ref{tab:estimators-lambda}. 

Whether we use $\bLambda$ or not, we still need to calculate the augmented component $\bPsi (y,\bz; \btheta)$, which is not straightforward since it involves computing at least one expectation (ratio of two expectations for the right-censored covariate problem). We propose using $\bPsi_{\rm close}(y,\bz;\btheta) = E_{X|Y,\bZ}\{ \bS_\btheta^F(y,X,\bz;\btheta) \}$ to define $\bPsi_{\rm updated}(y,\bz;\btheta) = \bLambda \bPsi_{\rm close}(y,\bz;\btheta)$ across all augmented estimators. This choice has a closed form when $X$ is linear in $m(\cdot)$ and when we assume $X$ follows a normal distribution (shown in Section   \ref{sec:closed-form}). We are allowed to make this choice since we prove it does not affect the consistency of the augmented estimators  (Sections \ref{sec:thm1-proof-part1} and \ref{sec:thm1-proof-part2}).



\begin{table}[!th]
\caption{Changes to estimating equations for augmented estimators that guarantee increased efficiency used in the right-censored covariate problem when censoring is noninformative (i.e., independent censoring) and in the missing covariate problem when missingness is at random (i.e., independent missingness).}
\small
\scalebox{0.9}{
 \begin{tabular}{>{\raggedright}p{1.05\linewidth}c}
 \hline
\makecell{\uline{Updating the augmented components $\bPsi(y,\bz;\btheta)$ of Table \ref{tab:estimators}}}\\ [0.3em]
 \makecell{$\bPsi_{\rm updated}(y,\bz; \btheta)  = \bLambda \bPsi(y,\bz; \btheta)$, where $\bLambda$ is as follows:}  \\[1em]
\makecell{\uline{Augmented complete case (ACC) estimator}}\\ [-0.8em]
 \makecell{\footnotesize $\bLambda_{\rm ACC,eff}^{\rm cens,ind} = -E [ \{\Delta - \highlight{\pi_{Y,\bZ}^{\rm cens, ind}(Y,\bZ)}  \}  \bPhi_{\rm CC}^{\rm cens,ind}(\bO^{\rm cens}; \btheta) \highlight{\bPsi(Y,\bZ;\btheta)^T} ] E ( [ \{\Delta - \highlight{\pi_{Y,\bZ}^{\rm cens, ind}(Y,\bZ)}  \} \highlight{\bPsi(Y,\bZ;\btheta)^T} ]^{\otimes 2} )^{-1}$} \\ [-0.8em]
 \makecell{\footnotesize $\bLambda_{\rm ACC,eff}^{\rm miss,ind} = -E [ \{R - \pi_{\bZ}^{\rm miss, ind}(\bZ)  \} \bPhi_{\rm CC}^{\rm miss,ind}(\bO^{\rm miss}; \btheta) \bPsi(Y,\bZ;\btheta)^T ] E ( [ \{R - \pi_{\bZ}^{\rm miss, ind}(\bZ)  \} \bPsi(Y,\bZ;\btheta)^T ]^{\otimes 2} )^{-1}$
} \\ [0.5em]
\makecell{\uline{Modified augmented complete case estimator}} \\ [-0.8em]
\makecell{\footnotesize $\bLambda_{\rm MACC,eff}^{\rm cens,ind} = -E [ \{1 - \Delta/ \highlight{\pi_{X,\bZ}^{\rm cens, ind}(W,\bZ)}  \}\bPhi_{\rm CC}^{\rm cens,ind}(\bO^{\rm cens}; \btheta) \highlight{\bPsi(Y,\bZ;\btheta)^T} ] E ( [ \{1-\Delta/\highlight{\pi_{X,\bZ}^{\rm cens, ind}(X,\bZ)}  \} \highlight{\bPsi(Y,\bZ;\btheta)^T} ]^{\otimes 2} )^{-1}$}\\ [-1em]
\makecell{\footnotesize $\bLambda_{\rm MACC,eff}^{\rm miss,ind} = -E [ \{1 - R/\pi_{\bZ}^{\rm miss, ind}(\bZ)  \}  \bPhi_{\rm CC}^{\rm miss,ind}(\bO^{\rm miss}; \btheta) \bPsi(Y,\bZ;\btheta)^T ] E ( [ \{1 - R/\pi_{\bZ}^{\rm miss, ind}(\bZ)  \} \bPsi(Y,\bZ;\btheta)^T ]^{\otimes 2} )^{-1}$} \\ [0.5em]
\makecell{\uline{Augmented IPW estimator}}\\ [-0.8em]
 \makecell{\footnotesize $\bLambda_{\rm AIPW,eff}^{\rm cens,ind} = -E [ \{1 - \Delta/ \highlight{\pi_{X,\bZ}^{\rm cens, ind}(W,\bZ)}  \} \highlight{\bPhi_{\rm IPW}^{\rm cens,ind}(\bO^{\rm cens}; \btheta) \bPsi(Y,\bZ;\btheta)^T} ] E ( [ \{1-\Delta/\highlight{\pi_{X,\bZ}^{\rm cens, ind}(W,\bZ)}  \} \highlight{\bPsi(Y,\bZ;\btheta)^T} ]^{\otimes 2} )^{-1}$} \\ [-1em]
\makecell{\footnotesize $\bLambda_{\rm AIPW,eff}^{\rm miss,ind} = -E [ \{1 - R/\pi_{\bZ}^{\rm miss, ind}(\bZ)  \} \bPhi_{\rm IPW}^{\rm miss,ind}(\bO^{\rm miss}; \btheta) \bPsi(Y,\bZ;\btheta)^T ] E ( [ \{1 - R/\pi_{\bZ}^{\rm miss, ind}(\bZ)  \} \bPsi(Y,\bZ;\btheta)^T ]^{\otimes 2} )^{-1}$} & \\ [0.8in]
\hline
\multicolumn{2}{p{\linewidth}}{
{\footnotesize  \textbf{Note:} While the form of $\bPsi(y,\bz;\btheta)$ can be that of Table \ref{tab:estimators}, any function of the same length as $\btheta$ and only a function of $(Y,\bZ; \btheta)$ can be used to define $\bPsi_{\rm updated}(y,\bz;\btheta)$.  Major differences from the missing covariate estimating equations are highlighted in $\highlight{\rm gray}$. Definition of notation and terms are as in Table \ref{tab:estimators}.
}} \\ 
\hline

  \end{tabular}
  }
  \label{tab:estimators-lambda-independent}
  \end{table}

\section{Adjustment of estimators to handle dependent right-censoring and dependent missingness}
\label{sec:dependence}

\subsection{Dependency assumptions}
\label{sec:dependence-assumptions}

%
Until now, we have assumed independent covariate right-censoring (i.e., $X\independent C|\bZ$), also known as noninformative covariate right-censoring. This assumption is violated when the event rate differs between individuals whose age at diagnosis is observed and those whose age is right-censored. For instance, in Huntington disease, this scenario could occur if the event rate is higher among individuals whose age is right-censored, who may possess an unknown predisposition (associated with $X$) that is not present among individuals whose age at diagnosis was observed. Ignoring informative covariate right-censoring may result in incorrectly estimating how impairments change  as a function of time to diagnosis. Establishing estimators that adjust for informative covariate right-censoring (i.e., $X\notindependent C|\bZ$) will help avoid this incorrect estimation in Huntington disease studies.

The analog to informative covariate right-censoring in the missing covariate problem is missingness not at random (i.e., $X\notindependent R|\bZ$). In this type of missingness, there is a pattern to the missingness that depends on $X$; in our example, those with the unknown predisposition (associated with $X$) are the ones whose data are more likely to be missing.  Whether the covariate censoring is informative or the missingness is not at random, there is dependency  between $X$ and  $(C,R)$ that is not explained by the observed data. We will therefore refer to these assumptions as dependent censoring and dependent missingness. The superscript ``dep'' will be used to denote when a estimator uses these dependency assumptions. Moreover, we clarify that the dependency is with $X$ \textit{and not with $Y$}, meaning that the right-censoring and missingness remains conditionally independent of the symptoms given time at diagnosis and the set of the fully observed covariates, i.e.,  $(C,R) \independent Y | (X,\bZ)$.

\subsection{Impact of dependency on the distinction between right-censoring and missingness}
\label{sec:dependence-impact}

In both the right-censored and missing covariate  problems, an underlying mechanism that depends on $X$ explains why $X$ is not observed. What distinguishes the right-censored covariate problem from the missing covariate problem is the partial information about $X$ (i.e., when $X$ is right-censored, its true value is larger than $C$). 
Under the dependent censoring assumption, the probability that $X$ is observed when it is subject to right-censoring  is 
\bse
\pi_{X,\bZ}^{\rm cens, dep}(x,\bz)&\equiv&\pr(\Delta=1|X=x,\bZ=\bz)=\int_{x<c}f_{C|X,\bZ}(c,x,\bz)dc.
\ese
Under the dependent missingness assumption, the probability that $X$ is observed is
\bse
  \pi_{X,\bZ}^{\rm miss, dep}(x,\bz)&\equiv&\pr(R=1|X=x,\bZ=\bz). 
\ese

When we assumed independent censoring and missingness, the probability that $X$ is known was only a function of $\bZ$ in the missing covariate problem and a function of $(X,\bZ)$ in the right-censored covariate problem. Now, under dependent censoring and missingness, the probability that $X$ is known is a function of $(X,\bZ)$ in both problems. This change occurs because both problems now need to account for the dependency between $X$ and $(C,R)$. Thus, under dependent censoring and missingness problems, the probability $X$ is observed is the same. However, the partial information provided by the right-censored covariate problem provides more information, reducing the number of unverifiable assumptions required to estimate these probabilities. In Section \ref{sec:implement}, we explain what assumptions are needed to correctly model $f_{C|X,\bZ}$ and $f_{R|X,\bZ}$, i.e., the distributions needed to estimate these probabilities. 

\subsection{Impact of dependency on the estimators}
\label{sec:dependence-methods}

\subsubsection{CC estimator}
\label{sec:dependence-cc}

%
The estimating equations for the CC estimator rely only on the distribution $f_{Y|X,\bZ}$, so the CC estimator stays the same whether $X$ is right-censored or missing, and whether the covariate right-censoring and missingness is independent or dependent.

\subsubsection{IPW estimator}
\label{sec:dependence-ipw}

%
The probability in the IPW estimator changes to  $\pi_{X,\bZ}^{\rm cens, dep}$ in the right-censored covariate problem and to $\pi_{X,\bZ}^{\rm miss, dep}$ in the missing covariate problem (Table \ref{tab:estimators-dependent}). 

\begin{table}[!t]
\caption{Estimating equations used in the right-censored covariate problem when censoring is informative (i.e., dependent censoring) and in the missing covariate problem when missingness is not at random (i.e., dependent missingness).}
\small
\scalebox{0.89}{
 \begin{tabular}{>{\raggedright}p{1.05\linewidth}c}
\hline
\makecell{\uline{Inverse probability weighting (IPW) estimator}}\\
\makecell{ \footnotesize $\sumi \bPhi_{\rm IPW}^{\rm cens, dep}(\bO_i^{\rm cens}; \btheta) \equiv \sumi \delta_i \bS_{\btheta}^F (y_i, w_i, \bz_i)/\pi_{X,\bZ}^{\rm cens, dep}(w_i,\bz_i) = \bzero
$}  & \\ [2em]
\makecell{ \footnotesize $\sumi \bPhi_{\rm IPW}^{\rm miss, dep}(\bO_i^{\rm miss}; \btheta) \equiv \sumi r_i \bS_{\btheta}^F (y_i, x_i, \bz_i)/\pi_{X,\bZ}^{\rm miss, dep}(x_i,\bz_i) = \bzero$} &\\[1em]
\makecell{\uline{Maximum likelihood estimator (MLE)}}\\ [-0.8em]
\makecell{ \footnotesize $\sumi \bPhi_{\rm MLE}^{\rm cens, dep}(\bO_i^{\rm cens};\btheta) \equiv \sum_{i=1}^n  \biggr[ \delta_i\bS_\btheta^F(y_i,w_i,\bz_i) + (1- \delta_i) \frac{\partial}{\partial \btheta^T}  \log \biggr\{ \highlight{\int_{w_i < x}} f_{Y|X,\bZ} (y_i, x, \bz_i; \btheta) f_{X|C,\bZ} (x,w_i,\bz_i) dx \biggr\}   \biggr]=\0$} &\\ [-1.5em]
\makecell{ \footnotesize $\sumi \bPhi_{\rm MLE}^{\rm miss, dep}(\bO_i^{\rm miss};\btheta) \equiv \sum_{i=1}^n  \biggr[ r_i\bS_\btheta^F(y_i,x_i,\bz_i)  + (1- r_i) \frac{\partial}{\partial \btheta^T}  \log \biggr\{ \int_{\D(x)} f_{Y|X,\bZ} (y_i, x, \bz_i; \btheta) f_{X|R,\bZ} (x,r_i=0,\bz_i) dx \biggr\}   \biggr]=\0$} &\\[1em]
\makecell{\uline{Augmented complete case (ACC) estimator}}\\
 \makecell{ \footnotesize $\sumi \bPhi_{\rm ACC}^{\rm cens,dep}(\bO_i^{\rm cens}; \btheta) \equiv \sum_{i=1}^n \left[ \delta_i \bS_{\btheta}^F(y,w_i,\bz_i)+ \{\delta_i- \pi_{Y,\bZ}^{\rm cens,dep}(y_i, \bz_i) \} \bPsi_{\rm ACC, eff}^{\rm cens,dep}(y_i,\bz_i; \btheta)  \right] = \bzero$} &\\ [2em]
 \makecell{ \footnotesize $\sumi \bPhi_{\rm ACC}^{\rm miss, dep}(\bO_i^{\rm miss};\btheta)\equiv \sum_{i=1}^n \left[ r_i \bS_{\btheta}^F(y,x_i,\bz_i) + \{r_i- \pi_{Y,\bZ}^{\rm miss,dep}(y_i, \bz_i) \} \bPsi_{\rm ACC,eff}^{\rm miss,ind}(y_i,\bz_i; \btheta)  \right] = \bzero$
}&\\[1em]
\makecell{\uline{Modified augmented complete case (MACC) estimator}}\\
\makecell{ \footnotesize $
\sumi \bPhi_{\rm MACC}^{\rm cens,dep}(\bO_i^{\rm cens};\btheta) \equiv \sum_{i=1}^n \left[ \delta_i \bS_{\btheta}^F(y,w_i,\bz_i) +  \{1- \delta_i/\pi_{X,\bZ}^{\rm cens,dep}(w_i, \bz_i) \} \bPsi_{\rm MACC,eff}^{\rm cens,dep}(y_i,\bz_i; \btheta)  \right] = \bzero$}&\\ 
\makecell{ \footnotesize $\sumi \bPhi_{\rm MACC}^{\rm miss,dep}(\bO_i^{\rm miss};\btheta) \equiv \sum_{i=1}^n \left[ r_i \bS_{\btheta}^F(y,x_i,\bz_i) +\{1- r_i/\pi_{X,\bZ}^{\rm miss,dep}(x_i,\bz_i) \} \bPsi_{\rm MACC,eff}^{\rm miss,dep}(y_i,\bz_i; \btheta)  \right] = \bzero$} &\\[1em]
\makecell{\uline{Augmented inverse probability weighting (AIPW) estimator}}&\\  [-0.8em]
 \makecell{ \footnotesize $\sumi\bPhi_{\rm AIPW}^{\rm cens,dep}(\bO_i^{\rm cens};\btheta)\equiv\sum_{i=1}^n \left[ \delta_i\bS_\btheta^F(y_i,w_i,\bz_i) / \pi_{X,\bZ}^{\rm cens,dep}(w_i,\bz_i) + \{1- \delta_i/\pi_{X,\bZ}^{\rm cens,dep}(w_i, \bz_i) \} \bPsi_{\rm AIPW,eff}^{\rm cens,dep}(y_i,\bz_i; \btheta) \right] = \bzero
$}&\\ [-1.8em]
\makecell{ \footnotesize $\sumi\bPhi_{\rm AIPW}^{\rm miss,dep}(\bO_i^{\rm miss};\btheta)\equiv\sum_{i=1}^n \left[ r_i\bS_\btheta^F(y_i,x_i,\bz_i)/\pi_{X,\bZ}^{\rm miss,dep}(x_i,\bz_i) + \{1- r_i/\pi_{X,\bZ}^{\rm miss,dep}( x_i,\bz_i)\} \bPsi_{\rm AIPW,eff}^{\rm miss,dep}(y_i,\bz_i; \btheta) \right] = \bzero
$}&\\
  \hline

\multicolumn{2}{p{\linewidth}}{
{\footnotesize \textbf{Note:} 
%
dep: dependent censoring (i.e., $X\notindependent C|\bZ$) for the right-censored covariate problem and dependent missingness (i.e., $X\notindependent R|\bZ$) for the missing covariate problem; definitions of cens, miss, and $\bS_{\btheta}^F(y,x,\bz)$ are as defined in Table \ref{tab:estimators}. Major differences from the missing covariate estimating equations are highlighted in $\highlight{\rm gray}$. Specific terms are defined below. 
}}\\

\multicolumn{1}{p{\linewidth}}{
{\footnotesize 
$\pi_{X,\bZ}^{\rm cens, dep}(x,\bz)\equiv \pr(\Delta=1|X=x,\bZ=\bz)=\int_{x<c}f_{C|X,\bZ}(c,x,\bz)dc$, 
$\pi_{X,\bZ}^{\rm miss, dep}(x,\bz)\equiv \pr(R=1|X=x, \bZ=\bz)$;  
}}\\ 
\multicolumn{1}{p{\linewidth}}{
{\footnotesize 
$\pi_{Y,\bZ}^{\rm cens, dep}(y,\bz)\equiv \pr(\Delta=1|Y=y,\bZ=\bz)$, 
$\pi_{Y,\bZ}^{\rm miss, dep}(y,\bz)\equiv \pr(R=1|Y=y,\bZ=\bz)$;  
}}\\ 
\multicolumn{2}{p{\linewidth}}{
{\footnotesize 
$\bPsi_{\rm ACC,eff}^{\rm cens,dep}(y, \bz
; \btheta)= -E_{X|Y,\bZ, \Delta=1} \{ \bS_{\btheta}^F(y,X,\bz)\}$, 
$\bPsi_{\rm ACC, eff}^{\rm  miss,dep}(y, \bz; \btheta)= 
-E_{X|Y,\bZ, R=1} \{ \bS_{\btheta}^F(y,X,\bz)\}$;  
}}\\
\multicolumn{2}{p{1.05\linewidth}}{
{\footnotesize 
$\bPsi_{\rm MACC,eff}^{\rm cens,dep}(y, \bz
; \btheta)= \frac{E_{X|Y,\bZ}[\{\pi_{X, \bZ}^{\rm cens,dep}(X,\bz)-1\} \bS^F_{\btheta} (y,X,\bz)]}{ E_{X|Y,\bZ}\{1-1/\pi_{X,\bZ}^{\rm cens,dep}(X,\bz)}\}$,
$\bPsi_{\rm MACC,eff}^{\rm miss,dep}(y, \bz
; \btheta)= \frac{E_{X|Y,\bZ}[\{\pi_{X,\bZ}^{\rm miss,dep}(X,\bz)-1\} \bS^F_{\btheta} (y,X,\bz)]}{ E_{X|Y,\bZ}\{1-1/\pi_{X,\bZ}^{\rm miss,dep}(X,\bz)\}}$;  
}}\\
%
\multicolumn{2}{p{1.05\linewidth}}{
{\footnotesize 
$\bPsi_{\rm AIPW,eff}^{\rm cens,dep}(y, \bz
; \btheta) = \frac{E_{X|Y,\bZ}[\{1-1/\pi_{X,\bZ}^{\rm cens,dep}(X,\bz)\} \bS^F_{\btheta} (y,X,\bz)]}{ E_{X|Y,\bZ}\{1 - 1/\pi_{X,\bZ}^{\rm cens,dep}(X,\bz) \} }$,
$\bPsi_{\rm AIPW,eff}^{\rm miss,dep}(y, \bz
; \btheta) = \frac{E_{X|Y,\bZ}[\{1-1/\pi_{X,\bZ}^{\rm miss,dep}(X,\bz)\} \bS^F_{\btheta} (y,X,\bz)]}{ E_{X|Y,\bZ}\{1-1/\pi_{X,\bZ}^{\rm miss,ind}(X,\bz) \} }$.  
}}\\
\hline
  \end{tabular}
  }
  \label{tab:estimators-dependent}
  \vspace{-1em}
  \end{table}

\subsubsection{MLE}
\label{sec:dependence-mle}

The MLE is affected by assumptions of  dependent covariate right-censoring/missingness because the specification of the likelihood changes in two main ways. 
The first is that the estimating equations now depend on $f_{X|C,\bZ}$ when assuming dependent right-censoring, and on $f_{X|R,\bZ}$ when assuming dependent missingness. Estimating these two distributions requires modeling the dependency between variables that are not simultaneously observed. That estimation is difficult since the two distributions may not be identifiable, unless certain assumptions are made (see Section \ref{sec:implement}). 

The second is that the domains of the integral with respect to $X$ vary between the right-censored covariate problem and the missing covariate problem. The domain for the right-censored covariate problem is the same as when we assumed independent covariate right-censoring: the domain is all $X$  greater than $C$. In the missing covariate problem, the domain of integration, denoted by $\D(X)$, varies. With dependent missingness, there may be times where the dependency between $X$ and $R$ informs a specific region $\D(X)$, such as all $X$  higher or lower than a certain cutoff. However, more often than not, the integration domain $\D(X)$ is not known. For example, $X$ can have a non-monotone missingness pattern that makes it difficult to discern the exact domain $\D(X)$  \parencite{Little1992}. In these settings, it may be impossible to compute the MLE , as its computation depends on unverifiable information.
 
\subsubsection{Augmented estimators}
\label{sec:dependence-acc}

For the ACC estimator,  we still use the probability $\pi_{Y,\bZ}^{\rm cens, dep}(y,\bz)$ in the right-censored covariate problem, but now use  $ \pi_{Y,\bZ}^{\rm miss, dep}(y,\bz)$ for the missing covariate problem, which does not simplify to $\pi_{\bZ}^{\rm miss, ind}(\bz)$. For the MACC and AIPW estimators, we now use $\pi_{X,\bZ}^{\rm cens, dep}(x,\bz)$ and $ \pi_{X,\bZ}^{\rm miss, dep}(x,\bz)$. After applying these changes, the estimating equations for the ACC, MACC, and AIPW estimators under dependent right-censoring and missingness leads to identical estimating equations by interchanging $(W,\Delta)$ with $(XR, R)$ (Table \ref{tab:estimators-dependent}). 
Therefore, applying the augmented estimators for the dependent missing covariate problem directly to the right-censored covariate problem lead to consistent results. Still, the partial information provided by the right-censored covariate requires fewer unverifiable assumptions.

In Sections \ref{sec:thm1-proof-part1-efficiency} and \ref{sec:thm1-proof-part2-efficiency}, we derive the augmented components the augmented estimators under dependent right-censoring, which match the representations found in \textcite{Bartlettetal2014} for the ACC estimator and \textcite{rotnitzky1997analysis} for the AIPW estimator. To our knowledge, the MACC had not been previously proposed, making it a new addition to the  estimators handling covariates under dependent right-censoring and missingness. 

\subsubsection{Guaranteeing statistical efficiency and reducing computation cost}
\label{sec:efficiency-computational-augmented-estimators-dependent}

%
Just as with independent covariate right-censoring and missingness, we can guarantee higher efficiency of the augmented estimators by updating the augmented component using a fixed matrix $\bLambda$, i.e., $\bPsi_{\rm updated}(y,\bz;\btheta) = \bLambda \bPsi(y,\bz;\btheta)$ (Table \ref{tab:estimators-lambda-dependent}). We propose using $\bPsi_{\rm close}(y,\bz;\btheta) = E_{X|Y,\bZ}\{ \bS_\btheta^F(y,X,\bz;\btheta) \}$ to define $\bPsi_{\rm updated}(y,\bz;\btheta) = \bLambda \bPsi_{\rm close}(y,\bz;\btheta)$ across all augmented estimators. This choice grants us a closed form solution that reduces the computational burden associated with the estimation process and ensures the augmented estimators remain consistent---a result we prove in Theorem \ref{thm:all-robust-consistent}.

\begin{table}[!th]
\caption{Changes to estimating equations for augmented estimators that guarantee increased efficiency used in the right-censored covariate problem when right-censoring is informative (i.e., dependent censoring) and in the missing covariate problem when missingness is not at random (i.e., dependent missingness).}
\small
\scalebox{0.9}{
 \begin{tabular}{>{\raggedright}p{1.05\linewidth}c}
 \hline
\makecell{\uline{Updating the augmented components $\bPsi(y,\bz;\btheta)$ of Table \ref{tab:estimators-dependent}}}\\ 
 \makecell{ $\bPsi_{\rm updated}(y,\bz; \btheta)  = \bLambda \bPsi(y,\bz; \btheta)$, where $\bLambda$ is as follows:}  \\[1em]
\makecell{\uline{Augmented complete case (ACC) estimator}}\\ [-1em]
 \makecell{ $\bLambda_{\rm ACC,eff}^{\rm cens,dep} = -E [ \{\Delta - {\pi_{Y,\bZ}^{\rm cens, dep}(Y,\bZ)}  \}  \bPhi_{\rm CC}^{\rm cens,ind}(\bO^{\rm cens}; \btheta) {\bPsi(Y,\bZ;\btheta)^T} ] E ( [ \{\Delta - {\pi_{Y,\bZ}^{\rm cens, dep}(Y,\bZ)}  \} {\bPsi(Y,\bZ;\btheta)^T} ]^{\otimes 2} )^{-1}$} \\ [-2em]
 \makecell{$\bLambda_{\rm ACC,eff}^{\rm miss,dep} = -E [ \{R - \pi_{Y,\bZ}^{\rm miss, dep}(Y,\bZ)  \} \bPhi_{\rm CC}^{\rm miss,ind}(\bO^{\rm miss}; \btheta) \bPsi(Y,\bZ;\btheta)^T ] E ( [ \{R - \pi_{Y,\bZ}^{\rm miss, dep}(Y,\bZ)  \} \bPsi(Y,\bZ;\btheta)^T ]^{\otimes 2} )^{-1}$
} \\ 
\makecell{\uline{Modified augmented complete case (MACC) estimator}} \\ [-1em]
\makecell{$\bLambda_{\rm MACC,eff}^{\rm cens,dep} = -E [ \{1 - \Delta/ {\pi_{X,\bZ}^{\rm cens, dep}(W,\bZ)}  \}\bPhi_{\rm CC}^{\rm cens,ind}(\bO^{\rm cens}; \btheta) {\bPsi(Y,\bZ;\btheta)^T} ] E ( [ \{1-\Delta/{\pi_{X,\bZ}^{\rm cens, dep}(W,\bZ)}  \} {\bPsi(Y,\bZ;\btheta)^T} ]^{\otimes 2} )^{-1}$}\\ [-2em]
\makecell{$\bLambda_{\rm MACC,eff}^{\rm miss,dep} = -E [ \{1 - R/\pi_{X,\bZ}^{\rm miss, dep}(X,\bZ)  \}  \bPhi_{\rm CC}^{\rm miss,ind}(\bO^{\rm miss}; \btheta) \bPsi(Y,\bZ;\btheta)^T ] E ( [ \{1 - R/\pi_{X,\bZ}^{\rm miss, dep}(X,\bZ)  \} \bPsi(Y,\bZ;\btheta)^T ]^{\otimes 2} )^{-1}$} \\ 
\makecell{\uline{Augmented inverse probability weighting (AIPW) estimator}}\\ [-1em]
 \makecell{$\bLambda_{\rm AIPW,eff}^{\rm cens,dep} = -E [ \{1 - \Delta/ {\pi_{X,\bZ}^{\rm cens, dep}(W,\bZ)}  \} {\bPhi_{\rm IPW}^{\rm cens,dep}(\bO^{\rm cens}; \btheta) \bPsi(Y,\bZ;\btheta)^T} ] E ( [ \{1-\Delta/{\pi_{X,\bZ}^{\rm cens, dep}(W,\bZ)}  \} {\bPsi(Y,\bZ;\btheta)^T} ]^{\otimes 2} )^{-1}$} \\ [-2em]
\makecell{$\bLambda_{\rm AIPW,eff}^{\rm miss,dep} = -E [ \{1 - R/\pi_{X,\bZ}^{\rm miss, dep}(X,\bZ)  \} \bPhi_{\rm IPW}^{\rm miss,dep}(\bO^{\rm miss}; \btheta) \bPsi(Y,\bZ;\btheta)^T ] E ( [ \{1 - R/\pi_{X,\bZ}^{\rm miss, dep}(X,\bZ)  \} \bPsi(Y,\bZ;\btheta)^T ]^{\otimes 2} )^{-1}$}&\\ [-0.5em]
\hline
\multicolumn{2}{p{\linewidth}}{
{\footnotesize  \textbf{Note:} While the form of $\bPsi(y,\bz;\btheta)$ can be that of Table \ref{tab:estimators-dependent}, any function of the same length as $\btheta$ and only a function of $(Y,\bZ;\btheta)$ can be used to define $\bPsi_{\rm updated}(y,\bz;\btheta)$. Definition of notation and terms are as in Table \ref{tab:estimators-dependent}.  Aside from minor notation differences, there are no major differences between the right-censored and missing covariate problems.
}} \\ 
\hline

  \end{tabular}
  }
  \label{tab:estimators-lambda-dependent}
  \end{table}

\vskip -3em
\section{Implementing the estimators}
\label{sec:implement}


Implementing the IPW, MLE, ACC, MACC, and AIPW estimators require specification of various distributions prior to estimating $\btheta$: $f_{X,C|\bZ}$ and $f_{\Delta|Y,\bZ}$ for the right-censored covariate problem  and $f_{X,R|\bZ}$ and $f_{R|Y,\bZ}$ for the  missing covariate problem, respectively. For pedagogical purposes, we assume that the form of these distributions (e.g., $f_{X,C|\bZ} \sim$ bivariate normal) are known but the set of finite-dimensional parameters indexing these distributions are unknown. We refer to the parameters of these distributions as \textit{nuisance parameters} since they are not of main interest, but knowing them is needed to estimate $\btheta$. There are two approaches to obtaining these nuisance parameters: a mechanistic approach in which the analyst \emph{specifies} them, and an empirical approach in which the analyst \emph{estimates} them. 

\subsection{Mechanistic approach to nuisance parameters}

A mechanistic approach fully specifies the finite-dimensional parameters of the nuisance distributions (i.e., the nuisance parameters).  For example, if $f_{X,C|\bZ}$ (similar approach with $f_{X,R|\bZ}$) follows a bivariate normal distribution, the mean and covariance parameters are explicitly specified. If $f_{\Delta|Y,\bZ}$  (and $f_{R|Y,\bZ}$) follows a logistic regression model, its parameters are specified. A mechanistic approach is convenient. It reduces the work needed to implement the estimators and simplifies the asymptotic variance for $\btheta$, as it avoids the additional variability introduced by estimating the nuisance parameters. Still, correctly specifying a nuisance distribution and its parameters is not easy; even after decades of research to find and validate a model, the model may not be completely accurate. In Huntington disease, extensive research has focused on developing alternative measures to time to diagnosis for use in risk-group classifications. Yet, \textcite{Zhangetal2011} cautioned researchers against over-interpreting these alternative measures and using them as substitute measurements for the true time to diagnosis. 

\subsection{Empirical approach to nuisance paramaters} 

Let $\balpha$ index the distribution $f_{X,C|\bZ}$.  Ensuring $\balpha$ is identifiable requires, for example, that the covariance of $(X,C)$ given $\bZ$ is known and $f_{X,C|\bZ}$ is sufficiently smooth \parencite{ebrahimi2003identifiability}. This covariance is zero for the case of independent covariate right-censoring,  but not zero under covariate dependent right-censoring. When $\balpha$ is identifiable, an empirical approach to estimating $\balpha$ is to maximize the log-likelihood of $(W,\Delta)$  given $\bZ$:
\bse
\wh\balpha = \underset{\balpha}{\mathrm{argmax}} \sumi\left\{\delta_i\log\int_{w_i}^{\infty}f_{C,X|\bZ}(c,x,\bz;\balpha)dc + (1-\delta_i)\log\int_{w_i}^{\infty}f_{C,X|\bZ}(c,x,\bz;\balpha)dx \right\}.
\ese
The resulting $\wh\balpha$ is a  $\sqrt{n}$-consistent estimator (Section \ref{sec:thm1-proof-part1}). 


%
A typical model for $f_{\Delta|Y,\bZ} (\delta=1, y,\bz) \equiv \pr(\Delta=1|y,\bz)$ (or $\pr(R=1|y,\bz)$) is a generalized linear model with $\Delta$ (or $R$) as the outcome and $(y,\bz)$ as covariates, such as a logistic regression model with mean  $\mu(y,\bz;\btau)$. An empirical approach to estimating $\btau$ is using any software that implements logistic regression. 

Let $\bxi$ index the distribution $f_{XR,R|\bZ}$. Under dependent missingness,  $\bxi$ is non-identifiable based solely on data $(XR,R,\bZ)$; additional assumptions are needed. For example, $\bxi$ is identifiable if  the conditional covariance of $X$ and $R$ given $\bZ$ is known,  and the domain of integration $\D(x)$ is known. These are quite strong assumptions, however. When $X$ is missing, positing a correct form of the conditional covariance of $X$ and $R$ given $\bZ$ is nearly impossible. Also, knowing the correct domain of integration $\D(x)$ (i.e., the possible missing domain) is difficult when there is no particular pattern to the missingness.  

With empirical approaches, the asymptotic variance of $\btheta$ must account for the added variability from estimating nuisance parameters. Our proof of Theorem \ref{thm:all-robust-consistent} (Sections \ref{sec:thm1-proof-part1} and \ref{sec:thm1-proof-part2}) shows the necessary adjustments to include this uncertainty.

\section{Robustness, efficiency, and asymptotic  properties of the estimators}
\label{sec:summary}

\begin{figure}[!ht]
\caption{Can the nuisance distribution be misspecified?}
    \centering
    \label{fig:nuisance-conditions}
\includegraphics[width=0.95\textwidth]{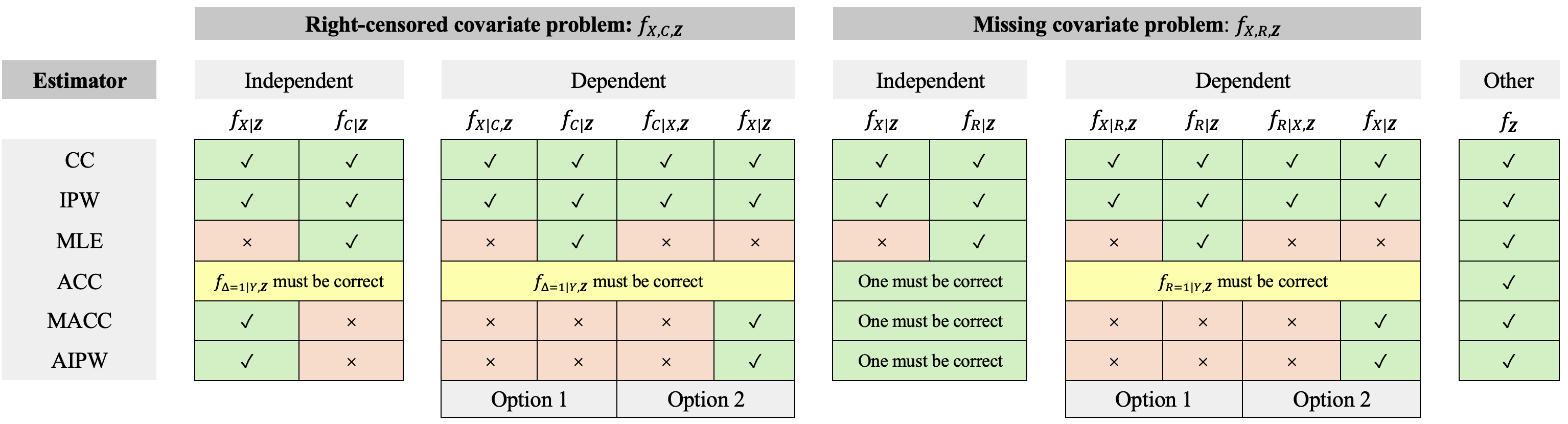}
    \caption*{ \footnotesize \textbf{Note:} ``\checkmark" indicates that the density can misspecified, whereas $\times$ signifies that the density needs to be correctly specified. Under dependent right-censoring and missingness, two options of decomposing $f_{C,X,\bZ}$ and $f_{R,X,\bZ}$ are provided.}
    \vskip -2em
\end{figure}

All estimators are consistent and asymptotically normal in both the right-censored and missing covariate problems (Theorem \ref{thm:all-robust-consistent}). The consistency and asymptotic normality hold even when some of the nuisance distributions are misspecified---a result that means the estimators are robust to misspecification. This robustness is crucial to correctly estimating the progression patterns of Huntington disease impairments. Proofs of the consistency, asymptotic normality, and robustness to misspecification are in Sections  \ref{sec:thm1-proof-part1} and \ref{sec:thm1-proof-part2}.
\begin{Th}
\label{thm:all-robust-consistent} Assume $(C,R) \independent Y | X, \bZ$ for the right-censored and missing covariate problems. Let $\wh{\btheta}$ be the solution to $\sumi \bPhi(\bO_i; \btheta) = \bzero$, where $\bPhi(\cdot)$ is an estimating function corresponding to one of the six estimators evaluated at observed data $\bO_i$. Under regularity conditions stated in \ref{thm1:consistency}, the estimator $\wh\btheta$ is  consistent and  asymptotically normal:
\bse
\sqrt{n}(\wh{\btheta}-\btheta_0) \rightarrow \Normal(\bzero, \bA^{-1}\bB\bA^{\rm -T}),
\ese
where $\bA = E\{\partial \bPhi(\bO;\btheta_0)/\partial\btheta^{\trans}\}$ and $\bB =  E\{\bPhi(\bO;\btheta_0)^{\otimes2}\}$. The  influence function corresponding to $\wh\btheta$ is $\bUpsilon=-\bA^{-1}\bPhi(\bO;\btheta_0)$. The specific forms of $\bA$, $\bB$, and $\bUpsilon$ are listed in 
Sections \ref{sec:thm1-proof-part1} and \ref{sec:thm1-proof-part2}; forms are given for each estimator,  when either a mechanistic or an empirical approach is used to specify or model the nuisance  distributions.  The estimators are shown to yield consistency and asymptotic normality under the misspecifications listed in Figure \ref{fig:nuisance-conditions}. 
\end{Th}

Having established when consistency and asymptotically normality holds, an important next step is determining how the estimators compare in terms of statistical efficiency. 

\begin{Th}
\label{thm:all-efficiency} 
Assume correct specification of distributions required for consistency and asymptotic normality listed in Theorem \ref{thm:all-robust-consistent}. The statistical efficiency of the estimators, from most to least efficient,  is
\bse
\var(\wh{\btheta}_{\text{MLE}}) < 
\begin{array}{c}
\var(\wh{\btheta}_{\text{ACC}}) \\
\var(\wh{\btheta}_{\text{MACC}})
\end{array}
< \var(\wh{\btheta}_{\text{CC}}) \leq \var(\wh{\btheta}_{\text{AIPW}}) < \var(\wh{\btheta}_{\text{IPW}}).
\ese
The order of statistical efficiency is the same for the right-censored covariate problem under dependent censoring and for the missing covariate problem. 
\end{Th}
The proof of Theorem \ref{thm:all-efficiency} is in Section \ref{sec:thm2-proof} and rests on computing the variances of the estimators and establishing the order of increasing variance. Results from Theorems \ref{thm:all-robust-consistent} and \ref{thm:all-efficiency} together highlight the benefits and trade-offs of each estimator. 

Whether the covariate right-censoring and missingness are independent or dependent, the CC and IPW estimators exhibit the highest level of robustness, maintaining consistency regardless of how all distributions, aside from $f_{Y|X,\bZ}$, are specified. That robustness means we can misspecify the probability $\pi_{X,\bZ}$ in the IPW estimator as any function of $(X,\bZ)$ and still achieve a consistent estimator.  This ability to misspecify is especially helpful when the covariate right-censoring and missingness are dependent, as correctly specifying the dependency between $C$ and $X$ (or $R$ and $X$) in  $\pi_{X,\bZ}$ is difficult. 

A few caveats about the IPW estimator are worth noting. When  $(C,R)\independent Y |(X,\bZ)$, we proved in Theorem \ref{thm:all-robust-consistent}  that even if we misspecify the probability $\pi_{X,\bZ}$, the IPW estimator will remain consistent in both the right-censored and missing covariate covariate problems. The consistency of the IPW estimator, however, is up to a point. One may believe that because we are free to misspecify the probability $\pi_{X,\bZ}$, we may also use the probability $\pi_{Y,\bZ}^{\rm cens}$.  Modeling the probability  $\pi_{Y,\bZ}^{\rm cens}$ rather than  $\pi_{X,\bZ}$ is easier because  we always observe the data $(Y,\Delta, \bZ)$, whereas we do not always observe the data $(X,\Delta,\bZ)$. However, we prove in Section \ref{sec:thm1-proof-part1-robustness} that, for the right-censored covariate problem, using   $\pi_{Y,\bZ}^{\rm cens}$  instead of $\pi_{X,\bZ}$ in the IPW estimator will yield an inconsistent estimator regardless of whether the covariate right-censoring is independent or dependent. 

That bias persists in the missing covariate problem when the missingness is dependent, but not when the missingness is independent. When the missingness is independent,  $\pi_{Y,\bZ}^{\rm miss}(y,\bz)$ reduces to $\pi_{\bZ}^{\rm miss,ind}(\bz)$ because $R\independent Y |(X,\bZ)$; with that simplification, the IPW estimator remains consistent (Section \ref{sec:thm1-proof-part2}). That same simplification does not occur when the missingness is dependent, and the resulting IPW estimator with probability $\pi_{Y,\bZ}^{\rm miss}$ is inconsistent.  These results further highlight that when adapting estimators from the missing to the right-censored covariate problem, some properties carry over and others do not.

Still, the high robustness makes the CC and IPW estimators a safe choice, yet both have the lowest efficiency among all estimators. In stark contrast, the MLE is much more efficient, but less robust. When the covariate right-censoring or missingness are independent,  the distribution $f_{X|\bZ}$ must be correctly specified to ensure the MLE yields a  consistent estimator. That requirement persists in the dependent setting, where now the correct specification of $f_{X|C,\bZ}$ for the right-censored covariate problem and of $f_{X|R,\bZ}$ for the missing covariate problem is essential to ensure the MLE produces a consistent estimator. When these distributions are correct, the MLE achieves optimal efficiency (Theorem \ref{thm:all-efficiency}). This optimal efficiency makes the MLE still sought after, even when correctly specifying $f_{X|\bZ}$, $f_{X|C,\bZ}$, and $f_{X|R,\bZ}$   is difficult.

\begin{figure}[!ht]
  \centering
   \caption{Comparing the robustness and efficiency of augmented estimators with and without $\bLambda$ under misspecification. 
   } \label{fig:lambda-comparisons}
 \includegraphics[width=16cm]{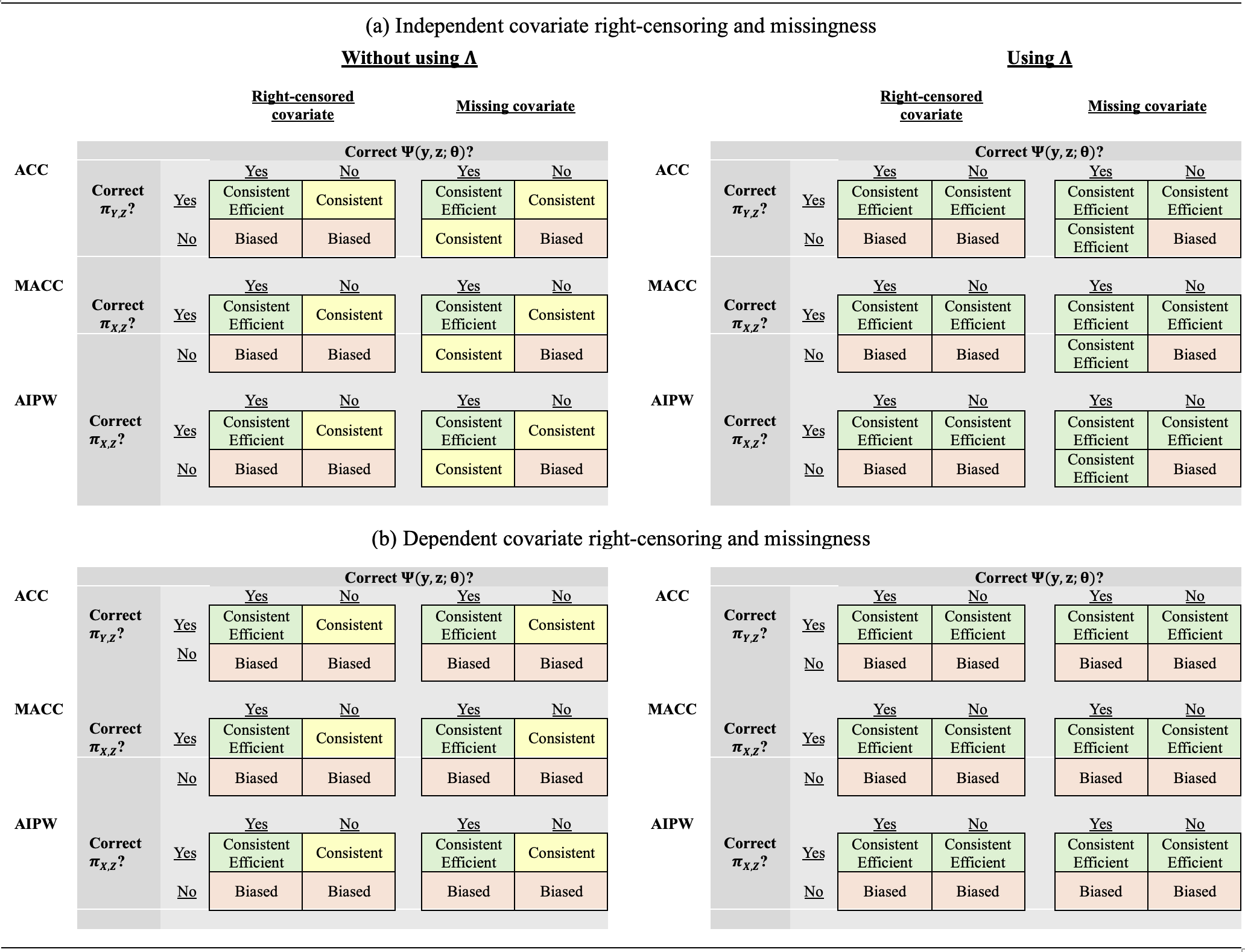}
  \caption*{ \footnotesize {\textbf{Note:} Independent covariate right-censoring and missingness are defined as $X \independent (C,R) |\bZ$, and dependent covariate right-censoring and missingness are defined as $X \notindependent (C,R) |\bZ$. The probabilities $\pi_{Y,\bZ}$ and $\pi_{X,\bZ}$  are reduced to $\pi_{\bZ}$ for the missing covariate problem under independent covariate missingness.  ``Consistent" indicates that the estimator is consistent, ``Efficient" indicates that it obtains higher efficiency when compared to the original estimator, and ``Biased" indicates that it is inconsistent.}}
  \vspace{-3em}
\end{figure}

While none of the augmented estimators is more efficient than the MLE, the ACC and MACC estimators are more efficient than the CC estimator, just as the AIPW is more efficient than the IPW estimator.  The robustness of the augmented estimators differs depending on the assumptions made about covariate right-censoring and missingness.  Typically, augmented estimators are doubly robust, meaning they produce consistent estimators even if the probabilities (i.e., $\pi_{Y,\bZ}$ or $\pi_{X,\bZ}$ ) or the augmented component (i.e.,  $\bPsi(y,\bz)$) are misspecified, though not both. However, the ACC, MACC, and AIPW estimators are \emph{only} doubly robust when $X$ is missing and the covariate missingness is independent. Otherwise, these three estimators are singly robust, meaning they yield a consistent estimator only when the probabilities are correctly specified (i.e., $\pi_{Y,\bZ}$ for ACC and $\pi_{X,\bZ}$ for MACC and AIPW). The single robustness property results from the augmented component $\bPsi (y,\bz; \btheta)$ not always having mean zero, so correctly specifying the probabilities is necessary for consistency. The double or single robustness of the estimators persists with or without using $\bLambda$ in both independent and dependent  covariate right-censoring and missingness (Figure \ref{fig:lambda-comparisons}). This difference between when we have single and double robustness further underscores that augmented estimators designed for the missing covariate problem cannot  be blindly applied to the right-censored covariate problem.

\section{Simulation study}
\label{sec:simulations_paper}

%
We conducted a simulation study to assess the robustness and efficiency of the estimators in the right-censored covariate problem, specifically examining the effects of correctly and incorrectly specifying $f_{X,C|\bZ}$. We did not study the impact of correctly and incorrectly specifying $f_{X,R|Z}$ in the missing covariate problem as those results are well documented; see, for example, 
\textcite{Little1992}. 

\subsection{Data generation and metrics for comparison}
\label{sec:sims_data_paper}

In all $N=3,000$ simulations, we generated data with a sample size of $n = 1,000$ and a censoring rate of $50\%$. Data were generated from the regression model in Equation \eqref{eqn:equation_paper2}, where  $m(X,Z;\btheta) = \beta_0 + \beta_{AX}(A-X) + \beta_ZZ$,  $\epsilon\sim\Normal(0,1)$, and $A\sim\Normal(0,1)$.  The true parameter values were $\beta_0=1$, $\beta_{AX}=3$, and $\beta_Z=2$. We used $(A-X)$ as opposed to $X$ alone to mimic the case when time to diagnosis is of interest; here, $A$ is the current age and $X$ is the age of diagnosis, so $A-X$ is the time to diagnosis. We generated data $(X,C,Z)$ from a trivariate normal distribution where the covariance of $(C,X)$ conditional on $Z$ was zero and not zero to guarantee independent and dependent right-censoring, respectively. 


We applied all estimators, plus two others: the oracle estimator, which estimates $\btheta$ using the full but unobserved data, and the naive estimator, which uses all observations but incorrectly treats $W$ as $X$. The oracle estimator serves as a gold standard for comparison, and the naive estimator illustrates the worst-case scenario. We handled the unknown parameters of $f_{X,C|\bZ}$ in two ways: a mechanistic approach in which the nuisance parameters are specified and an empirical approach in which the nuisance parameters are estimated using maximum likelihood. When the nuisance parameters were specified, all estimators were applied under independent and dependent right-censoring. When the nuisance parameters were estimated, all estimators were considered for the independent right-censoring case and only correctly specified scenarios were evaluated: the probability $\pi_{X,\bZ}(w,\bz)$ was calculated using $f_{C|\bZ}(c,\bz; \wh\boldeta)$, and $f_{X|\bZ}(x,\bz; \wh\bgamma)$ was used for the augmented estimators and the MLE. For dependent right-censoring, $f_{X,C|\bZ}$ is not always identifiable; therefore only estimators that do not require  this distribution were evaluated: oracle, naive, CC, and ACC with $\bLambda$. To showcase that the ACC estimator with $\bLambda$ in the dependent missing covariate problem can be used for the right-censored covariate problem, we modified the simulation study from  \textcite{Bartlettetal2014} by letting $C = X - beta\{\exp(Z), 1\}$ when $X$ was not observed. We lay out more explicit details of our simulation study in Section \ref{sec:additional_simulations}.

We evaluated all estimators by computing the mean estimate of $\btheta$ (i.e., $N^{-1}\sum_{i=1}^N\wh{\btheta}_i$)  and its percent bias (i.e.,  $N^{-1}\sum_{i=1}^N (\wh{\btheta}_i-\btheta_0)/\btheta_0$); the empirical standard deviation of $\wh\btheta$ across all simulations; and the empirical mean of the estimated standard errors (i.e.,  $N^{-1}\sum_{i=1}^{N} \wh{\rm SE}_i$).  Estimated standard errors were computed using the asymptotic variances derived in our theorems, with all expectations replaced by empirical averages.  Lastly, we calculated the empirical coverage of the estimated 95\% confidence intervals.

\subsection{Simulation results}

The majority of the bias was present in the intercept coefficient, $\beta_0$,  when bias was expected, as illustrated in Figures \ref{fig:indepdent-boxplot} and \ref{fig:depdent-boxplot}. Therefore, our discussion focuses on results for $\beta_0$ under independent right-censoring  (Table \ref{tab:independent_known_1000_small}: Known nuisance parameters). For more simulation results, refer to  Tables \ref{tab:independent_known_1000} -  \ref{tab:dependent_estimated_1000} and Figures \ref{fig:indepdent-boxplot} - \ref{fig:dependent-efficiency}. 

All estimators behaved as according to the claims in Theorems \ref{thm:all-robust-consistent} and \ref{thm:all-efficiency} whether $f_{X,C|\bZ}$ was estimated or assumed known. The CC estimator was consistent both when the covariate right-censoring was both independent and dependent. That consistency is preferred over the naive estimator, which is highly biased because it erroneously assumes $W$ is the true $X$ value when it is not. As expected, however, the CC estimator was inefficient: by deleting data, it was up to 43\% less efficient compared to the oracle estimator (i.e., ${\rm (SD_{\rm oracle} - SD_{CC})/SD_{CC}}$). The IPW estimator remained as robust as the CC estimator, but incorrect specification of the probability $\pi_{X,\bZ}$ led to 13\% reduced efficiency (i.e., ${\rm (SD_{\rm IPW, incorrect} - SD_{\rm IPW, correct})/SD_{\rm IPW, correct}}$). Furthermore, the choice of the probability $\pi_{Y,\bZ}$ for the IPW estimator led to bias.

The MLE was unbiased and was the most efficient estimator when the density $f_{X|\bZ}$ ($f_{X|C,\bZ}$ for dependent right-censoring) was correctly specified, but when it was incorrectly specified, the MLE was biased, and its coverage level was far from the 95\% nominal level. These results show that while the MLE may achieve the highest efficiency of all estimators, it is sensitive to misspecification and careful modeling of nuisance distributions is necessary. All augmented estimators were consistent when the probabilities were correctly specified, regardless of whether the augmented component $\bPsi(y,\bz;\btheta)$ was correctly specified. Incorrect specification of the probabilities, however, led to bias even when the augmented component was correctly specified. These results corroborate our theoretical claims in Theorem \ref{thm:all-robust-consistent}, showing that augmented estimators are singly robust in the right-censored covariate problem.

The augmented estimators without $\bLambda$ were only more efficient when both the probability of $X$ being observed and the augmented component were correctly specified. When the augmented component was incorrectly specified but the probability was correctly specified, the augmented estimators were less efficient. For example, compared to the CC estimator, the MACC estimator was about 21\% \emph{less} efficient when only the probability was correctly specified but 3\% \emph{more} efficient when both were correctly specified. The flip-flopping between efficiency loss and gains was corrected when using $\bLambda$; with $\bLambda$, the MACC estimator was 1\% more efficient than the CC estimator even when using the incorrect form of $\bPsi(y,\bz; \btheta)$. Aside from the benefits of guaranteed efficiency gains, 
the MACC estimator with $\bLambda$ was also much more computationally efficient than without $\bLambda$, thanks to the use of the closed-form augmented component, $\bPsi_{\rm updated}(y,\bz;\btheta)$. On an M1 MacBook Air with 16 GB RAM, the MACC estimator with $\bLambda$ required 2 minutes per simulation to converge, while the MACC estimator without $\bLambda$ required up to 2 hours.
Similar results were observed for the ACC and AIPW estimators. These results show that using $\bLambda$ improves computational efficiency since, with it, we may use the closed-form $\bPsi_{\rm updated}(y,\bz;\btheta)$.

Among the three augmented estimators, the ACC estimator with $\bLambda$ was the most straightforward to implement.  It resulted in nearly unbiased and efficient estimates for $\btheta$, with coverage probabilities close to 95\%  for both independent and dependent covariate right-censoring. This strong performance makes the ACC estimator with $\bLambda$ our preferred choice.

\begin{table}[h!]
\caption{\label{tab:independent_known_1000_small}Independent covariate right-censoring with $f_{C,X|\bZ}(c,x,\bz; \balpha)$ assumed known. Simulation results of the mean estimate, mean estimated standard errors (SE) and empirical standard deviations (SD) scaled by 100, and confidence interval coverage  (95\% Cov) for each estimator.}
\centering
\resizebox{0.8\linewidth}{!}{
\begin{tabular}[t]{llrrrrr}
\toprule
\textbf{Estimator} & \textbf{Specification} & \textbf{Estimate} & \textbf{Bias} & \textbf{SE} & \textbf{SD} & \textbf{95\% Cov}  \\
\midrule
\addlinespace[0.3em]
&&\multicolumn{5}{c}{\textbf{Intercept: $\beta_0 = 1$}} \\
\hspace{1em}Oracle & &  1.00 & 0.00 & 3.16 & 3.19 & 94.10 \\
\addlinespace
\hspace{1em}Naive &  & -0.38 & -1.38 & 9.10 & 9.07 & 0.00  \\
\addlinespace
\hspace{1em}CC & &    1.00 & -0.00 & 4.57 & 4.57 & 94.96   \\
\addlinespace
 & correct $\pi_{X,\bZ}(w,\bz)$ &  1.00 & -0.00 & 4.77 & 4.83 & 94.61 \\
\multirow[t]{-2}{*}{\raggedright\arraybackslash \hspace{1em}IPW} 
& incorrect $\pi_{X,\bZ}(w,\bz)$ & 1.00 & -0.00 & 5.50 & 5.54 & 95.07 \\
& correct $\pi_{Y,\bZ}(y,\bz)$ & 0.88 & -0.12 & 4.87 & 4.94 & 29.91  \\
\addlinespace
 & correct $f_{X|\bZ}$ & 1.00 & -0.00 & 4.42 & 4.14 & 94.81   \\
\multirow[t]{-2}{*}{\raggedright\arraybackslash \hspace{1em}MLE} 
& incorrect $f_{X|\bZ}$  &  0.06 & -0.94 & 8.94 & 8.97 & 0.10 \\
\addlinespace
&\multicolumn{5}{l}{\underline{without using $\bLambda$}}\\
\multirow[t]{-2}{*}{\raggedright\arraybackslash \hspace{1em}ACC} 
 & correct $\pi_{X,\bZ}(w,\bz)$ and correct $\bPsi_{\rm ACC}(y,\bz)$&  1.00 & -0.00 & 4.40 & 4.40 & 94.51  \\
& incorrect $\pi_{X,\bZ}(w,\bz)$ and correct $\bPsi_{\rm ACC}(y,\bz)$ & 0.84 & -0.16 & 5.10 & 5.11 & 10.27   \\
 & correct $\pi_{X,\bZ}(w,\bz)$ and incorrect $\bPsi_{\rm ACC}(y,\bz)$  & 1.00 & -0.00 & 5.18 & 5.24 & 94.66  \\
& incorrect $\pi_{X,\bZ}(w,\bz)$ and incorrect $\bPsi_{\rm ACC}(y,\bz)$  &   0.67 & -0.33 & 6.08 & 6.24 & 0.00   \\
&\multicolumn{5}{l}{\underline{using $\bLambda$}}\\
 & correct $\pi_{X,\bZ}(w,\bz)$ &  1.00 & -0.00 & 4.40 & 4.44 & 94.76  \\
& incorrect $\pi_{X,\bZ}(w,\bz)$ & 0.91 & -0.09 & 4.43 & 4.58 & 47.20 \\
\addlinespace
&\multicolumn{5}{l}{\underline{without using $\bLambda$}}\\
\multirow[t]{-2}{*}{\raggedright\arraybackslash \hspace{1em}MACC} 
 & correct $\pi_{X,\bZ}(w,\bz)$ and correct $\bPsi_{\rm MACC}(y,\bz)$&  1.00 & -0.00 & 4.43 & 4.43 & 95.02  \\
& incorrect $\pi_{X,\bZ}(w,\bz)$ and correct $\bPsi_{\rm MACC}(y,\bz)$ &  0.72 & -0.29 & 5.63 & 5.77 & 0.25  \\
 & correct $\pi_{X,\bZ}(w,\bz)$ and incorrect $\bPsi_{\rm MACC}(y,\bz)$  & 1.00 & -0.00 & 5.42 & 5.52 & 94.40   \\
& incorrect $\pi_{X,\bZ}(w,\bz)$ and incorrect $\bPsi_{\rm MACC}(y,\bz)$  &  0.67 & -0.33 & 6.08 & 6.24 & 0.00  \\
&\multicolumn{5}{l}{\underline{using $\bLambda$}}\\
 & correct $\pi_{X,\bZ}(w,\bz)$ &  1.00 & -0.00 & 4.46 & 4.53 & 94.76 \\
& incorrect $\pi_{X,\bZ}(w,\bz)$ & 0.93 & -0.07 & 4.51 & 4.67 & 64.29 \\
\addlinespace
&\multicolumn{5}{l}{\underline{without using $\bLambda$}}\\
\multirow[t]{-2}{*}{\raggedright\arraybackslash \hspace{1em}AIPW} 
 & correct $\pi_{X,\bZ}(w,\bz)$ and correct $\bPsi_{\rm AIPW}(y,\bz)$ & 1.00 & -0.00 & 4.69 & 4.78 & 94.15 \\
& incorrect $\pi_{X,\bZ}(w,\bz)$ and correct $\bPsi_{\rm AIPW}(y,\bz)$ & 0.57 & -0.43 & 6.19 & 6.97 & 0.00  \\
 & correct $\pi_{X,\bZ}(w,\bz)$ and incorrect $\bPsi_{\rm AIPW}(y,\bz)$  & 1.00 & -0.00 & 4.95 & 5.07 & 94.20 \\
& incorrect $\pi_{X,\bZ}(w,\bz)$ and incorrect $\bPsi_{\rm AIPW}(y,\bz)$ &  0.73 & -0.27 & 5.90 & 6.15 & 0.41 \\
&\multicolumn{5}{l}{\underline{using $\bLambda$}}\\
 & correct $\pi_{X,\bZ}(w,\bz)$ & 1.00 & -0.00 & 4.60 & 4.76 & 93.95 \\
& incorrect $\pi_{X,\bZ}(w,\bz)$ &   0.88 & -0.12 & 5.35 & 5.97 & 37.44 \\
\addlinespace

\bottomrule

\end{tabular}}
\end{table}

\vskip -0.5em

\section{Application to Huntington disease}
\label{sec:real_data}

%
Huntington disease is an inherited neurodegenerative disease caused by a triplet repeat expansion of cytosine-adenine-guanine (CAG)  in the HTT gene; any individual with $\geq 40$ CAG repeats will develop the disease with 100\% certainty \parencite{HDCRG1993}.  An individual is clinically diagnosed (or ``diagnosed,'' for short) when their motor abnormalities are unequivocal signs of Huntington disease \parencite{Kieburtz1996}. The hallmark impairment of the disease is motor dysfunction, but cognitive dysfunction often appears years earlier and is a major concern for individuals affected by the disease, as it can impact their ability to function, drive, work, and maintain a quality of life \parencite{McAllisteretal2021}.  An ongoing research goal is to quantify the progression of cognitive dysfunction in the period leading up to a diagnosis, known as the prodromal stage, since with that knowledge, researchers can test in clinical trials whether an experimental therapy can slow or stop cognitive dysfunction before irreparable damage is done \parencite{tabrizi2022potential}.

Toward this goal, several observational studies have collected data on the extent and timing of cognitive dysfunction before a diagnosis. However, these studies face a significant limitation: some individuals leave a study early or the study concludes before all individuals meet the  diagnosis criteria, so the time of diagnosis is right-censored since it will occur after (to the right of) the individual's last study visit, we just do not know when. Modeling cognitive dysfunction before diagnosis is thus a right-censored covariate problem since  time \emph{to} diagnosis, defined as the difference between observation time and the right-censored time \emph{of} diagnosis, is used to capture the period before diagnosis.

We addressed this right-censored covariate problem using  data from Enroll-HD, a non-interventional multi-centered (North America, Latin America, Europe, Asia, Australia, and New Zealand) observational study conducted to identify features that indicate the earliest onset of Huntington disease symptoms \parencite{landwehrmeyer2017data}. The study is ongoing, and new participants are being accepted, but we analyzed data only up to the fourth revision; this includes patients who were admitted between 2012 and 2018.  Local ethics approvals were obtained from all study sites and written informed consent was obtained from all patients. 


%
Individually, we analyzed scores from the Symbol Digit Modality Test (SDMT), Stroop Word Reading Test (SWRT), Stroop Color Naming Test (SCNT), and the Categorical Verbal Fluency Test (CFT) \parencite{Smith1973, Stroop1935, Bentonetal1994}. Together, these scores reflect the coordination of visual scanning, working memory, fine motor speed, and concentration where higher scores reflect better cognitive health. We modeled each of the scores as a function of time to diagnosis while adjusting for disease burden, education, and sex (male or female) to account for the lifetime pathology associated with Huntington disease. The disease burden is quantified by the CAG-Age-Product (CAP), which is calculated using the formula CAP = (Age at Study Entry)$\times$(CAG-30) / 6.49 \parencite{warner2022standardizing}. Educational attainment was recorded using the International Standard Classification of Education (ISCED), and it was dichotomized as (i) General Educational Development (GED)/High School (HS) equivalency or higher educational attainment (ISCED level $\leq$ 3) and (ii) lower educational attainment (ISCED level 0-2).

\subsection{Analytical sample selection and statistical analysis}

We analyzed data only from individuals who were genetically confirmed to have the Huntington gene mutation and who were 100\% guaranteed to be diagnosed with the disease (i.e., $\geq$ 40 CAG repeats). We removed individuals who, at study entry, had already met the criteria for a diagnosis, as we focused on the prodromal stage and those who had missing values for the outcome test scores (SDMT, SWRT, SCNT, CFT), sex, CAP, and/or education at baseline ($n=129, 2.5\%$). The analytical sample further excluded patients from Latin America due to the small sample size from this group ($n=36, 1\%$). Ultimately, our analysis was based on data from 4,943 individuals. Among them, 852 individuals had received a diagnosis, resulting in a censoring rate of time to diagnosis of 83\%. The analytical sample was then analyzed cross-sectionally at baseline using the model:
\bse
{\rm Cognitive\ Score}_i = \beta_0 + \beta_{\rm TTD} {\rm TTD}_i + \beta_{\rm CAPs} {\rm CAPs}_i + \beta_{\rm HS} {\rm I(HS)}_i + \beta_{\rm Female} {\rm I(Female)}_i + \epsilon_i,
\ese
where for the $i^{th}$ individual, ${\rm Cognitive\ Score }_i$ represents one of the four cognitive test scores (i.e., scores on SDMT, SWRT, SCNT, or CFT); ${\rm TTD}_i$ refers to time to diagnosis in years defined as age at baseline minus age at diagnosis; ${\rm CAPs}_i$ is the CAP score; ${\rm I(HS)}_i$ equals 1 if educational attainment is at least GED/High School and 0 if lower;  ${\rm I(Female)}_i$ is equal to 1 if the individual is female and 0 if male; and $\epsilon_i$ is the random error. In this model, ${\rm TTD}_i$ is the right-censored covariate, and having ${\rm TTD}_i = -1$ indicates that the individual is one year away from their diagnosis.

We used the CC, IPW, MLE, ACC, MACC, and AIPW estimators to estimate the model parameters. All augmented estimators used $\bLambda$. A bivariate normal distribution was used for $f_{X,C|\bZ}$ where the conditional covariance between $(X,C)$ given $\bZ$ was zero to impose independent right-censoring. Distribution parameters $\balpha$ for $f_{X,C|\bZ}$ were estimated using maximum likelihood, and these were used to define the probability $\pi_{X,\bZ}$ and the distribution $f_{X|\bZ}$ for the MLE. The probability $\pi_{Y,\bZ}$ used for the ACC estimator was estimated using logistic regression. The regression estimates of the CC and ACC estimators are valid for both independent and dependent right-censoring, whereas the IPW, AIPW, and MACC estimators assume independent right-censoring. The dependent right-censoring versions of the of IPW, MLE, MACC, and AIPW estimators were omitted from the analysis as they require stronger unverifiable assumptions than those made under independent right-censoring. 

\subsection{Results}
\label{sec:data_results}


Overall, the mean age at baseline was 40 years, and the majority of participants were female (59\%), European (59\%), Caucasian (94\%), and had obtained a GED/High School diploma or higher educational level (89\%). Similar to that of other studies, the conditional mean of age at  diagnosis was 45 years \parencite{tabrizi2022potential}.


%
All estimators estimated a decrease in the cognitive test scores for each year closer to diagnosis (Table \ref{tab:hd_app_table})---a result that agrees with existing clinical studies that also showed cognitive dysfunction worsens as diagnosis approaches \parencite{Heimetal2020, McAllisteretal2021}. However, the exact decrease---the effect estimate---differed across estimators, raising the question, which estimator(s) can we trust?

Our theoretical claims and simulation results lead us to trust the CC and ACC estimators most.  The CC estimator is consistent regardless of whether the nuisance distributions are misspecified, and the ACC estimator is consistent as long as the probability $\pi_{Y,\bZ}$ is correctly specified, which is possible since $(\Delta,Y,\bZ)$ is fully observable.  In fact, the ACC estimator produced effect estimates akin to those of the CC estimator, suggesting a correct specification of  $\pi_{Y,\bZ}$. Between these two estimators, though both provide similarly reliable effect estimates, the higher efficiency of the ACC estimator makes it the preferred choice.

Compared to the CC and ACC estimators, the other estimators -- IPW, AIPW, MACC, and MLE -- yielded different effect estimates for time to diagnosis, all closer to the null. The differences in effect estimates likely result from the sensitivity of misspecifying nuisance distributions. For example, the expected decrease in SWRT score was $-3.06$ (95\% confidence interval: $-4.10, -2.02$) and $-2.48$ ($-3.99, -0.97$) for the CC and IPW estimators, respectively. The CC and IPW estimators are equally robust to misspecification so long as the probability $\pi_{X,\bZ}$ in the IPW estimator is a function of $(X,\bZ)$. When $\pi_{X,\bZ}$ is misspecified as a function of $(Y,\bZ)$, we proved the IPW estimator is no longer consistent (Section \ref{sec:thm1-proof-part1-robustness}). Our specification of $\pi_{X,\bZ}$ included education level, but education level can be considered a proxy measure for SWRT score (i.e., the outcome $Y$). Therefore, our $\pi_{X,\bZ}$ may actually be a function of $(Y,\bZ)$, not just $(X,\bZ)$. This specification will bias the IPW estimator, which would explain why the effect estimates from the IPW estimator differed from those of the CC estimator. 

Consistency of the AIPW  and MACC estimators also depends on correct specification of $\pi_{X,\bZ}$. Both will be inconsistent when $\pi_{X,\bZ}$ is a function of $(Y,\bZ)$, following a similar argument as in Section  \ref{sec:thm1-proof-part1-robustness}. Possible misspecification of $\pi_{X,\bZ}$ combined with the fact that multiple probabilities $\pi_{X,\bZ}$ were near $10^{-3}$ at times may explain why the effect estimates from AIPW  and MACC estimators differed from those of CC and ACC estimators. Finally, the different effect estimates from the MLE were likely due to the possible misspecification of  $f_{X|\bZ}$.   

Our applied analysis is strong due to the large, well-characterized sample and the comparison of multiple estimators. However, there are some limitations. First, using the ISCED to define education level may not accurately equate years of education across different countries. Second, cognitive tests were conducted in the participants’ preferred languages, which could introduce variability. Third, our analysis is cross-sectional, based on baseline data, rather than longitudinal. As such, while our findings are consistent with previous studies that indicate cognitive decline as adults approach clinical diagnosis (e.g., \cite{Paulsenetal2014}), they are limited to associations and may not be interpreted as cognitive test score trajectories over time.   

\begin{table}[h!]
\caption{\label{tab:hd_app_table} Linear regression parameter estimates (standard errors) of various core cognitive performance scores, Enroll-HD ($n=4,943$).
}
\centering
\resizebox{0.99\linewidth}{!}{
\begin{tabular}[t]{lccccc}
\toprule
\textbf{Estimator} & \textbf{Intercept} & \textbf{Time to Diagnosis (Years)} & \textbf{CAP} & \makecell{\textbf{ Education (GED/HS or higher)}} & \textbf{Sex (Female)} \\
\midrule
\multicolumn{5}{l}{\textbf{\underline{Outcome 1: Symbol Digit Modality Test (SDMT)}}} \\
\addlinespace[0.3em]
CC & 62.62 (2.82)$^{**}$ & -1.74 (0.32)$^{**}$ & -0.37 (0.03)$^{**}$ & 6.42 (1.04)$^{**}$ & 0.32 (0.75) \\
IPW & 63.83 (4.07)$^{**}$ & -1.46 (0.36)$^{**}$ & -0.35 (0.04)$^{**}$ & 5.51 (1.41)$^{**}$ & -0.93 (1.16) \\
MACC & 62.68 (2.36)$^{**}$ & -1.23 (0.25)$^{**}$ & -0.35 (0.02)$^{**}$ & 5.31 (0.91)$^{**}$ & 1.24 (0.70) \\
AIPW & 60.57 (2.04)$^{**}$ & -0.98 (0.14)$^{**}$ & -0.29 (0.02)$^{**}$ & 2.82 (0.93)$^{**}$ & 0.07 (0.75) \\
ACC &  63.63 (2.21)$^{**}$ &  -1.51 (0.32)$^{**}$ &  -0.37 (0.02)$^{**}$ &  6.31 (0.69)$^{**}$ &  0.81 (0.47)\\
MLE & 57.89 (1.07)$^{**}$ & -0.82 (0.06)$^{**}$ & -0.27 (0.01)$^{**}$ & 6.27 (0.52)$^{**}$ & 1.85 (0.32)$^{**}$ \\
\addlinespace[0.3em]
\multicolumn{5}{l}{\textbf{\underline{Outcome 2: Stroop Word Reading Test (SWRT)}}} \\
\addlinespace[0.3em]
CC &  103.21 (4.41)$^{**}$ & -3.06 (0.53)$^{**}$  & -0.42 (0.04)$^{**}$ & 10.25 (1.72)$^{**}$ & 0.25 (1.29) \\
IPW & 98.10 (6.83)$^{**}$ & -2.48 (0.77)$^{**}$ &  -0.35 (0.06)$^{**}$ & 11.51 (2.71)$^{**}$ & -1.35 (2.09) \\
MACC & 104.33 (4.03)$^{**}$ & -2.04 (0.41)$^{**}$  & 0.40 (0.03)$^{**}$ & 8.03 (1.55)$^{**}$ & 1.41 (1.17) \\
AIPW & 104.66 (5.88)$^{**}$ & -0.80 (0.27)$^{**}$ & -0.35 (0.05)$^{**}$ & 5.51 (5.51)$^{**}$ & 2.82 (0.75)$^{**}$ \\
ACC &  103.27 (3.52)$^{**}$ & -2.85 (0.53)$^{**}$ & -0.42 (0.03)$^{**}$ & 9.83 (1.21)$^{**}$ & 0.87 (0.80) \\
MLE & 101.06 (1.72)$^{**}$ & -1.14 (0.10)$^{**}$ & -0.29 (0.02)$^{**}$ & 6.15 (0.86)$^{**}$ & 1.09 (0.52)$^{**}$ \\
\addlinespace[0.3em]
\multicolumn{5}{l}{\textbf{\underline{Outcome 3: Stroop Color Naming Test (SCNT)}}} \\
\addlinespace[0.3em]
CC &  81.75 (3.63)$^{**}$ & -2.42 (0.47)$^{**}$ & -0.33 (0.03)$^{**}$ & 6.22 (1.40)$^{**}$ & -0.11 (1.04) \\
IPW & 77.72 (6.54)$^{**}$ & -2.18 (0.74)$^{**}$ & -0.25 (0.06)$^{**}$ & 4.54 (2.46)$^{**}$ & -2.31 (2.04) \\
MACC &  85.51 (3.14)$^{**}$ & -0.91 (0.21)$^{**}$ & -0.34 (0.03)$^{**}$ & 4.96 (1.16)$^{**}$ & 1.33 (0.88) \\
AIPW & 86.62 (8.49)$^{**}$ & -0.53 (0.37) & -0.31 (0.08)$^{**}$ & 2.64 (1.01)$^{**}$ & 1.68 (0.67)$^{**}$ \\
ACC &  81.70 (2.82)$^{**}$ &  -2.24 (0.47)$^{**}$ &  -0.33 (0.03)$^{**}$ &  5.92 (0.92)$^{**}$ &  -0.65 (0.61) \\
MLE & 79.20 (1.38)$^{**}$ & -0.94 (0.08)$^{**}$ & -0.25 (0.01)$^{**}$ & 5.59 (0.68)$^{**}$ & 0.95 (0.42)$^{*}$ \\
\addlinespace[0.3em]
\multicolumn{5}{l}{\textbf{\underline{Outcome 4: Categorical Verbal Fluency Test (CFT)}}} \\
\addlinespace[0.3em]
CC &  25.39 (1.31)$^{**}$ & -0.69 (0.16)$^{**}$ & -0.12 (0.01)$^{**}$ & 2.33 (0.48)$^{**}$ & -0.25 (0.41) \\
IPW & 25.19 (1.71)$^{**}$ & -0.34 (0.15)$^{*}$ & -0.11 (0.02)$^{**}$ & 2.26 (1.63) & -0.41 (0.51) \\
MACC &  24.83 (1.20)$^{**}$ & -0.28 (0.08)$^{**}$ & -0.11 (0.01)$^{**}$ & 1.75 (0.39)$^{**}$ & 0.71 (0.32)$^{*}$ \\
AIPW & 16.09 (1.37)$^{**}$ & -0.53 (0.08)$^{**}$ & -0.02 (0.01)$^{*}$ & 1.36 (0.43)$^{**}$ & 1.26 (0.35)$^{**}$ \\
ACC &  25.46 (1.05)$^{**}$ & -0.49 (0.15)$^{**}$ &  -0.12 (0.01)$^{**}$ & 2.31 (0.31)$^{**}$ &  -0.08 (0.21) \\
MLE &  21.99 (0.51)$^{**}$ & -0.30 (0.03)$^{**}$ & -0.07 (0.00)$^{**}$ & 2.29 (0.25)$^{**}$ & 0.24 (0.16) \\
\bottomrule
\multicolumn{5}{l}{\textbf{Note:} Significant at the 0.01 ($^{**}$) or 0.05 ($^{*}$) level; CAP = (Age at Study Entry)$\times$(CAG-Age-Product - 30) / 6.49 } \\
\multicolumn{5}{l}{General Educational Development (GED); High School (HS)} \\
[-1em]
\end{tabular}}
\end{table}

\section{Discussion}
\label{sec:discussion}

We highlight in this paper the similarities and  differences between six estimators---CC, IPW, MLE, ACC, MACC, and AIPW---applicable to the right-censored and missing covariate problems. All estimators except the MACC estimator have  been  proposed for the right-censored covariate problem, but only under independent covariate right-censoring.  We establish that these estimators, including the MACC estimator, are not limited to just this setting and can be easily modified to handle dependent covariate right-censoring---a discovery that widens their use in the censored covariate literature.

The robustness and efficiency properties outlined in Theorems \ref{thm:all-robust-consistent} and \ref{thm:all-efficiency} can help analysts  make an informed decision about which estimator to choose in practice. That decision is affected by what distributions the analyst is comfortable misspecifying and how much statistical efficiency is desired as estimators with higher statistical efficiency provide more precise statistical inference. Under the assumption that $(C,R)\independent Y|(X,\bZ)$, our recommendations for selecting an estimator, ordered from highest to lowest preference, are as follows:
\begin{enumerate}[noitemsep]
    \item \textbf{ACC estimator with $\bLambda$:} Among  all estimators, we recommend the ACC estimator with $\bLambda$ as it is more efficient than the CC, AIPW, and IPW estimators. Although it requires correctly specifying the distribution $f_{\Delta|Y,\bZ}$ in the right-censored covariate problem, but this is actually feasible given that we \emph{always} observe the data $(\Delta,Y,\bZ)$. Moreover, the ACC estimator is consistent under independent and dependent covariate right-censoring.

    \item \textbf{MACC estimator with $\bLambda$:} The MACC estimator also recovers some of the efficiency lost in the CC estimator, but its consistency  hinges on the correct specification of $f_{C|\bZ}$ (or $f_{C|X,\bZ}$ for dependent covariate right-censoring). The parameters governing $f_{C|X,\bZ}$ are not identifiable without unverifiable assumptions; therefore, ensuring $f_{C|\bZ}$ (or $f_{C|X,\bZ}$) is correctly specified is difficult.  
    \item \textbf{CC estimator:} While deleting data is usually a statistical faux pas, the CC estimator remains consistent in both right-censored and missing covariate problems, regardless of whether the covariate right-censoring or missingness is independent or dependent. Therefore, we recommend the CC estimator when asymptotic unbiasedness is the main concern, but not when statistical efficiency is the primary interest.
    \item \textbf{MLE:} The MLE is less robust than the CC and IPW estimators when the covariate right-censoring and missingness is assumed independent or dependent. However, when there is confidence in correctly specifying $f_{X|\bZ}$  (or  $f_{X|C,\bZ}$), the MLE is the preferred choice, as it will be a consistent estimator with optimal statistical efficiency.
    
    \item \textbf{IPW and AIPW estimators:} We discourage using the IPW and AIPW estimators, as both are less efficient than the CC estimator, which is simpler to implement. Moreover, the AIPW estimator is prone to bias when $\pi_{X,\bZ}$ is misspecified. Accurately specifying  $\pi_{X,\bZ}$ in the dependent covariate right-censoring and missingness is not straightforward as the parameters governing this probability are not identifiable without additional, often unverifiable assumptions.
\end{enumerate}   

No matter which estimator is chosen, we recommend also using the CC estimator, as it is the most robust estimator so long as $(C,R) \independent Y | (X,\bZ)$. Under this assumption, any differences in effect estimates between the CC estimator and another estimator would indicate that the other estimator is inconsistent. A search can then take place to pinpoint the source of the inconsistency, such as misspecified nuisance distributions.

While the IPW estimator is the least of our preferred choices, it has been shown to reduce bias and/or improve efficiency over the CC estimator under \emph{different} assumptions than we consider.
When $(\Delta,R)\notindependent Y |(X,\bZ)$---an assumption implied by $(C,R)\independent Y|(X,\bZ)$ \parencite{AshnerGarcia2023}---the CC and IPW estimators are inconsistent for both right-censored and missing covariate problems. However, for the right-censored covariate problem, \textcite{MatsouakaAtem2020} showed that including $Y$ in the probability for the IPW estimator corrects the  bias present in the CC estimator. For the missing covariate problem, \textcite{little2022comparison} also describes how including $Y$ in the probability for the IPW estimator corrects the bias present in the CC estimator and increases efficiency. However, this choice would lead to the IPW estimator producing bias when in fact $(C,R) \independent Y | (X,\bZ)$.

Using our preferred ACC estimator  in future Huntington disease data analyses could benefit ongoing research. Preventative clinical trials, which aim to slow or stop the disease before irreversible damage occurs, depend on understanding impairment progression patterns before diagnosis.  Yet most analyses currently use either a CC estimator  \parencite{Longetal2014}, which sacrifices efficiency,  or an imputation estimator, which conveniently replaces time to diagnosis with an estimated value \parencite{scahill_biological_2020}, but whose consistency requires correct specification of $f_{X|\bZ}$ (or $f_{X|C,\bZ}$) when covariate right-censoring is independent (or dependent)  \parencite{Bernhardtetal2015}. The ACC estimator overcomes these limitations and can be used without testing whether the covariate right-censoring is independent or dependent, as its form remains the same in both scenarios. Not needing that type of test is advantageous in practice because  when individuals drop out of Huntington observational  studies,  researchers often do not know why, except in rare instances. Without knowing why individuals drop out, it is difficult to know whether we should use  estimators that assume independent or dependent covariate right-censoring. With the ACC estimator, we bypass this guessing game.  

Our analysis of Enroll-HD data using the ACC estimator confirmed earlier findings that cognitive dysfunction worsens before diagnosis, but we can have more confidence in this conclusion because we now have more efficient estimates. These results, along with the proven properties of the ACC estimator, give us confidence to use it in future analyses where the progression patterns of impairment before diagnosis are unknown---insights that could help pinpoint when and how to best intervene in Huntington disease.

By thoroughly exploring right-censored and missing covariate problems, we clarified the necessary modifications for applying estimators to both. This exploration offered an unprecedented comparison of the estimators' robustness to misspecification and statistical efficiency. While we assumed $f_{Y|X,\bZ}$ followed a normal distribution, other distributions could be used, and even greater flexibility could be achieved by assuming a restricted mean model (i.e., $E(\epsilon|X,\bZ)=0$). Our work thus equips analysts to understand the nuances and similarities of these problems, preventing blind decisions and helping them choose the correct estimator in practice.

\section*{Acknowledgements}
The authors thank the staff and adults of Enroll-HD study for their important contributions. Data are available upon request from \url{https://www.enroll-hd.org}, and code is available at  \url{https://github.com/jesusepfvazquez/right-censored-covariates}. This work was supported by the National Institute of Neurological Disorders and Stroke under Grant R01NS131225 (JEV, YM, KM, TPG); National Institute of Environmental Health Sciences under Grant T32ES007018 (JEV); and the National Science Foundation (MCA). The authors report that there are no competing interests to declare.

\printbibliography


\setcounter{equation}{0}
\setcounter{section}{0}
\setcounter{table}{0}
\setcounter{figure}{0}
\renewcommand{\thesection}{S.\arabic{section}}
\renewcommand{\thefigure}{S.\arabic{figure}}
\renewcommand{\thetable}{S.\arabic{table}}
\renewcommand{\theequation}{S.\arabic{equation}}

\newpage
\setcounter{page}{1}
\begin{center}
{\LARGE{\bf 
Supplementary Material for \textit{Establishing the Parallels and Differences Between Right-Censored and Missing Covariates}}}
\end{center}

\baselineskip=12pt
\setlength{\parskip}{0.1in}

\begin{center}
Jesus E. Vazquez, Marissa C. Ashner, Yanyuan Ma, Karen Marder,  and Tanya P. Garcia
\end{center}

\noindent Throughout the Supplementary Material, we denote different (conditional) densities $f$ with subscripts to differentiate between densities. For example,  $f_{Y|X,\bZ}(y,w,\bz;\btheta)$  denotes the conditional  density function of $Y$ given $(X,\bZ)$ evaluated at $Y=y$, $X=w$, and $\bZ=\bz$.  Similarly, $E_{X|Y,\bZ}(\cdot)$ will be used to denote the conditional expectation with respect to the density $f_{X|Y,\bZ}$. For a vector or matrix $\bM$,  we define $\bM^{\otimes2}\equiv\bM\bM\trans$. When unspecified, $E(\cdot) \equiv E_{Y,W,\Delta,\bZ}(\cdot)$ for the right-censored covariate problem and  $E(\cdot) \equiv E_{Y,XR,R,\bZ}(\cdot)$ for the missing covariate problem.


\section{Closed form solution of $\bPsi_{\rm closed}(y,\bz;\btheta)$}
\label{sec:closed-form}

The augmentation term $\bPsi_{\rm closed}(y,\bz;\btheta) = - E_{X|Y,\bZ}\{ \bS_{\btheta}^F(y, X, \bz;\btheta) \}$ was used in the augmented estimators to increase computational speed in estimating $\btheta$ (Tables \ref{tab:estimators-lambda} and \ref{tab:estimators-lambda-dependent}). That augmentation  term can be written as
\bse
E_{X|Y,\bZ}\{\bS_{\btheta}^F(y, X, \bz;\btheta)\}
&=& \int \bS_{\btheta}^F(y, x,
\bz;\btheta)f_{X|Y,\bZ}(x,y,\bz)dx\\
&=& \int \bS_{\btheta}^F(y, x,
\bz;\btheta)\frac{f_{X,Y,\bZ}(x,y,\bz)}{f_{Y,\bZ}(y,\bz)}dx\\
&=& \frac{\int \bS_{\btheta}^F(y, x,
\bz;\btheta)f_{X,Y,\bZ}(x,y,\bz)dx}{\int f_{Y,X,\bZ}(y,x,\bz)dx}\\
&=& \frac{\int \bS_{\btheta}^F(y, x,
\bz;\btheta) f_{Y|X,\bZ}(y,x,\bz;\btheta)f_{X|\bZ}(x,\bz) f_\bZ(\bz)dx}{\int  f_{Y|X,\bZ}(y,x,\bz;\btheta)f_{X|\bZ}(x,\bz)f_\bZ(\bz)dx}\\
&=& \frac{\int \bS_{\btheta}^F(y, x,
\bz;\btheta) f_{Y|X,\bZ}(y,x,\bz;\btheta)f_{X|\bZ}(x,\bz) dx}{\int  f_{Y|X,\bZ}(y,x,\bz;\btheta)f_{X|\bZ}(x,\bz)dx}. \\
\ese
In the following subsections, we derive the closed-form expression. We consider the special case when $m\{X=x,\bZ=(a,z); \bbeta \} =  \beta_0 + \beta_1 (a-x) + \beta_2 z$ and make the following assumptions:
\begin{enumerate}
    \item $X$ is normally distributed with mean equal to $\mu_x$ and variance equal to $\sigma^2_x$,
    \item $\epsilon$ is normally distributed with mean equal to zero and variance equal to $\sigma^2$. 
\end{enumerate}

\subsection{Denominator}

The denominator is

\bse
\D&\equiv& \int  f_{Y|X,\bZ}(y,x,\bz;\btheta)f_{X|\bZ}(x,\bz)dx \\
&=&  \int \frac{1}{\sqrt{2\pi\sigma^2}} \exp\left(-\frac{[y- \{\beta_0 + \beta_1 (a-x) +  \beta_2 z \} ]^2}{2\sigma^2} \right) \times \frac{1}{\sqrt{2\pi\sigma_x^2}} \exp\left\{-\frac{(x- \mu_x)^2}{2\sigma_x^2} \right\} dx \\
&=& \frac{1}{2\pi \sqrt{\sigma^2 \sigma_x^2}} \int \exp\left[-\frac{\{ (y- \beta_0 - \beta_1a - \beta_2z) + \beta_1x\}^2}{2\sigma^2}\right] \exp\left\{- \frac{(x- \mu_x)^2}{2\sigma_x^2} \right\} dx \\
& =& \frac{1}{2\pi \sqrt{\sigma^2 \sigma_x^2}} \int \exp\left( -\frac{x^2\beta_1^2 + 2x\beta_1e_* + e^{*2} }{2\sigma^2} \right) \exp\left(- \frac{x^2- 2x\mu_x + \mu_x^2}{2\sigma_x^2} \right) dx\\
&=&  \frac{1}{2\pi \sqrt{\sigma^2 \sigma_x^2}} \int \exp \biggr\{ x^2\left(-\frac{\beta_1^2}{2\sigma^2} - \frac{1}{2\sigma_x^2} \right)  +  x\left(-\frac{\beta_1e_*}{\sigma^2} + \frac{\mu_x}{\sigma_x^2} \right) +  \left(- \frac{e_*{2}}{2\sigma^2} -\frac{\mu_x^2}{2\sigma_x^2}\right) \biggr\} dx,   
\ese
where $e_* = y- (\beta_0 + \beta_1a + \beta_2z)$ and $\epsilon = \beta_1 x  + e_*$. The error $\epsilon$ represents the model error from Equation (\ref{eqn:equation_paper2}) for the special case considered in this section.
At this step, let
\bse
a_* &=& \left(-\frac{\beta_1^2}{2\sigma^2} - \frac{1}{2\sigma_x^2} \right); \quad b_* = \left(-\frac{\beta_1e_*}{\sigma^2} + \frac{\mu_x}{\sigma_x^2} \right); \quad c_* = \left(- \frac{e^{*2}}{2\sigma^2} -\frac{\mu_x^2}{2\sigma_x^2}\right).
\ese
Then, continuing with our calculation of the denominator $\D$, we can complete the square to obtain
\bse
\D&=& \frac{1}{2\pi \sqrt{\sigma^2 \sigma_x^2}} \int \exp ( x^2a_*  +  xb_* + c_* ) dx \\
&=& \frac{1}{2\pi \sqrt{\sigma^2 \sigma_x^2}} \int \exp \left\{ a_*\left(x + b_*/2a_*\right)^2 + c_* - b_*^2/4a_* \right\} dx \\
&=& \frac{1}{2\pi \sqrt{\sigma^2 \sigma_x^2}} \exp\left( c_* - b_*^2/4a_* \right) \int \exp \{ -(x - \mu_*)^2/2 \sigma_*^2\} dx \\
&=& \frac{\sqrt{2\pi\sigma_*^2}}{2\pi \sqrt{\sigma^2 \sigma_x^2}} \exp\left( c_* - b_*^2/4a_* \right) \int \frac{1}{\sqrt{2\pi\sigma_*^2}} \exp \{ -(x - \mu_*)^2/2 \sigma_*^2\} dx,
\ese
where $\mu_* = -b_*/(2a_*)$ and $\sigma_*^2 = \sigma_x^2 \sigma^2/(\sigma^2 + \beta_1^2\sigma_x^2)$.
Therefore, the denominator is 
\bse
\D &=&  \frac{1}{\sqrt{2\pi (\sigma^2 + \beta_1^2\sigma_x^2)}}  \exp\left( c_* - \frac{b_*^2}{4a_*} \right).
\ese

\subsection{Numerator}
The numerator is a vector, with each element corresponding to the integral of the partial derivative of the log-likelihood with respect to $\btheta \in \{\beta_0,  \beta_1, \beta_2, \sigma \}$. The components of $\bS_{\btheta}^F(y, x, \bz;\btheta)^T = \{\dot{\beta_0} l(y,x,z; \btheta), \dot{\beta_1} l(y,x,z; \btheta), \dot{\beta_1} l(y,x,z; \btheta), \dot{\sigma} l(y,x,z; \btheta)\}^T$ are equal to
\bse
\dot{\beta_0} l(y,x,z; \btheta) = \epsilon/\sigma^2; \quad \dot{\beta_1} l(y,x,z; \btheta) = (a-x)\epsilon/\sigma^2; \\
\dot{\beta_2} l(y,x,z; \btheta) = z\epsilon/\sigma^2; \quad \dot{\sigma} l(y,x,z; \btheta) = -1/\sigma + \epsilon^2/\sigma^3.
\ese
For $E_{X|Y,\bZ}\{\dot{\beta_0} l(y,x,z; \btheta) \}$, the numerator takes the form
\bse
&& \int  \dot{\beta_0} l(y,x,z; \btheta) f_{Y|X,\bZ}(y,x,\bz;\btheta)f_{X|\bZ}(x,\bz)dx \\
&& = \int \frac{(\beta_1x + e_*)}{\sigma^2}  f_{Y|X,\bZ}(y,x,\bz;\btheta)f_{X|\bZ}(x,\bz)   dx\\
&& = \frac{1}{\sigma^2} \left\{ \beta_1\int  x f_{Y|X,\bZ}(y,x,\bz;\btheta)f_{X|\bZ}(x,\bz)dx + e_*\int  f_{Y|X,\bZ}(y,x,\bz;\btheta)f_{X|\bZ}(x,\bz)dx \right\} \\
&& = \frac{1}{\sigma^2} \left\{ \beta_1 \int  x f_{Y|X,\bZ}(y,x,\bz;\btheta)f_{X|\bZ}(x,\bz)dx + e_* \D \right\} \\
&& = \frac{1}{\sigma^2} \left( \beta_1\D \mu_* + e_*\D \right).
\ese
The last line follows since 
\bse
&& \beta_1 \int  x f_{Y|X,\bZ}(y,x,\bz;\btheta)f_{X|\bZ}(x,\bz)dx \\
&& = \beta_1 \frac{\sqrt{2\pi\sigma_*^2}}{2\pi \sqrt{\sigma^2 \sigma_x^2}} \exp\left( c_* - \frac{b_*^2}{4a_*} \right) \int x \frac{1}{\sqrt{2\pi\sigma_*^2}} \exp \biggr\{ \frac{\left(x - \mu_*\right)^2}{-2 \sigma_*^2}\biggr\} dx, 
\\
&& = \beta_1 \frac{\sqrt{2\pi\sigma_*^2}}{2\pi \sqrt{\sigma^2 \sigma_x^2}} \exp\left( c_* - \frac{b_*^2}{4a_*} \right) \mu_* = \beta_1\D \mu_*.
\ese
Since $z$ is a constant, then $E_{X|Y,\bZ}\{\dot{\beta_2} l(y,x,z; \btheta) \} =  (z/\sigma^2) \left( \beta_1\D \mu_* + e_*\D \right)$. $E_{X|Y,\bZ}\{\dot{\beta_1} l(y,x,z; \btheta) \}$ takes the following form
\bse
&& \int  \dot{\beta_1} l(y,x,z; \btheta) f_{Y|X,\bZ}(y,x,\bz;\btheta)f_{X|\bZ}(x,\bz)dx \\
&& = \int f_{Y|X,\bZ}(y,x,\bz;\btheta)f_{X|\bZ}(x,\bz) \left\{ \frac{(a-x)(\beta_1x + e_*)}{\sigma^2} \right\} dx \\
&& = \frac{1}{\sigma^2} \int f_{Y|X,\bZ}(y,x,\bz;\btheta)f_{X|\bZ}(x,\bz) \left\{ -\beta_1x^2+x(a\beta_1- e_*)+ae_* \right\} dx
\ese
since $(a-x)(\beta_1x + e_*) = a\beta_1x+ ae_* - \beta_1x^2 - xe_* = -\beta_1x^2+x(a\beta_1- e_*)+ae_*$. This simplification leads to
\bse
&& \int  \dot{\beta_1} l(y,x,z; \btheta) f_{Y|X,\bZ}(y,x,\bz;\btheta)f_{X|\bZ}(x,\bz)dx \\
&& = \frac{1}{\sigma^2} \biggr\{ -\beta_1 \int x^2 f_{Y|X,\bZ}(y,x,\bz;\btheta)f_{X|\bZ}(x,\bz) dx + (a\beta_1- e_*)\int x f_{Y|X,\bZ}(y,x,\bz;\btheta)f_{X|\bZ}(x,\bz) dx \\
&& \quad \quad + ae_* \int f_{Y|X,\bZ}(y,x,\bz;\btheta)f_{X|\bZ}(x,\bz) dx \biggr\} \\ 
&& = \frac{1}{\sigma^2} \biggr\{ -\beta_1 \int x^2 f_{Y|X,\bZ}(y,x,\bz;\btheta)f_{X|\bZ}(x,\bz) dx + (a\beta_1- e_*)\D\mu_* + ae_*\D \biggr\} \\
&& = \frac{1}{\sigma^2} \left\{-\beta_1 \D ( \sigma_*^2 + \mu_*^2 ) + (a\beta_1- e_*)\D\mu_* + ae_*\D \right\}.
\ese
The last line follows since
\bse
\int x^2 f_{Y|X,\bZ}(y,x,\bz;\btheta)f_{X|\bZ}(x,\bz) dx = \D E(X^2) = \D (\sigma_*^2 + \mu_*^2).
\ese
Finally, $E_{X|Y,\bZ}\{\dot{\sigma} l(y,x,z; \btheta) \}$ equals
\bse
&& \int  \dot{\sigma} l(y,x,z; \btheta) f_{Y|X,\bZ}(y,x,\bz;\btheta)f_{X|\bZ}(x,\bz)dx \\
&& = \int f_{Y|X,\bZ}(y,x,\bz;\btheta)f_{X|\bZ}(x,\bz) \left\{ -\frac{1}{\sigma} + \frac{(x\beta_1 + e_*)^2}{\sigma^3} \right\} dx \\
&& = \int f_{Y|X,\bZ}(y,x,\bz;\btheta)f_{X|\bZ}(x,\bz) \left\{x^2\left(\frac{\beta_1^2}{\sigma^3}\right) + x\left(\frac{2\beta_1e_*}{\sigma^3}\right) + \left( \frac{{e_*}^2}{\sigma^3} -\frac{1}{\sigma} \right) \right\} dx \\
&& = \biggr\{ \left(\frac{\beta_1^2}{\sigma^3}\right) \int x^2 f_{Y|X,\bZ}(y,x,\bz;\btheta)f_{X|\bZ}(x,\bz) dx \biggr\} + \biggr\{ \left(\frac{2\beta_1e_*}{\sigma^3}\right) \int x f_{Y|X,\bZ}(y,x,\bz;\btheta)f_{X|\bZ}(x,\bz) dx \biggr\} \\
&& + \biggr\{ \left( \frac{{e_*}^2}{\sigma^3} -\frac{1}{\sigma} \right) \int f_{Y|X,\bZ}(y,x,\bz;\btheta)f_{x|\bZ}(x,\bz) dx \biggr\} \\
&& = \left(\frac{\beta_1^2}{\sigma^3}\right) \D \left( \sigma_*^2 + \mu_*^2 \right) + \left(\frac{2\beta_1e_*}{\sigma^3}\right) \D \mu_* + \left( \frac{{e_*}^2}{\sigma^3} -\frac{1}{\sigma} \right) \D.
\ese

\subsection{Final expression of $E_{X|Y,\bZ}\{\bS_{\btheta}^F(y, X,\bz;\btheta)\}$}

With the closed forms of the numerator and denominator, the components of $E_{X|Y,\bZ}\{\bS_{\btheta}^F(y, x,
\bz;\btheta)\}$ take the  form

\bse
E_{X|Y,\bZ}\{\dot{\beta_0} l(y,X,z; \btheta) \} &=&  \frac{1}{\sigma^2} \left( \beta_1 \mu_* + e_* \right), \\
E_{X|Y,\bZ}\{\dot{\beta_1} l(y,X,z; \btheta) \} &=& \frac{1}{\sigma^2} \left\{ -\beta_1 ( \sigma_*^2 + \mu_*^2 ) + (a\beta_1- e_*)\mu_* + ae_* \right\},
\ese

\bse
E_{X|Y,\bZ}\{\dot{\beta_2} l(y,X,z; \btheta) \} &=&  \frac{z}{\sigma^2} \left( \beta_1 \mu_* + e_* \right), \\
E_{X|Y,\bZ}\{\dot{\sigma} l(y,X,z; \btheta) \} &=& \left(\frac{\beta_1^2}{\sigma^3}\right) \left[ \sigma_*^2 + \mu_*^2 \right] + \left(\frac{2\beta_1e_*}{\sigma^3}\right) \mu_* + \left( \frac{{e_*}^2}{\sigma^3} -\frac{1}{\sigma} \right).
\ese
The remaining step multiplies each of the above components by $-1$ to obtain $\bPsi_{\rm closed}(y,\bz;\btheta)$. For an additional covariate $z_p$, the closed form solution is $E_{X|Y,\bZ}\{\dot{\beta_p} l(y,X,z; \btheta) \} = z_p \left( \beta_1 \mu_* + e_* \right)/ \sigma^2$. 

\section{Proof of Theorem \ref{thm:all-robust-consistent} for the right-censored covariate problem}
\label{sec:thm1-proof-part1}

\noindent We first prove Theorem \ref{thm:all-robust-consistent} for the right-censored covariate problem. The proof is  divided into multiple parts. First, we address the specification \textit{or} estimation of nuisance distributions used in the estimators. Then, we  prove consistency, and asymptotic normality. We first present results under independent covariate right-censoring, and then explain modifications made to adjust for dependent covariate right-censoring. To minimize notation, we drop the superscripts (e.g., ``cens, ind"). 

\subsection{Specification or estimation of nuisance distributions $f_{X|\bZ}$, $f_{C|\bZ}$}
\label{sec:est_nuisance_distributions}  

Under independent covariate right-censoring, estimation of $\btheta$ with the estimators in Table \ref{tab:estimators} requires specifying or estimating nuisance distributions---$f_{X|\bZ}$, $f_{C|\bZ}$, and/or $f_{\Delta|Y,\bZ}$---as discussed in Section \ref{sec:implement}. For example,  
\begin{itemize}
    \item IPW estimator: Requires the density $f_{C|\bZ}$ to compute the probability 
    $\pi_{X,\bZ}(x,\bz)$;
    \item MLE: Requires the density $f_{X|\bZ}(x,\bz)$;
    \item ACC estimator: Requires the density  $f_{\Delta|Y,\bZ}$ to compute the probability $\pi_{Y,\bZ}(y, \bz)$ and the density of $f_{X|\bZ}$ to compute $\bPsi(y,\bz;\btheta)$;
    \item MACC and AIPW estimators: Requires the density $f_{C|\bZ}$ to compute the probability $\pi_{X,\bZ}(x,\bz)$ and the density $f_{X|\bZ}$ to compute $\bPsi(y,\bz;\btheta)$.

\end{itemize}

Throughout, we consider parametric models $f_{X|\bZ}(x,\bz;\bgamma)$, $f_{C|\bZ}(c,\bz;\boldeta)$, and $f_{\Delta|Y,\bZ}(\delta,y,\bz;\bkappa)$, where $(\bgamma, \boldeta, \bkappa)$ is a set of finite-dimensional parameters; we refer to them as nuisance parameters. Under a \emph{mechanistic} approach, the nuisance parameters are assumed known and are specified by the analyst. Under an \emph{empirical} approach, the nuisance parameters need to be estimated. We show how these parameters are estimated and present the corresponding asymptotic properties. 
We describe a logistic regression model for estimating $\bkappa$ in Section \ref{sec:implement}, therefore, we focus on the parameter set $\balpha = (\boldeta^T, \bgamma^T)^T$.


Estimating the parameters governing $f_{X|\bZ}(x,\bz;\bgamma)$ and $f_{C|\bZ}(c,\bz;\boldeta)$ separately is not possible, as they are not identifiable \parencite{ebrahimi2003identifiability}. Yet, under the assumption of independent covariate right-censoring, the parameters are identifiable when $\balpha = (\boldeta^T, \bgamma^T)^T$ is jointly estimated using the density of $(W,\Delta)$ conditional on $\bZ$. The log-likelihood for one individual is given by,
\bse
&& \log f_{W,\Delta| \bZ}(w_i,\delta_i, \bz_i; \balpha) \\
&& = \delta_i \log\biggr\{ \int_{w_i\leq C} f_{C,X|\bZ}(c,w_i,\bz_i;\balpha) dc \biggr\} + (1-\delta_i) \log\biggr\{ \int_{w_i\leq X} f_{C,X|\bZ}(w_i,x,\bz_i\balpha) dx \biggr\} \\
&& = \ \delta_i \log\biggr\{ \int_{w_i\leq C} f_{C|\bZ}(c,\bz_i;\boldeta) f_{X|\bZ}(w_i,\bz_i;\bgamma) dc \biggr\}  \\
&& \sextant \sextant \sextant + (1-\delta_i) \log\biggr\{ \int_{w_i\leq X} f_{C|\bZ}(w_i,\bz_i; \boldeta) f_{X|\bZ}(x_i, \bz_i;\bgamma) dx \biggr\}. 
\ese
The second equality follows since $X \independent C | \bZ$. The corresponding score equation equals
\bse
 \bS_{\balpha}(w_i,\delta_i,\bz_i; \balpha) &=& \frac{\partial}{\partial \balpha^T} \log f_{W,\Delta|\bZ}(w_i,\delta_i,\bz_i; \balpha) \\
 &=& \delta_i \frac{\int_{w_i\leq C} \{ \partial \log f_{C,X|\bZ}(c,w_i,\bz_i; \balpha) /\partial \balpha^T  \}  f_{C,X|\bZ}(c,w_i,\bz_i; \balpha) dc}{\int_{w_i\leq C} f_{C,X|\bZ}(c,w_i,\bz_i; \balpha) dc} \\
&& + (1-\delta_i) \frac{\int_{w_i\leq X} \{ \partial \log f_{C,X|\bZ}(w_i,x,\bz_i; \balpha) /\partial \balpha^T  \} f_{C,X|\bZ}(w_i,x,\bz_i; \balpha) dx}{\int_{w_i\leq X} f_{C,X|\bZ}(w_i,x,\bz_i; \balpha) dx}. 
\ese
Finally, let $\bO^* = (W,\Delta,\bZ)$ and $\wh\balpha$ be the solution to the estimating equation 
\be
\label{eqn:alpha_est}
 \sumi \bPhi_{\balpha}(\bO^*_i; \balpha) = \sumi \bS_{\balpha}(w_i,\delta_i,\bz_i; \balpha) = \bzero.
\ee

%
\subsection{Consistency}
\label{thm1:consistency}


Each estimator we consider is a solution to so-called $m$-estimating equations. Specifically, the estimator $\wh{\btheta}$ is the solution to $\sumi \bPhi_{\rm est}(\bO_i; \btheta, \balpha) = \bzero$, where $\bPhi_{\rm est}(\cdot)$ is an estimating function evaluated at observed data $\bO_i$, and $\balpha$ represents the set of nuisance parameters. For the case of the CC estimator, we consider $\sumi \bPhi_{\rm est}(\bO_i; \btheta) = \bzero$, as it does not depend on any nuisance parameters. We use the following regularity conditions when proving the consistency and asymptotic normality of $\wh\btheta$ (Section \ref{sec:thm1-proof-part1-normality}):
\begin{enumerate}[label=(A\arabic*),ref=(A\arabic*)]
    \item\label{reg:compact}
    $\btheta_0\in\bOmega$, and $\bOmega$ is compact.
    \item\label{reg:bounded}
    $E[\sup_{\btheta\in\Omega}\|\bPhi_{\rm est}(\bO;  \btheta, \balpha)\|_2]<\infty$. 
    \item\label{reg:smooth-A} $E \{ \partial \bPhi_{\rm est}(\bO;  \btheta_0, \balpha_0)/ \partial \btheta^T \}$ has bounded eigenvalues and is invertible.
    \item\label{reg:smooth-A-alpha}
    $E \{ \partial \bPhi_{\rm est}(\bO;  \btheta_0, \balpha_0)/ \partial \balpha^T \}$ has bounded eigenvalues for $\btheta \in \bOmega$.
\end{enumerate}

The first two conditions, \ref{reg:compact} and \ref{reg:bounded}, are conditions that show consistency of the estimator and are typically assumed in large sample theory \parencite{newey1994large}. Condition \ref{reg:smooth-A} assumes that the solution of the $m$-estimating equation is well-behaved and unique, two other standard conditions under large sample theory. Finally, condition \ref{reg:smooth-A-alpha} regulates the influence of the nuisance parameters by assuming finite eigenvalues. This assumption is reasonable since we consider a finite-dimensional nuisance parameter set, and in our implementation, we consider a parametric model for estimating $\balpha$ that ensures that the function is sufficiently smooth. Additionally, we assume that all probabilities are bounded (not including) 0 and 1. For the case of the CC estimator, condition \ref{reg:smooth-A-alpha} is not necessary since the $m$-estimating equation does not depend on $\balpha$. Throughout, we work to prove that $\wh\btheta$ is a consistent estimator of $\btheta$ by showing its estimating equation is unbiased; i.e., $E \{ \bPhi_{\rm est}(\bO;  \btheta_0; \balpha_0) \} = \0$. By showing that the estimator is unbiased and assuming standard regularity conditions \ref{reg:compact}-\ref{reg:smooth-A-alpha}, the estimator $\wh\btheta$ is consistent and asymptotically normal by following Theorem 2.6 of \textcite{newey1994large}.

\begin{enumerate}   

\item {\bf Nuisance distribution estimators:}

Since $\int f_{W,\Delta|\bZ}(w,\delta,\bz; \balpha_0) dw d\delta = 1$, it follows that 
\be
\bzero &=& \frac{\partial}{\partial \balpha^T} \int  f_{W,\Delta|\bZ}(w,\delta,\bz; \balpha_0) dw d\delta \nonumber\\
&=& \int \biggr\{ \frac{\partial}{\partial \balpha^T}  \log f_{W,\Delta|\bZ}(w,\delta,\bz; \balpha_0) \biggr\} f_{W,\Delta|\bZ}(w,\delta,\bz; \balpha_0) dw d\delta \nonumber
\\
&=& \int \bS_{\balpha}(w,\delta,\bz;\balpha_0) f_{W,\Delta|\bZ}(w,\delta,\bz; \balpha_0) dw d\delta \nonumber
\\
&=& E_{W,\Delta|\bZ}\{\bS_{\balpha}(W,\Delta,\bZ;\balpha_0)\}. \label{eqn:proof-zero}
\ee
The second equality holds by standard regularity conditions that allow us to interchange the order of the integral and partial derivative. It therefore follows that 
\bse
E_{W,\Delta,\bZ}\{\bS_{\balpha}(W,\Delta,\bZ;\balpha_0)\} &=& E_{\bZ} [ E_{W,\Delta|\bZ} \{\bS_{\balpha}(W,\Delta,\bZ;\balpha_0)\} ] = \bzero.
\ese

\item {\bf CC estimator:}

Note that $E_{Y|X,\bZ} \{  \bS_{\btheta}^F (Y,x,\bz; \btheta_0) \} =\bzero$ following a similar argument as in equation \eqref{eqn:proof-zero}, where  $\bS_{\balpha}$ is replaced by   $\bS_{\btheta}^F$. Then,  
\bse
E\left\{\Delta \bS_{\btheta}^F (Y,W,\bZ; \btheta_0)\right\} &=& E_{Y,\bZ} [ E_{W,\Delta|Y,\bZ} \left\{\Delta \bS_{\btheta}^F (Y,W,\bZ; \btheta_0)\right\} ] \\
&=& E_{Y,\bZ} [ E_{X|Y,\bZ} \left\{\pi_{X,\bZ}(
X,\bZ; \boldeta_0) \bS_{\btheta}^F (Y,X,\bZ; \btheta_0)\right\} + E_{C|Y,\bZ} ( \0 ) ] \\
&=&  E_{X,\bZ} [ \pi_{X,\bZ}(X,\bZ; \boldeta_0) E_{Y|X,\bZ} \{  \bS_{\btheta}^F (Y,X,\bZ; \btheta_0) \} ]  = \bzero.
\ese

\item {\bf IPW estimator:}
\bse
E\left\{\frac{\delta \bS_{\btheta}^F (Y,W,\bZ; \btheta_0)}{ \pi_{X,\bZ}(W,\bZ; \boldeta_0)}\right\} &=& E_{Y,\bZ} \biggr[ E_{W,\Delta|Y,\bZ} \left\{\frac{\delta \bS_{\btheta}^F (Y,W,\bZ; \btheta_0)}{ \pi_{X,\bZ}(W,\bZ; \boldeta_0)}\right\} \biggr] \\
&=& E_{Y,\bZ} \biggr[ E_{X|Y,\bZ} \left\{\frac{ \pi_{X,\bZ}(X,\bZ; \boldeta_0)}{ \pi_{X,\bZ}(X,\bZ; \boldeta_0)} \times \bS_{\btheta}^F (Y,X,\bZ; \btheta_0)\right\} + E_{C|Y,\bZ} ( \0 ) \biggr] \\
&=& E_{Y,X,\bZ} \left\{ \bS_{\btheta}^F (Y,X,\bZ; \btheta_0)\right\}  = \bzero.
\ese

\item {\bf MLE:}
\bse
E\{\bPhi_{\rm{MLE}}(\bO; \btheta_0, \bgamma_0) \} &=& E\left\{{\delta\bS^F_{\btheta}(Y, W, \bZ; \btheta_0)} \right\} \\
&& +
 E\left\{ \left(1-{\delta}\right)\frac{\int_{c < X} \bS^F_{\btheta}(y,x,\bz; \btheta_0) f_{Y|X,\bZ}(y,x,\bz; \btheta_0)f_{X|\bZ}(x,\bz; \bgamma_0) dx}{\int_{c < X} f_{Y|X,\bZ}(y,x,\bz; \btheta_0)f_{X|\bZ}(x,\bz; \bgamma_0) dx}
\right\}.
\ese
The first term has a mean of zero, as it corresponds to the estimating equation for the CC estimator. The second term also has a mean of zero, as shown by first noting it is equivalent to the following: 
\bse
\int_{Y,W,\Delta,\bZ} \biggr\{ \left(1-{\delta}\right)\frac{\int_{c < X} \bS^F_{\btheta}(y,x,\bz; \btheta_0)f_{Y|X,\bZ}(y,x,\bz; \btheta_0)f_{X|\bZ}(x,\bz; \bgamma_0) dx}{\int_{c < X} f_{Y|X,\bZ}(y,x,\bz; \btheta_0)f_{X|\bZ}(x,\bz; \bgamma_0) dx}
f_{Y,W,\Delta,\bZ}(y, w, \delta, \bz) \biggr\} dydwd\delta d\bz.
\ese 
Using the assumptions that $C \independent Y|X,\bZ$, it follows that 
\bse
f_{Y,W,\Delta,\bZ}(y, w, \delta, \bz)  &=& \biggr\{ \int_{x \leq C} f_{Y|X,\bZ}(y,w,\bz;\btheta_0)f_{C,X|\bZ}(c,w,\bz;\balpha_0) dc \biggr\}^{\delta}  \\
&& \quad \quad + \biggr\{ \int_{c < X} f_{Y|X,\bZ}(y,x,\bz; \btheta_0)f_{C,X|\bZ}(w,x,\bz;\balpha_0)  dx \biggr\}^{1-\delta}.
\ese
When $\delta=1$, the expectation is $\bzero$. When $\delta=0$ the expectation is equal to 
\bse
&=& \int_{Y,C,\bZ} \biggr\{ \frac{\int_{c \leq X} \bS^F_{\btheta}(y,x,\bz; \btheta_0)f_{Y|X,\bZ}(y,x,\bz; \btheta_0)  dx}{\int_{c \leq X} f_{Y|X,\bZ}(y,x,\bz; \btheta_0)f_{X|\bZ}(x,\bz; \bgamma_0) dx} \biggr\} 
\\
&& \quad \times \biggr\{ \int_{c \leq X} f_{Y|X,\bZ}(y,x,\bz; \btheta_0)f_{X|\bZ}(x,\bz; \bgamma_0)  dx \biggr\} f_{C|\bZ}(c, \bz; \boldeta_0) f_\bZ(\bz)  dydcd\bz \\
&=& \int_{Y,C,\bZ} \int_{c \leq X} \bS^F_{\btheta}(y,x,\bz; \btheta_0) f_{Y|X,\bZ}(y,x,\bz; \btheta_0)f_{X|\bZ}(x,\bz; \bgamma_0) dx f_{C|\bZ}(c, \bz; \boldeta_0) f_\bZ(\bz) dydcd\bz
\\
&=& \int_{C,\bZ} \int_{c \leq X} \biggr\{ \int_Y \bS^F_{\btheta}(y,x,\bz; \btheta_0)f_{Y|X,\bZ}(y,x,\bz; \btheta_0) dy \biggr\} f_{X|\bZ}(x,\bz; \bgamma_0)dx  f_{C,\bZ}(c, \bz) dcd\bz \\
&=& \int_{C,\bZ} \int_{c \leq X}  \bzero \times  f_{X|\bZ}(x,\bz; \bgamma)dx  f_{C|\bZ}(c, \bz; \boldeta_0) f_\bZ(\bz) dcd\bz = \bzero.
\ese 
The last line follows since $\int \bS^F_{\btheta}(y,x,\bz; \btheta_0)f_{Y|X,\bZ}(y,x,\bz; \btheta_0) dy = E_{Y|X,\bZ} \{ \bS^F_{\btheta}(Y,x,\bz; \btheta_0)\} = \bzero$. Therefore, $E\{\bPhi_{\rm{MLE}}(\bO; \btheta_0, \bgamma_0) \} = \bzero$.

\item {\bf ACC estimator:} 
It suffices to show that the expectation of the augmentation part of the ACC estimator is zero since we already showed its first part, the estimating equation for the CC estimator,  has mean zero.  Observe  that
\bse
&& E_{Y,W,\Delta,\bZ} [ \{\Delta- \pi_{Y,\bZ}(Y,\bZ; \bkappa_0) \}\bPsi_{\rm ACC}(Y, \bZ; \btheta_0)]  \\
&& = E_{Y,\bZ}[ E_{W,\Delta|Y,\bZ}\{\Delta - \pi_{Y,\bZ}(Y,\bZ; \bkappa_0) \}\bPsi_{\rm ACC}(Y, \bZ; \btheta_0)].
\ese 
Consistency is achieved if $E_{W,\Delta|Y,\bZ}\{\Delta - \pi_{Y,\bZ}(y,\bz; \bkappa_0) \} = \0$. That result follows since
\bse
E_{W,\Delta|Y,\bZ}\{\Delta - \pi_{Y,\bZ}(y,\bz; \bkappa_0) \} &=& E_{W,\Delta|Y,\bZ}(\Delta) - \pi_{Y,\bZ}(y,\bz; \bkappa_0) \\
&=& \pi_{Y,\bZ}(y,\bz; \bkappa_0) - \pi_{Y,\bZ}(y,\bz; \bkappa_0) = \0.
\ese 

Since $\bLambda_{\rm ACC}$ is a non-random matrix, the proof of consistency stays the same when modifying the ACC estimator to include   $\bLambda_{\rm ACC}$.

\item {\bf MACC estimator:}

Like the ACC estimator, it suffices to show that the augmentation part of the MACC estimator has mean zero. Observe  that
\bse
&& E_{Y,W,\Delta,\bZ} [ \{1 - \Delta/\pi_{X,\bZ}(W,\bZ; \boldeta_0) \}\bPsi_{\rm MACC}(Y, \bZ; \btheta_0)] \\
&& = E_{Y,\bZ}[ E_{W,\Delta|Y,\bZ}\{1-\Delta/\pi_{X,\bZ}(W,\bZ; \boldeta_0) \}\bPsi_{\rm MACC}(Y, \bZ; \btheta_0)],
\ese 
where we achieve consistency if $E_{W,\Delta|Y,\bZ}\{1-\Delta/\pi_{X,\bZ}(W,\bz; \boldeta_0) \} = 0$. Using rules of conditional expectation, it follows that
\bse
E_{W,\Delta|Y,\bZ}\{1-\delta/\pi_{X,\bZ}(W,\bz; \boldeta_0) \} &=& 1-E_{W,\Delta|Y,\bZ}\{\Delta/\pi_{X,\bZ}(W,\bz; \boldeta_0) \} \\
&=& 1-E_{X|Y,\bZ}\{\pi_{X,\bZ}(X,\bz; \boldeta_0) / \pi_{X,\bZ}(X,\bz; \boldeta_0) \}\\
&=& 1 - 1 = 0.
\ese 
Since $E_{W,\Delta|Y,\bZ}\{1-\Delta/\pi_{X,\bZ}(W,\bz; \boldeta_0) \} = 0$, then it follows that $E_{Y,W,\Delta,\bZ} \{ \bPhi_{\rm MACC}(\bO; \btheta_0, \boldeta_0) \} = \bzero$. 

Since $\bLambda_{\rm MACC}$ is assumed to be a non-random matrix, the same consistency properties apply when the MACC estimator uses $\bLambda_{\rm MACC}$.

\item {\bf AIPW estimator:}

The estimating equation for the AIPW estimator combines the IPW estimating equations with an augmentation part. We already showed that the IPW estimating equations have mean zero. To prove consistency, it remains to be shown that the augmentation part of the AIPW estimating equations also has a mean of zero. In that augmentation part, $\bPsi_{\rm AIPW}(y, \bz; \btheta_0)$ is any function of $(Y,\bZ;\btheta_0)$. By interchanging the roles of $\bPsi_{\rm AIPW}(y, \bz; \btheta_0)$ with $\bPsi_{\rm MACC}(y, \bz; \btheta_0)$, the proof of consistency and robustness is the same as that for the MACC estimator. 

Since $\bLambda_{\rm AIPW}$ is a non-random matrix, the same consistency properties hold when the AIPW estimator uses $\bLambda_{\rm AIPW}$.

\end{enumerate}

\subsection{Robustness}
\label{sec:thm1-proof-part1-robustness}

We explain when each estimator is robust to the misspecification of the nuisance distributions. 
\begin{enumerate}
    \item {\bf CC estimator:} 

The CC estimator is robust to the misspecification of $f_{C|X,\bZ}$, $f_{X|\bZ}$, and $f_{\bZ}$, as these nuisance distributions  are not used in estimating $\btheta$. 

\item {\bf IPW estimator:}

Consider the case that $f_{C|\bZ}$ is incorrectly specified as $f_{C|\bZ}^*$. Let  $\pi^*_{X,\bZ}(x,\bz; \boldeta) = \int_{x \leq C} f^*_{C|\bZ} (c,\bz; \boldeta) dc$. Then,
\bse
E\left\{\frac{\delta \bS_{\btheta}^F (Y,W,\bZ; \btheta_0)}{ \pi^*_{X,\bZ}(W,\bZ; \boldeta)}\right\}
&=& E_{Y,\bZ} \biggr[ E_{X|Y,\bZ} \left\{\frac{ \pi_{X,\bZ}(X,\bZ; \boldeta_0)}{ \pi^*_{X,\bZ}(X,\bZ; \boldeta)} \times \bS_{\btheta}^F (Y,X,\bZ; \btheta_0)\right\} + E_{C|Y,\bZ} ( \0 ) \biggr] \\
&=& E_{X,\bZ} \biggr[ \frac{ \pi_{X,\bZ}(X,\bZ; \boldeta_0)}{ \pi^*_{X,\bZ}(X,\bZ; \boldeta)} \times  E_{Y|X,\bZ} \left\{ \bS_{\btheta}^F (Y,X,\bZ; \btheta_0)\right\} \biggr] \\
&=& E_{X,\bZ} \biggr[ \frac{ \pi_{X,\bZ}(X,\bZ; \boldeta_0)}{ \pi^*_{X,\bZ}(X,\bZ; \boldeta)} \times  \bzero \biggr] = \bzero.
\ese
The last line follows since $E_{Y|X,\bZ} \left\{ \bS_{\btheta}^F (Y,x,\bz; \btheta_0)\right\} = \bzero$. Therefore, the IPW estimator is robust to the misspecification of $f_{C|\bZ}$, $f_{X|\bZ}$, and $f_{\bZ}$.

Now consider the case when the probability of being observed is defined as $\pi_{Y,\bZ}(y,\bz; \bkappa_0) \equiv \pr(\Delta=1 | Y=y, \bZ = \bz; \bkappa_0)$. Then,
\bse
E\left\{\frac{\delta \bS_{\btheta}^F (Y,W,\bZ; \btheta_0)}{ \pi_{Y,\bZ}(Y,\bZ; \bkappa_0)}\right\}
&=& E_{Y,\bZ} \biggr[ E_{X|Y,\bZ} \left\{\frac{ \pi_{X,\bZ}(X,\bZ; \boldeta_0)}{ \pi_{Y,\bZ}(Y,\bZ; \bkappa_0)} \times \bS_{\btheta}^F (Y,X,\bZ; \btheta_0)\right\} + E_{C|Y,\bZ} ( \0 ) \biggr] \\
&=& E_{X,\bZ} \biggr[  \pi_{X,\bZ}(X,\bZ; \boldeta_0) \times  E_{Y|X,\bZ} \left\{ \frac{\bS_{\btheta}^F (Y,X,\bZ; \btheta_0)} {\pi_{Y,\bZ}(Y,\bZ; \bkappa_0)}\right\} \biggr] \\
&\ne & E_{X,\bZ} \left\{  \pi_{X,\bZ}(X,\bZ; \boldeta_0) \times \bzero  \right\}.
\ese
Since $E_{Y|X,\bZ} \left\{ \bS_{\btheta}^F (Y,x,\bz; \btheta_0)/\pi_{Y,\bZ}(Y,\bz; \bkappa_0)\right\} \ne \bzero$, using the probability $\pi_{Y,\bZ}(y,\bz;\bkappa_0)$ will result in an inconsistent estimator of $\btheta$ for our setting.
%

\item {\bf MLE:}

Suppose we misspecify the nuisance distribution $f_{X|\bZ}(x,\bz; \bgamma)$ as  $f_{X|\bZ}^*(x,\bz; \bgamma)$. The part corresponding to the CC estimator remains consistent, but the second part is now
\bse
&& \int_{Y,C,\bZ} \biggr\{ \frac{\int_{c < X} \bS^F_{\btheta}(y,x,\bz; \btheta_0)f_{Y|X,\bZ}(y,x,\bz; \btheta_0)f^*_{X|\bZ}(x,\bz; \bgamma) dx}{\int_{c < X} f_{Y|X,\bZ}(y,x,\bz; \btheta_0)f^*_{X|\bZ}(x,\bz; \bgamma) dx} \biggr\} 
\\
&& \quad \times \biggr\{ \int_{c \leq X} f_{Y|X,\bZ}(y,x,\bz; \btheta_0)f_{X|\bZ}(x,\bz; \bgamma)  dx \biggr\} f_{C,\bZ}(c, \bz) dydcd\bz \\
&& \ne \int_{Y,C,\bZ} \int_{c < X} \bS^F_{\btheta}(y,x,\bz; \btheta_0) f_{Y|X,\bZ}(y,x,\bz; \btheta_0)f^*_{X|\bZ}(x,\bz; \bgamma) f_{C,\bZ}(c, \bz)dxdydcd\bz.
\ese
In this case, the integral in the denominator on the first line would no longer cancel with the integral with respect to $X$ on the second line. Therefore, the conditional integral with respect to $Y$ is not equal to $\bzero$. There is no way to generally simplify this expression as is. Therefore, the MLE is not robust to the misspecification of $f_{X|\bZ}$, but it is robust to the misspecification of $f_{C|\bZ}$ and $f_{\bZ}$.

\item {\bf ACC estimator:}

To show that the ACC estimator is only singly robust, it suffices to show that the estimating equation is unbiased only when the probabilities are correctly specified. We have already proved that the ACC estimator is consistent when the probabilities are correctly specified. Now, consider the case that $\pi_{Y,\bZ}(y,\bz; \bkappa)$ is incorrectly specified as $\pi_{Y,\bZ}^*(y,\bz; \bkappa_0)$ and $\bPsi_{\rm ACC} (y,\bz;\btheta_0) = \bPsi_{\rm ACC}^{\rm eff} (y,\bz;\btheta_0)$, where $\bPsi_{\rm ACC}^{\rm eff} (y,\bz;\btheta_0) = - E_{X|Y,\bZ,\Delta=1}\{ \bS^F_{\btheta}(y,X,\bz;\btheta_0)\}$. We later show that this choice guarantees higher efficiency over the CC estimator. Then, it follows
\bse
&& E_{Y,W,\Delta,\bZ}[\{\Delta - \pi_{Y,\bZ}^*(Y,\bZ; \bkappa) \}\bPsi_{\rm ACC}(Y,\bZ;\btheta_0)] \\
&& = E_{Y,\bZ}[\{E_{W,\Delta|Y,\bZ}(\Delta) - \pi_{Y,\bZ}^*(Y, \bZ; \bkappa) \}\bPsi_{\rm ACC}(Y, \bZ;\btheta_0)] \\
&& = -E_{Y,\bZ}[\{ \pi_{Y,\bZ}(Y, \bZ; \bkappa_0) - \pi_{Y,\bZ}^*(Y, \bZ; \bkappa) \} E_{X|Y,\bZ,\Delta=1}\{ \bS^F_{\btheta}(Y,X,\bZ;\btheta_0)\}] \\
&& = -E_{Y,\bZ}[ E_{X|Y,\bZ,\Delta=1}\{ \pi_{Y, \bZ}(Y, \bZ; \bkappa_0) \bS^F_{\btheta}(Y,X,\bZ;\btheta_0)\}] \\
&& \quad \quad + E_{Y,\bZ}[ E_{X|Y,\bZ,\Delta=1}\{ \pi_{Y,\bZ}^*(Y, \bZ; \bkappa) \bS^F_{\btheta}(Y,X,\bZ;\btheta_0)\}] \\
&& \ne \bzero,
\ese 
where the last line follows since $\pi_{Y,\bZ}(y,\bz; \bkappa_0) \ne \pi_{Y,\bZ}^*(y,\bz; \bkappa)$. This result shows that the ACC estimator is consistent only when $\pi_{Y,\bZ}(y,\bz; \bkappa_0) \equiv f_{\Delta|Y,\bZ}(\delta=1, y,\bz; \bkappa_0)$ is correctly specified. Therefore, the ACC estimator is not robust to the misspecification of $f_{\Delta|Y,\bZ}$    but is robust to the misspecification of $f_{C,X|\bZ}$ and $f_{\bZ}$.

Since $\bLambda_{\rm ACC}$ is a non-random matrix, the same robustness properties apply as with the ACC estimator without $\bLambda_{\rm ACC}$.

\item \textbf{MACC estimator:} 

To show that the MACC estimator is only singly robust, it suffices to show that the estimating equation is unbiased only when the probabilities are correctly specified. We have already proved that the MACC estimator is consistent when the probabilities are correctly specified. Now, consider the case when $f_{C|X,\bZ}$ is incorrectly specified as $f_{C|X,\bZ}^*$ and $\bPsi_{\rm MACC} (y,\bz;\btheta_0) = \bPsi_{\rm MACC}^{\rm eff} (y,\bz;\btheta_0)$, where $\bPsi_{\rm ACC}^{\rm eff} (y,\bz;\btheta_0) = E_{X|Y,\bZ}[\{\pi_{X, \bZ} (X,\bZ; \boldeta_0)-1\} \bS^F_{\btheta} (Y,X,\bZ;\btheta_0)]/ E_{X|Y,\bZ}\{1-1/\pi_{X,\bZ}(X,\bZ; \boldeta_0) \}$. We later show that this choice  guarantees higher efficiency over the CC estimator. Let  $\pi^*_{X,\bZ}(x,\bz; \boldeta) = \int_{x \leq C} f^*_{C|\bZ} (c,\bz; \boldeta^*) dc$. Then,
\bse
&& E_{Y,W,\Delta,\bZ}[\{1 - \Delta/\pi_{X,\bZ}^*(X,\bZ; \boldeta) \}\bPsi^{\rm eff}_{\rm MACC}(Y,\bZ;\btheta_0)] \\
&& = E_{Y,\bZ} ( [ 1-E_{W,\Delta|Y,\bZ}\{\Delta/\pi^*_{X,\bZ}(W,\bZ; \boldeta) \} ] \bPsi^{\rm eff}_{\rm MACC}(Y,\bZ;\btheta_0) )\\
&& = E_{Y,\bZ} ( [ 1-E_{X|Y,\bZ}\{\pi_{X,\bZ}(X,\bZ; \boldeta_0)/\pi^*_{X,\bZ}(X,\bZ; \boldeta) \} ] \bPsi^{\rm eff}_{\rm MACC}(Y,\bZ;\btheta_0) ) \\
&& = E_{Y,\bZ} \left( \left[ 1-E_{X|Y,\bZ} \left\{ \frac{\pi_{X,\bZ}(X,\bZ; \boldeta_0)}{\pi_{X,\bZ}^*(X,\bZ; \boldeta)} \right\} \right] \frac{E_{X|Y,\bZ}[\{\pi_{X, \bZ} (X,\bZ; \boldeta_0)-1\} \bS^F_{\btheta} (Y,X,\bZ;\btheta_0)]}{ E_{X|Y,\bZ}\{1-1/\pi_{X,\bZ}(X,\bZ; \boldeta_0) \}} \right) \\
&& \ne \bzero,
\ese 
where the last line follows since $E_{X|Y,\bZ}\{\pi_{X,\bZ}(X,\bZ; \boldeta_0) / \pi^*_{X,\bZ}(X,\bZ; \boldeta) \} \ne 1$. Therefore, the MACC estimator is only consistent when $f_{C|\bZ}$ is correctly specified, but is robust to misspecification to $f_{X|\bZ}$ and $f_{\bZ}$.

Since $\bLambda_{\rm MACC}$ is assumed to be a non-random matrix, the same robustness properties apply as in the regular MACC.

\item {\bf AIPW estimator:} 

The AIPW estimator has the same robustness as the MACC estimator.   The augmented components  $\bPsi_{\rm AIPW}(y, \bz; \btheta_0)$ and $\bPsi_{\rm MACC}(y, \bz; \btheta_0)$ can be interchanged as these vectors can be any vector that is a function of $(Y,\bZ)$ and of the same length as $\btheta$. Therefore, the AIPW estimator is only consistent when $f_{C|\bZ}$ is correctly specified but is robust to the misspecification to $f_{X|\bZ}$ and $f_{\bZ}$. 

When we use $\bLambda_{\rm AIPW}$ in the AIPW estimator, the robustness stays the same since $\bLambda_{\rm AIPW}$ remains a fixed matrix like $\bLambda_{\rm MACC}$.

\end{enumerate}
\subsection{Asymptotic Normality} 
\label{sec:thm1-proof-part1-normality}

%
We use the theory of $m$-estimating equations to prove asymptotic normality. 

\noindent\emph{Only unknown parameter is $\btheta$:} 
Since $\wh\btheta$ is the solution to $\sumi \bPhi_{\rm est}(\bO_i; \btheta) = \bzero$ and we proved that $E\{ \bPhi_{\rm est}(\bO;  \btheta) \} = \0$ for all $\btheta$, then we can  perform a Taylor series expansion about the true value of $\btheta$, denoted by $\btheta_0$: 
\bse
\bzero &=&  n^{-1/2}\sumi \bPhi_{\rm est}(\bO_i; \wh\btheta)\\
&=& n^{-1/2}\sumi \bPhi_{\rm est}(\bO_i; \btheta_0) + n^{-1}\left\{  \sumi \frac{\partial\bPhi_{\rm est}(\bO_i; \wt\btheta)}{\partial\btheta^T} \right\} n^{1/2}(\wh\btheta - \btheta_0)\\
&=&n^{-1/2}\sumi \bPhi_{\rm est}(\bO_i; \btheta_0) + E\left\{ \frac{\partial\bPhi_{\rm est}(\bO; \btheta_0)}{\partial\btheta^T} \right\} n^{1/2}(\wh\btheta - \btheta_0)+o_p(1),
\ese
where $\wt\btheta$ is an intermediate value between $\wh\btheta$ and $\btheta_0$ and $o_p(1)$ converges to zero in probability. The last equality follows by the Law of Large Numbers.  Under regularity condition \ref{reg:smooth-A}, we have
\bse
n^{1/2}(\wh\btheta - \btheta_0) &=& -\left[E\left\{ \frac{\partial\bPhi_{\rm est}(\bO; \btheta_0)}{\partial\btheta^T} \right\}\right]^{-1}n^{-1/2}\sumi \bPhi_{\rm est}(\bO_i; \btheta_0) +o_p(1). \ese
This result, combined with  the Central Limit Theorem and Law of Large Numbers, implies that 
\bse
n^{1/2}(\wh\btheta - \btheta_0) \rightarrow_d \Normal (\bzero, \bA^{-1} \bB \bA^{-T} ),
\ese
where $\bA_{\rm est} = E\biggr\{ \frac{\partial}{\partial \btheta^T} \bPhi_{\rm est}(\bO;\btheta_0)\biggr\} $ and $\bB_{\rm est}= E\biggr\{ \bPhi_{\rm est}(\bO;\btheta_0) ^{\otimes 2} \biggr\}$. The corresponding $i^{th}$ influence function of $\wh\btheta$ is $\bUpsilon_{\rm est}(\bO_i) = - \bA_{\rm est}^{-1} \bPhi_{\rm est}(\bO;\btheta_0)$, which is estimated by substituting $\wh\btheta$ for $\btheta_0$. The sandwich estimator for the asymptotic variance of $\wh\btheta$ is  $\var(\wh\btheta) = \wh\bA_{\rm est}\wh\bB_{\rm est} \wh\bA_{\rm est}^{-T}$, where $\wh\bA_{\rm est}$ and $\wh\bB_{\rm est}$ are $\bA_{\rm est}$ and $\bB_{\rm est}$ evaluated at $\wh\btheta$, respectively.


\noindent\emph{Additional unknown parameters besides $\btheta$:} Suppose we have additional unknown parameters $\balpha$ so that our $m$-estimating equation is of the form $\sumi\bPhi_{\rm est}(\bO^*_i;\btheta,\balpha)=\bzero$, where $\bO^* = (W,\Delta,\bZ)$. Suppose also that $\wh\balpha$ is the  solution to $\sumi\bPhi_{\balpha}(\bO^*_i;\balpha)=\bzero$. Using a first order Taylor series expansion, we have that 
\be
\label{eqn:alpha-expansion}
n^{1/2}(\wh{\balpha}-\balpha_0) =  -\left\{n^{-1}\sum_{i=1}^n\frac{\partial \bPhi_{\balpha}(\bO^*_i;\wt\balpha)}{\partial\balpha^T}\right\}^{-1}n^{-1/2}\sum_{i=1}^n \bPhi_{\balpha}(\bO^*_i;\balpha_0),
\ee
where $\wt\balpha$ is an intermediate value between $\wh\balpha$ and $\balpha_0$. Now, consider the first order Taylor expansion of $\bPhi_{\rm est}(\bO;\btheta,\balpha)$  about $\btheta_0$ and $\balpha_0$: 
\bse
\0 &=& n^{-1/2}\sumi \bPhi_{\rm est}(\bO_i;\wh\btheta_{\rm est;\balpha},\wh\balpha) \\
 &=& n^{-1/2}\sumi \bPhi_{\rm est}(\bO_i; \btheta_0,\balpha_0) +  n^{-1}  \sumi \frac{\partial\bPhi_{\rm est}(\bO_i; \wt\btheta,\wt\balpha)}{\partial\btheta^T}  n^{1/2}(\wh\btheta_{\rm est;\balpha} - \btheta_0)  \\
&& + n^{-1} \sumi\frac{\partial\bPhi_{\rm est}(\bO_i; \wt\btheta, \wt\balpha)}{\partial\balpha^T} n^{1/2}(\widehat{\balpha} - \balpha_{0}),
\ese
where $\wt\btheta$ is an intermediate value between $\wh\btheta_{\rm est;\balpha}$ and $\btheta_0$. Plugging in equation \eqref{eqn:alpha-expansion}, we have
\bse
\bzero&=&n^{-1/2}\sumi \bPhi_{\rm est}(\bO_i; \btheta_0,\balpha_0) +  n^{-1}  \sumi \frac{\partial\bPhi_{\rm est}(\bO_i; \wt\btheta,\wt\balpha)}{\partial\btheta^T}  n^{1/2}(\widehat{\btheta}_{\rm est; \balpha} - \btheta_0)  \\
&& - n^{-1} \sumi\frac{\partial\bPhi_{\rm est}(\bO_i; \wt\btheta, \wt\balpha)}{\partial\balpha^T} \left\{n^{-1}\sum_{i=1}^n\frac{\partial \bPhi_{\balpha}(\bO^*_i;\wt\balpha)}{\partial\balpha^T}\right\}^{-1}  n^{-1/2}\sumi \bPhi_{\balpha}(\bO^*_i; \balpha_0) +o_p(1).
\ese
Assuming regularity condition \ref{reg:smooth-A} for $\bPhi_{\rm est}(\bO_i;\btheta,\balpha)$, we have
\bse
n^{1/2}(\wh\btheta_{\rm est;\balpha}-\btheta_0)&=&-\left\{n^{-1}\sumi\frac{\partial \bPhi_{\rm est}(\bO_i;\wt\btheta,\wt\balpha)}{\partial\btheta^T}\right\}^{-1}\Bigg[n^{-1/2}\sumi \bPhi_{\rm est}(\bO_i; \btheta_0,\balpha_0) \\
&&- n^{-1} \sumi\frac{\partial\bPhi_{\rm est}(\bO_i; \wt\btheta, \wt\balpha)}{\partial\balpha^T} \left\{n^{-1}\sum_{i=1}^n\frac{\partial \bPhi_{\balpha}(\bO^*_i;\wt\balpha)}{\partial\balpha^T}\right\}^{-1} \\
&&\times n^{-1/2}\sumi \bPhi_{\balpha}(\bO^*_i; \balpha_0)
\Bigg] +o_p(1).
\ese
Further, using the Law of Large Numbers, the Central Limit Theorem, and Slutsky's Theorem, it follows that
\bse
n^{1/2}(\wh{\btheta}_{{\rm est; \balpha}}-\btheta_0) \rightarrow_d \Normal(\bzero, \bA_{\rm est;\balpha}^{-1}\bB_{\rm est;\balpha}\bA_{\rm est;\balpha}^{-T}), 
\ese
where $\bA_{\rm est;\balpha} = E\{\partial \bPhi_{\rm est}(\bO;\btheta_0,\balpha_0)/\partial\btheta^{\trans}\}$ and 
\bse
\bB_{\rm est;\balpha} &=&  E\{(\bPhi_{\rm est}(\bO; \btheta_0,\balpha_0) \\
&&- E\{\partial\bPhi_{\rm est}(\bO;\btheta_0,\balpha_0)/\partial\balpha^T\}[E\{\partial\bPhi_{\balpha}(\bO^*;\balpha_0)/\partial\balpha^T\}]^{-1}\bPhi_{\balpha}(\bO^*;\balpha_0) )^{\otimes2}\}.
\ese
The corresponding $i^{th}$ influence function of $\wh\btheta_{\rm est; \balpha}$  is 
\bse
\bUpsilon_{ \rm est; \balpha} (\bO_i) &=& \bA_{\rm est; \balpha}^{-1} \{ (\bPhi_{\rm est}(\bO_i; \btheta_0,\balpha_0) \\
&& \quad \quad - E\{\partial\bPhi_{\rm est}(\bO;\btheta_0,\balpha_0)/\partial\balpha^T\}[E\{\partial\bPhi_{\balpha}(\bO^*;\balpha_0)/\partial\balpha^T\}]^{-1}\bPhi_{\balpha}(\bO^*_i;\balpha_0) )\},
\ese
and the  estimate of the sandwich estimator for the asymptotic variance of $\wh\btheta_{\rm est;\balpha}$, under the presence of $\wh\balpha$, is $\var(\wh\btheta_{\rm est;\balpha}) \equiv \var\{ \bUpsilon_{ \rm est; \balpha} (\bO) \} = \wh\bA_{\rm est; \balpha}^{-1} \wh\bB_{\rm est; \balpha} \wh\bA_{\rm est; \balpha}^{-T}$, where $\wh\bA_{\rm est;\balpha}$ and $\wh\bB_{\rm est;\balpha}$ are $\bA_{\rm est;\balpha}$ and $\bB_{\rm est;\balpha}$ evaluated at $(\wh\btheta_{\rm est;\balpha}, \wh\balpha)$, respectively. We apply these results to show that the following estimators are asymptotically normally distributed. 

\begin{enumerate}
    \item {\bf Nuisance distribution estimators:}

Estimators of the nuisance distribution parameters are asymptotically normal: 
\bse
n^{1/2}(\widehat{\balpha} - \balpha_0) \rightarrow_d \Normal (\bzero, \bA_{\balpha}^{-1} \bB_{\balpha} \bA_{\balpha}^{-T} ),
\ese
where $\bA_{\balpha} = E_{W,\Delta,\bZ}\left\{ \frac{\partial}{\partial \balpha^T} \bPhi_{\balpha}(\bO^*; \balpha_0) \right\} $ and $\bB_{\balpha} = E_{W,\Delta,\bZ}\left\{ \bPhi_{\balpha}(\bO^*; \balpha_0) ^{\otimes 2} \right\}$.

\item {\bf CC estimator:} 

The CC estimator, $\wh\btheta_{\rm CC}$, is asymptotically normally distributed: 
\bse
n^{1/2}(\widehat{\btheta}_{\rm CC} - \btheta_0) \rightarrow_d \Normal \{\bzero, \bA_{\rm CC}^{-1} \bB_{\rm CC}  \bA_{\rm CC}^{-T}\},
\ese
where $\bA_{\rm CC} =  E \left\{ \frac{\partial}{\partial \btheta^T} \bPhi_{\rm CC}(\bO; \btheta_0) \right\}$ and $\bB_{\rm CC} = E\{ \bPhi_{\rm CC}(\bO;  \btheta_0)^{\otimes 2} \}$. 

\item {\bf IPW estimator:}

The IPW estimator, $\wh\btheta_{\rm IPW}$, is asymptotically normally distributed. When the nuisance distribution  $f_{C|\bZ}$ is known, we have that 
\bse
n^{1/2}(\wh{\btheta}_{{\rm IPW}}-\btheta_0) \rightarrow \Normal(\bzero, \bA_{\rm IPW}^{-1}\bB_{\rm IPW}\bA_{\rm IPW}^{-T}),
\ese
where $\bA_{\rm IPW} =  E \left\{ \frac{\partial}{\partial \btheta^T} \bPhi_{\rm IPW}(\bO; \btheta_0, \boldeta_0) \right\}$  and $\bB_{\rm IPW} =  E\{\bPhi_{\rm IPW}(\bO; \btheta_0,\boldeta_0)^{\otimes2}\}$.

When $f_{C|\bZ}$ is unknown, we propose a parametric nuisance distribution $f_{C|\bZ}(c,\bz;\boldeta)$ and estimate $\boldeta$ via estimating $\balpha=(\bgamma^T,\boldeta^T,)^T$, as in Section \ref{sec:est_nuisance_distributions}.
Adjusting for the added variability contributed by the unknown $\boldeta$ through $\balpha$, we have that 
\bse
n^{1/2}(\wh{\btheta}_{{\rm IPW;\balpha}}-\btheta_0) \rightarrow_d \Normal(\bzero, \bA_{\rm IPW;\balpha}^{-1}\bB_{\rm IPW;\balpha}\bA_{\rm IPW;\balpha}^{-T}), 
\ese
where $\bA_{\rm IPW;\balpha} = E\{\partial \bPhi_{\rm IPW}(\bO;\btheta_0,\balpha_0)/\partial\btheta^{\trans}\}$ and 
$\bB_{\rm IPW;\balpha} =  E\{(\bPhi_{\rm IPW}(\bO;\btheta_0,\balpha_0) - E\{\partial\bPhi_{\rm IPW}(\bO;\btheta_0,\balpha_0)/\partial\balpha^T\}[E\{\partial\bPhi_{\balpha}(\bO^*; \balpha)/\partial\balpha^T\}]^{-1}\bPhi_{\balpha}(\bO^*; \balpha) )^{\otimes2}\}$. 

\item {\bf MLE:}

The MLE, $\wh\btheta_{\rm MLE}$, is asymptotically normally distributed. When the nuisance distribution $f_{X|\bZ}$ is known, we have that 
\bse
n^{1/2}(\wh{\btheta}_{{\rm MLE}}-\btheta_0) \rightarrow \Normal\{\bzero, \bA_{\rm MLE}^{-1}\bB_{\rm MLE}\bA_{\rm MLE}^{-T}\}.
\ese
The components of the asymptotic variance are  
$\bA_{\rm MLE} =  E \biggr\{ \frac{\partial}{\partial \btheta^T} \bPhi_{\rm MLE}(\bO; \btheta_0, \bgamma_0) \biggr\}$ and 
$\bB_{\rm MLE} = E\{\bPhi_{\rm MLE}(\bO;\btheta_0,\bgamma_0)^{\otimes2}\}$.

When $f_{X|\bZ}$ is unknown, we propose a parametric nuisance distribution $f_{X|\bZ}(c,\bz;\bgamma)$ and estimate $\bgamma$ via estimating $\balpha=(\bgamma^T,\boldeta^T,)^T$, as in Section \ref{sec:est_nuisance_distributions}.
Adjusting for the added variability contributed by the unknown $\bgamma$ through $\balpha$, we have that 
\bse
n^{1/2}(\wh{\btheta}_{{\rm MLE;\balpha}}-\btheta_0) \rightarrow_d \Normal(\bzero, \bA_{\rm MLE;\balpha}^{-1}\bB_{\rm MLE;\balpha}\bA_{\rm MLE;\balpha}^{-T}), 
\ese
where $\bA_{\rm MLE;\balpha} = E\{\partial \bPhi_{\rm MLE}(\bO; \btheta_0,\balpha_0)/\partial\btheta^{\trans}\}$ and 
$\bB_{\rm MLE;\balpha} =  E\{(\bPhi_{\rm MLE}(\bO; \btheta_0,\balpha_0) - E\{\partial\bPhi_{\rm MLE}(\bO;\btheta_0,\balpha_0)/\partial\balpha^T\}[E\{\partial\bPhi_{\balpha}(\bO^*;\balpha_0)/\partial\balpha^T\}]^{-1}\bPhi_{\balpha}(\bO^*;\balpha_0) )^{\otimes2}\}$.

\item {\bf ACC estimator:}

The ACC estimator, $\wh\btheta_{\rm ACC}$, is asymptotically normally distributed. Suppose that   $f_{\Delta|Y,\bZ}$ is known
and specified as $f_{\Delta|Y,\bZ}(\delta,y,\bz;\bkappa)$.
When we do not use $\bLambda_{\rm ACC}$, we have that 
\bse
n^{1/2}(\wh{\btheta}_{{\rm ACC}}-\btheta_0) \rightarrow_d \Normal(\bzero, \bA_{\rm ACC}^{-1}\bB_{\rm ACC}\bA_{\rm ACC}^{-T}),
\ese
where  $\bA_{\rm ACC} = E \biggr\{ \frac{\partial}{\partial \btheta^T} \bPhi_{\rm ACC}(\bO; \btheta_0, \bkappa_0) \biggr\}$ and $\bB_{\rm ACC} = E\{\bPhi_{\rm ACC}(\bO;\btheta_0,\bkappa_0)^{\otimes2}\}$. When using $\bLambda_{\rm ACC}$, $\bA_{\rm ACC}$ and $\bB_{\rm ACC}$ become $\bA_{\rm ACC;\bLambda}$ and $\bB_{\rm ACC;\bLambda}$, respectively. The only difference is that $\bPhi_{\rm ACC}(\bO; \btheta_0, \bkappa_0)$ is replaced by $\bPhi_{\rm ACC;\bLambda}(\bO; \btheta_0, \bkappa_0)$.

Moreover, we can simplify $\bA_{\rm ACC}$ and  $\bA_{\rm ACC;\bLambda}$ since 
\bse
\bA_{\rm ACC} &=& E \biggr\{ \frac{\partial}{\partial \btheta^T} \bPhi_{\rm ACC}(\bO; \btheta_0, \bkappa_0) \biggr\} \\
&=& E \biggr[  \Delta  \frac{\partial}{\partial \btheta^T} \bS_\btheta^F(Y,W,\bZ;\btheta_0) + \left\{\Delta- \pi_{Y,\bZ}(Y, \bZ;\bkappa_0) \right\}  \frac{\partial}{\partial \btheta^T} \bPsi_{\rm ACC}(Y, \bZ; \btheta_0) \biggr]\\
&=& E \biggr[  \Delta  \frac{\partial}{\partial \btheta^T} \bS_\btheta^F(Y,W,\bZ;\btheta_0) \biggr] \ {\rm since\ the\ second\ part\ equals\ \0}
\ese
and
\bse
\bA_{\rm ACC;\bLambda} &=& E \biggr\{ \frac{\partial}{\partial \btheta^T} \bPhi_{\rm ACC; \bLambda}(\bO; \btheta_0, \bkappa_0) \biggr\} \\
&=& E \biggr[  \Delta  \frac{\partial}{\partial \btheta^T} \bS_\btheta^F(Y,W,\bZ;\btheta_0) + \left\{\delta- \pi_{Y,\bZ}(Y, \bZ;\bkappa_0) \right\} \bLambda_{\rm ACC}  \frac{\partial}{\partial \btheta^T} \bPsi_{\rm ACC}(Y, \bZ; \btheta_0) \biggr]\\
&=& E \biggr[  \Delta  \frac{\partial}{\partial \btheta^T} \bS_\btheta^F(Y,W,\bZ;\btheta_0) \biggr] \ {\rm since\ the\ second\ part\ equals\ \0.}
\ese
Therefore,  $\bA_{\rm ACC} = \bA_{\rm ACC;\bLambda}=\bA_{\rm CC}$, where $\bA_{\rm CC}$ is from the asymptotic distribution of the CC estimator.

Now suppose  $f_{\Delta|Y,\bZ}$ is unknown, and  we propose a parametric nuisance distribution $f_{\Delta|Y,\bZ}(\delta,y,\bz;\bkappa)$ and estimate $\bkappa$ using a logistic regression model with a logit-link function, as described in \textcite{Bartlettetal2014}. We have that
\bse
n^{1/2}(\wh{\btheta}_{{\rm ACC;\bkappa}}-\btheta_0) \rightarrow_d \Normal(\bzero, \bA_{\rm ACC;\bkappa}^{-1}\bB_{\rm ACC;\bkappa}\bA_{\rm ACC;\bkappa}^{-T}), 
\ese
where $\bA_{\rm ACC;\bkappa} = E\{\partial \bPhi_{\rm ACC}(\bO; \btheta_0,\bkappa_0)/\partial\btheta^{\trans}\}$ and 
$\bB_{\rm ACC;\bkappa} =  E\{(\bPhi_{\rm ACC}(\bO;\btheta_0,\bkappa_0) - E\{\partial\bPhi_{\rm ACC}(\bO;\btheta_0,\bkappa_0)/\partial\bkappa^T\}[E\{\partial\bS_{\bkappa}(Y,\bZ;\bkappa_0)/\partial\bkappa^T\}]^{-1}\bS_{\bkappa}(Y,\bZ;\bkappa_0) )^{\otimes2}\}$. The term $\bS_{\bkappa}(Y,\bZ;\bkappa_0)$ reflects the $m$-estimating equation with respect to the logistic regression model as a function of $(Y,\bZ)$ and indexed by $\bkappa$. 
%
When we do use $\bLambda_{\rm ACC}$, we have that  $\bA_{\rm ACC;\bkappa}$ and $\bB_{\rm ACC;\bkappa}$ become
$\bA_{\rm ACC;\bkappa,\bLambda}$ and 
$\bB_{\rm ACC;\bkappa,\bLambda}$, respectively. The only change is that $\bPhi_{\rm ACC}(\bO; \btheta_0, \bkappa_0)$ is replaced by $\bPhi_{\rm ACC;\bLambda}(\bO; \btheta_0, \bkappa_0)$. Similarly, it still follows that $\bA_{\rm ACC;\bkappa}=\bA_{\rm ACC;\bkappa,\bLambda}=\bA_{\rm CC}$.

\item \textbf{MACC estimator:} 

The MACC estimator, $\wh\btheta_{\rm MACC}$, is asymptotically normally distributed. Suppose that $f_{C|\bZ}$ is known and specified as $f_{C|\bZ}(c,\bz;\boldeta_0)$. When we do not use $\bLambda_{\rm MACC}$, 
\bse
n^{1/2}(\wh{\btheta}_{{\rm MACC}}-\btheta_0) \rightarrow_d \Normal\{\bzero, \bA_{\rm MACC}^{-1}\bB_{\rm MACC}\bA_{\rm MACC}^{-T}\}.
\ese
The components of the asymptotic variance are  $\bA_{\rm MACC} = E\{ \partial \bPhi_{\rm MACC}(\bO;\btheta_0,\boldeta_0)/ \partial \btheta^T\}$ and $\bB_{\rm MACC} = E\{\bPhi_{\rm MACC}(\bO;\btheta_0,\boldeta_0)^{\otimes2}\}$. When we do use $\bLambda_{\rm MACC}$, $\bA_{\rm MACC}$ and $\bB_{\rm MACC}$ become $\bA_{\rm MACC;\bLambda}$ and $\bB_{\rm MACC;\bLambda}$, respectively. The only change is that $\bPhi_{\rm MACC}(\bO;\btheta_0,\boldeta_0)$ is replaced by $\bPhi_{\rm MACC;\bLambda}(\bO;\btheta_0,\boldeta_0)$.

Moreover, $\bA_{\rm MACC} =\bA_{\rm MACC;\bLambda} = \bA_{\rm CC}$ since
\bse
\bA_{\rm MACC} &=& E \biggr\{ \frac{\partial}{\partial \btheta^T} \bPhi_{\rm MACC}(\bO; \btheta_0, \boldeta_0) \biggr\} \\
&=& E \biggr[  \Delta  \frac{\partial}{\partial \btheta^T} \bS_\btheta^F(Y,W,\bZ;\btheta_0) + \left\{ 1 - \frac{\Delta}{\pi_{X,\bZ}(W, \bZ;\boldeta_0)} \right\}  \frac{\partial}{\partial \btheta^T} \bPsi_{\rm MACC}(Y, \bZ; \btheta_0) \biggr]\\
&=& E \biggr[  \Delta  \frac{\partial}{\partial \btheta^T} \bS_\btheta^F(Y,W,\bZ;\btheta_0) \biggr] \ {\rm since\ the\ second\ part\ equals\ \0}
\ese
and
\bse
\bA_{\rm MACC; \bLambda} &=& E \biggr\{ \frac{\partial}{\partial \btheta^T} \bPhi_{\rm MACC; \bLambda}(\bO; \btheta_0, \boldeta_0) \biggr\} \\
&=& E \biggr[  \Delta  \frac{\partial}{\partial \btheta^T} \bS_\btheta^F(Y,W,\bZ;\btheta_0) + \biggr\{1- \frac{\Delta}{\pi_{X,\bZ}(W, \bZ; \boldeta_0)}\biggr\} \bLambda_{\rm MACC}  \frac{\partial}{\partial \btheta^T} \bPsi_{\rm MACC}(Y, \bZ; \btheta_0) \biggr]\\
&=& E \biggr[  \Delta  \frac{\partial}{\partial \btheta^T} \bS_\btheta^F(Y,W,\bZ;\btheta_0) \biggr] \ {\rm since\ the\ second\ part\ equals\ \0.}
\ese

Now suppose the nuisance distribution parameters $\boldeta$ are unknown in $f_{C|\bZ}(c,\bz;\boldeta_0)$, and they are estimated by estimating $\balpha$, as in Section \ref{sec:est_nuisance_distributions}. Then, when we do not use $\bLambda_{\rm MACC}$, 
\bse
n^{1/2}(\wh{\btheta}_{{\rm MACC; \balpha}}-\btheta_0) \rightarrow_d \Normal\{\bzero, \bA_{\rm MACC; \balpha}^{-1} \bB_{\rm MACC; \balpha} \bA_{\rm MACC; \balpha}^{-T}\}.
\ese
The components of the asymptotic variance are $\bA_{\rm MACC;\balpha} = \bA_{\rm CC}$ and $\bB_{\rm MACC; \balpha} =  E\{(\bPhi_{\rm MACC}(\bO;\btheta_0,\balpha_0) - E\{\partial\bPhi_{\rm MACC}(\bO;\btheta_0,\balpha_0)/\partial\balpha^T\} [E\{\partial\bPhi_{\balpha}(\bO^*;\balpha_0)/\partial\balpha^T\}]^{-1}  \bPhi_{\balpha}(\bO^*;\balpha_0) )^{\otimes2}\}$.  Let $\bA^*_{\rm MACC;\balpha} = E\{\partial\bPhi_{\rm MACC}(\bO;\btheta_0,\balpha_0)/\partial\balpha^T\}$, which becomes
\bse
&& \bA^*_{\rm MACC;\balpha} \\
&& = E\biggr\{ \frac{\partial}{\partial \balpha^T}    \bPhi_{\rm MACC}(\bO; \btheta_0, \boldeta_0) \biggr\} \\
&& = E \biggr(  \Delta  \frac{\partial}{\partial \balpha^T}  \bS_\btheta^F(Y,W,\bZ;\btheta_0) + \frac{\partial}{\partial \balpha^T} \biggr[ \biggr\{1- \frac{\delta}{\pi_{X,\bZ}(W, \bZ; \boldeta_0)} \biggr\} \bPsi_{\rm MACC}(Y, \bZ; \btheta_0) \biggr] \biggr)\\
&& = E \biggr( \biggr\{1- \frac{\Delta}{\pi_{X,\bZ}(W, \bZ; \boldeta_0)} \biggr\} \frac{\partial}{\partial \balpha^T} \bPsi_{\rm MACC}(Y, \bZ; \btheta_0) - \frac{\partial}{\partial \balpha^T}\biggr\{\frac{\delta}{\pi_{X,\bZ}(W, \bZ; \boldeta_0)} \biggr\}  \bPsi_{\rm MACC}(Y, \bZ; \btheta_0)  \biggr)\\
&& = E \biggr[  \frac{\Delta}{\pi_{X,\bZ}(W, \bZ; \balpha_0)^2}  \bPsi_{\rm MACC}(Y, \bZ; \btheta_0) \left\{  \frac{\partial}{\partial \balpha^T} \int_{w \leq c} f_{C|\bZ}(c,\bZ; \boldeta_0) dc \right\}  \biggr]. 
\ese
When we do use $\bLambda_{\rm MACC}$, $\bA_{\rm MACC;\balpha}$ becomes $\bA_{\rm MACC;\balpha,\bLambda}$ which still equals $\bA_{\rm CC }$. 
%
The matrix  $\bB_{\rm MACC; \balpha}$ becomes $\bB_{\rm MACC; \balpha,\bLambda}$. The only difference is that $\bPhi_{\rm MACC}(\bO;\btheta_0,\balpha_0)$ is replaced by $\bPhi_{\rm MACC; \bLambda}(\bO; \btheta_0,\balpha_0)$.
Finally, 
\bse
&& E\biggr\{ \frac{\partial}{\partial \balpha^T}    \bPhi_{\rm MACC; \bLambda}(\bO; \btheta_0, \boldeta_0) \biggr\} = \bLambda_{\rm MACC} \bA^*_{\rm MACC;\balpha}. 
\ese

\item \textbf{AIPW estimator} 

The AIPW estimator, $\wh\btheta_{\rm AIPW}$, is asymptotically normally distributed. Suppose that  $f_{C|\bZ}$ is known as $f_{C|\bZ}(c,\bz;\boldeta_0)$. Then, when $\bLambda_{\rm AIPW}$ is not used, 
\bse
n^{1/2}(\wh{\btheta}_{{\rm AIPW}}-\btheta_0) \rightarrow_d \Normal\{\bzero, \bA_{\rm AIPW}^{-1}\bB_{\rm AIPW}\bA_{\rm AIPW}^{-T}\}.
\ese
The components of the asymptotic variance are $\bA_{\rm AIPW} = E\{\partial \bPhi_{\rm AIPW}(\bO; \btheta_0, \boldeta_0)/\partial\btheta^{\trans}\}$ and $\bB_{\rm AIPW} = E\{\bPhi_{\rm AIPW}(\bO;\btheta_0,\boldeta_0)^{\otimes2}\}$. 
%
When we do use $\bLambda_{\rm AIPW}$, then $\bA_{\rm AIPW}$ and $\bB_{\rm AIPW}$ become  $\bA_{\rm AIPW;\bLambda}$ and $\bB_{\rm AIPW;\bLambda}$, respectively. The only difference is that $\bPhi_{\rm AIPW}(\bO;\btheta_0,\boldeta_0)$ is substituted by $\bPhi_{\rm AIPW;\bLambda}(\bO;\btheta_0,\boldeta_0)$.

Note that $\bA_{\rm AIPW} = \bA_{\rm AIPW;\bLambda}=\bA_{\rm IPW}$ since
\bse
\bA_{\rm AIPW} &=& E\biggr\{ \frac{\partial}{\partial \btheta^T}    \bPhi_{\rm AIPW}(\bO; \btheta_0, \boldeta_0) \biggr\} \\
&=& E \biggr[   \frac{\partial}{\partial \btheta^T}  \frac{\Delta \bS^F_\btheta(Y,W,\bZ;\btheta_0)}{\pi_{X,\bZ}(W, \bZ; \boldeta_0)} +  \biggr\{1- \frac{\delta}{\pi_{X,\bZ}(W, \bZ; \boldeta_0)} \biggr\}  \biggr\{ \frac{\partial}{\partial \btheta^T} \bPsi_{\rm AIPW}(Y, \bZ;\btheta_0) \biggr\} \biggr]\\
&=& E \biggr\{   \frac{\partial}{\partial \btheta^T}  \frac{\Delta \bS^F_\btheta(Y,W,\bZ;\btheta_0)}{\pi_{X,\bZ}(W, \bZ; \boldeta_0)} \biggr\}  \ {\rm since\ the\ second\ part\ equals\ \0}
\ese 
and
\bse
\bA_{\rm AIPW;\bLambda} &=& E\biggr\{ \frac{\partial}{\partial \btheta^T}    \bPhi_{\rm AIPW; \bLambda}(\bO; \btheta_0, \boldeta_0) \biggr\} \\
&=& E \biggr[   \frac{\partial}{\partial \btheta^T}  \frac{\Delta \bS^F_\btheta(Y,W,\bZ;\btheta)}{\pi_{X,\bZ}(W, \bZ; \boldeta_0)} +  \biggr\{1- \frac{\Delta}{\pi_{X,\bZ}(W, \bZ; \boldeta_0)} \biggr\} \bLambda_{\rm AIPW} \biggr\{ \frac{\partial}{\partial \btheta^T} \bPsi_{\rm AIPW}(Y, \bZ;\btheta) \biggr\} \biggr]\\
&=& E \biggr\{   \frac{\partial}{\partial \btheta^T}  \frac{\Delta \bS^F_\btheta(Y,W,\bZ;\btheta)}{\pi_{X,\bZ}(W, \bZ; \boldeta_0)} \biggr\}  \ {\rm since\ the\ second\ part\ equals\ \0.}
\ese 

When the nuisance distribution parameters $\boldeta$ need to be estimated by estimating $\balpha$ as in Section \ref{sec:est_nuisance_distributions}, then  
\bse
n^{1/2}(\wh{\btheta}_{{\rm AIPW;\balpha}}-\btheta_0) \rightarrow_d \Normal\{\bzero, \bA_{\rm AIPW}^{-1}\bB_{\rm AIPW; \balpha}\bA_{\rm AIPW}^{-T}\}.
\ese

The components of the asymptotic variance are $\bA_{\rm AIPW;\balpha} = \bA_{\rm IPW}$ and $\bB_{\rm AIPW; \balpha} =  E\{(\bPhi_{\rm AIPW}(\bO; \btheta_0,\balpha_0) - E\{\partial\bPhi_{\rm AIPW}(\bO;\btheta_0,\balpha_0)/\partial\balpha^T\} [E\{\partial\bPhi_{\balpha}(\bO^*;\balpha_0)/\partial\balpha^T\}]^{-1} \bPhi_{\balpha}(\bO^*;\balpha_0) )^{\otimes2}\}$. Let, $\bA^*_{\rm IPW;\balpha} = E\{\partial\bPhi_{\rm IPW}(\bO;\btheta_0,\balpha_0)/\partial\balpha^T\}$. Observe that 
\bse
&& E\biggr\{ \frac{\partial}{\partial \balpha^T}    \bPhi_{\rm AIPW}(\bO, \btheta_0, \balpha_{0}) \biggr\} \\
&& = E \left\{   \frac{\partial}{\partial \balpha^T}  \frac{\delta \bS^F_\btheta(y,w,\bz;\btheta_0)}{\pi_{X,\bZ}(w, \bz; \boldeta_0)} \right\} + E \left( \frac{\partial}{\partial \balpha^T} \biggr[\biggr\{1- \frac{\delta}{\pi_{X,\bZ}(w, \bz; \boldeta_0)} \biggr\} \bPsi_{\rm AIPW}(y, \bz; \btheta_0) \biggr] \right)\\
&& = \bA^*_{\rm IPW;\balpha} - \bA^*_{\rm AIPW; \balpha}.
\ese 

%
When we do use $\bLambda_{\rm AIPW}$, we now use $\bA_{\rm AIPW;\balpha,\bLambda}$ and $\bB_{\rm AIPW; \balpha,\bLambda}$. We have that $\bA_{\rm AIPW;\balpha,\bLambda}$ still equals $\bA_{\rm IPW}$. The matrix $\bB_{\rm AIPW; \balpha,\bLambda}$  is the same as $\bB_{\rm AIPW; \balpha}$, except that 
$\bPhi_{\rm AIPW}(\bO;\btheta_0,\balpha_0)$ changes to $\bPhi_{\rm AIPW; \bLambda}(\bO;\btheta_0,\balpha_0)$.
Finally, 
\bse
&& E\biggr\{ \frac{\partial}{\partial \balpha^T}    \bPhi_{\rm AIPW; \bLambda}(\bO; \btheta_0, \balpha_0) \biggr\}  = \bA^*_{\rm IPW;\balpha} - \bLambda_{\rm AIPW} \bA^*_{\rm AIPW; \balpha}.
\ese

\end{enumerate}

\subsection{Efficiency}
\label{sec:thm1-proof-part1-efficiency}

We now show how the augmented estimators were constructed to gain efficiency.

\begin{enumerate}
    \item {\bf ACC estimator:}

\emph{When only $\btheta$ is unknown and we do not use $\bLambda_{\rm ACC}$:}    
We choose $\bPsi_{\rm ACC}(y, \bz; \btheta)$ so the ACC estimator is more efficient than the CC estimator. To do so, we compare the influence functions of $\btheta_{\rm CC}$ and $\btheta_{\rm ACC}$, and we find the form of $\bPsi_{\rm ACC}(y, \bz; \btheta_0)$ that makes the efficiency condition in equation \eqref{eqn:efficiency-condition} true. Satisfying the efficiency condition is equivalent to 
\bse
&& \cov \biggr[\Delta \bS_\btheta^F(Y,X,\bZ;\btheta_0) +  \left\{ \Delta - \pi_{Y,\bZ}(Y, \bZ;\bkappa_0) \right\} \bPsi_{\rm ACC}(Y,\bZ; \btheta_0) \biggr] \\
&=& \cov\biggr( \Delta \bS_\btheta^F(Y,W,\bZ;\btheta_0), \Delta \bS_\btheta^F(Y,X,\bZ;\btheta_0) +  \left\{ \Delta - \pi_{Y,\bZ}(Y, \bZ;\bkappa_0) \right\} \bPsi_{\rm ACC}(Y,\bZ; \btheta_0) \biggr).
\ese
The left side of the equation is equal to
\bse
&& \cov \biggr[\Delta \bS_\btheta^F(Y,X,\bZ;\btheta_0) +  \left\{ \Delta - \pi_{Y,\bZ}(Y, \bZ;\bkappa_0) \right\} \bPsi_{\rm ACC}(Y,\bZ; \btheta_0) \biggr] \\
&& = \bB_{\rm CC} + \bH_{\rm ACC} + \bH_{\rm ACC}^T + \bD_{\rm ACC},
\ese
where 
\bse
\bB_{\rm CC} &=& E[\Delta \bS_\btheta^F(Y,W,\bZ;\btheta_0) \{\Delta \bS_\btheta^F(Y,W,\bZ;\btheta_0)\}^T], \\
\bH_{\rm ACC} &=& E \left(   \Delta \bS_\btheta^F(Y,W,\bZ;\btheta_0)  \left[  \left\{ \Delta - \pi_{Y,\bZ}(Y, \bZ;\bkappa_0) \right\} \bPsi_{\rm ACC}(Y,\bZ; \btheta_0) \right]^T \right), \\ 
\bD_{\rm ACC} &=&  E \left(  \left[  \left\{ \Delta - \pi_{Y,\bZ}(Y, \bZ;\bkappa_0) \right\} \bPsi_{\rm ACC}(Y,\bZ; \btheta_0) \right]^{\otimes 2} \right).
\ese
The right side of the equation is equal to
\bse
 && \cov\left[ \Delta \bS_\btheta^F(Y,W,\bZ;\btheta_0), \Delta \bS_\btheta^F(Y,W,\bZ;\btheta_0) +  \left\{ \Delta - \pi_{Y,\bZ}(Y, \bZ;\bkappa_0) \right\} \bPsi_{\rm ACC}(Y,\bZ;\btheta_0) \right] \\
 && = \bB_{\rm CC} + \bH_{\rm ACC}.
\ese
Therefore, $\bPsi(y,\bz;\btheta_0)$ should be chosen such that, $ \bH_{\rm ACC}^T + \bD_{\rm ACC} = \0$. Equivalently, we may find $\bPsi(y,\bz;\btheta_0)$ such that
\bse
\bzero &=& E_{W,\Delta|Y,\bZ} \left[  \left\{ \Delta - \pi_{Y,\bZ}(y, \bz;\bkappa_0) \right\} \left[ \Delta \bS_\btheta^F(y,W,\bz;\btheta_0) +  \left\{ \Delta - \pi_{Y,\bZ}(y, \bz;\bkappa_0) \right\} \bPsi_{\rm ACC}(y,\bz; \btheta_0) \right] ^T \right] \\
&=& E_{W,\Delta|Y,\bZ} \left[  \left\{ \Delta - \pi_{Y,\bZ}(y, \bz;\bkappa_0) \right\} \Delta \bS_\btheta^F(y,W,\bz;\btheta_0)^T  \right] \\
&& \quad \quad + E_{W,\Delta|Y,\bZ} \left[ \left\{ \Delta - \pi_{Y,\bZ}(y, \bz;\bkappa_0) \right\}^2 \right]  \bPsi_{\rm ACC}(y,\bz; \btheta_0)^T.
\ese
Solving for $\bPsi(y,\bz;\btheta_0)$ leads to
\bse
&& \bPsi_{\rm ACC, eff}(y,\bz; \btheta_0) \\
&& = - E_{W,\Delta|Y,\bZ} \left[  \left\{ \Delta - \pi_{Y,\bZ}(y, \bz;\bkappa_0) \right\} \Delta \bS_\btheta^F(y,W,\bz;\btheta_0)  \right] E_{W,\Delta|Y,\bZ} \left[ \left\{ \Delta - \pi_{Y,\bZ}(y, \bz;\bkappa_0) \right\}^2 \right]^{-T}  \\
&& = - E_{W,\Delta|Y,\bZ} \left\{ \Delta \bS_\btheta^F(y,W,\bz;\btheta_0)  \right\} \frac{1-\pi_{Y,\bZ}(y, \bz;\bkappa_0)}{\pi_{Y,\bZ}(y, \bz;\bkappa_0) \{ 1 -  \pi_{Y,\bZ}(y, \bz;\bkappa_0)\}}\\
&& = -  \frac{E_{W,\Delta|Y,\bZ} \left\{ \Delta \bS_\btheta^F(y,W,\bz;\btheta_0)  \right\}}{\pi_{Y,\bZ}(y, \bz;\bkappa_0)} \\
&& \equiv - E_{X|Y,\bZ, \Delta=1} \left\{ \bS_\btheta^F(y,X,\bz;\btheta_0) \right\}.
\ese 
For this choice of $\bPsi_{\rm ACC}(y,\bz; \btheta_0)$, the ACC estimator is more efficient than the CC estimator. Note that by interchanging the roles of $\Delta$ and $R$, we find that the form of $\bPsi_{\rm ACC}(y,\bz; \btheta_0)$ is equivalent to that of \textcite{Bartlettetal2014}.

\emph{When  $\btheta$ and $\bkappa$ are unknown and we do not use $\bLambda_{\rm ACC}$:} When $\bkappa$ is estimated using a logistic regression model with the logit-link function, \textcite{Bartlettetal2014} showed that the same $\bPsi_{\rm ACC}(y, \bz; \btheta_0)$, which is found when $\bkappa$ is assumed known, also ensures higher efficiency over the CC estimator. We refer the reader to \textcite{Bartlettetal2014} for more details. Note that the $\bPsi_{\rm ACC}(y, \bz; \btheta_0)$ that guarantees higher efficiency than the CC estimator may be different when estimating $\bkappa$ using a method other than logisitic regression.

\emph{When only $\btheta$ is unknown and we use $\bLambda_{\rm ACC}$:}  Instead of finding the form of $\bPsi_{\rm ACC}(y, \bz; \btheta_0)$ that will make the ACC estimator more efficient than the CC estimator, we find a fixed matrix $\bLambda_{\rm ACC}$ that will make the ACC estimator more efficient than the CC estimator for any choice of $\bPsi_{\rm ACC}(y, \bz; \btheta_0)$. Like before, we compare the influence functions of $\widehat{\btheta}_{\rm CC}$ and $\widehat{\btheta}_{\rm ACC; \bLambda}$. We choose a $\bLambda_{\rm ACC}$ that makes the efficiency condition in equation \eqref{eqn:efficiency-condition} true. That efficiency condition is equivalent to 
\bse
&& \cov \left[\Delta \bS_\btheta^F(Y,W,\bZ;\btheta_0) +  \left\{ \Delta - \pi_{Y,\bZ}(Y,\bZ; \bkappa_0) \right\} \bLambda_{\rm ACC} \bPsi_{\rm ACC}(Y,\bZ;\btheta_0)  \right]  \\
&&= \cov\left[ \Delta \bS_\btheta^F(Y,W,\bZ;\btheta_0), \Delta \bS_\btheta^F(Y,W,\bZ;\btheta_0) +  \left\{ \Delta - \pi_{Y,\bZ}(Y,\bZ; \bkappa_0) \right\} \bLambda_{\rm ACC} \bPsi_{\rm ACC}(Y,\bZ;\btheta_0) \right].
\ese
Solving for $\bLambda_{\rm ACC}$ leads to
\bse
\bLambda_{\rm ACC} && = -  E \left[ \left\{ \Delta - \pi_{Y,\bZ}(Y,\bZ; \bkappa_0) \right\} \bPsi_{\rm ACC}(Y,\bZ; \btheta_0) \Delta \bS_\btheta^F(Y,W,\bZ;\btheta_0)^T  \right] \\
&& \quad \quad \times E \left(  \left[ \left\{ \Delta - \pi_{Y,\bZ}(Y,\bZ; \bkappa_0) \right\} \bPsi_{\rm ACC}(Y,\bZ; \btheta_0) \right]^{\otimes 2} \right)^{-T}. 
\ese 
For any choice of $\bPsi_{\rm ACC}(y,\bz; \btheta_0)$, the above $\bLambda_{\rm ACC}$ guarantees higher efficiency over the CC estimator. The sample mean $\bLambda_{\rm ACC}$ is used in the estimation of $\btheta$. This is done by first obtaining an estimate of $\wh\btheta$ using a consistent estimator (e.g., the CC estimator), then using this estimate to compute 
\bse
\widehat{\bLambda}_{\rm ACC} &=&  \sumi   \biggr( \left[ \left\{ \delta_i - \pi_{Y,\bZ}(y_i, \bz_i; \bkappa_0) \right\} \bPsi_{\rm ACC}(y_i,\bz_i; \wh{\btheta}) \right] \delta_i \bS_\btheta^F(y_i,w_i,\bz_i;\wh{\btheta})^T  \biggr) \\
&& \quad \times \left(  \sumi   \left[ \left\{ \delta_i - \pi_{Y,\bZ}(y_i, \bz_i; \bkappa_0) \right\} \bPsi_{\rm ACC}(y_i,\bz_i; \wh{\btheta})  \right]^{\otimes 2}  \right)^{-T}.
\ese
Since $\bLambda_{\rm ACC}$ only affects the variance of $\btheta$ and not its consistency, then by Theorem 6.2 of \textcite{newey1994large}, the influence function $\bUpsilon_{\rm ACC; \bLambda}$ may be used in the estimation of $\btheta$ under the presence of $\widehat{\bLambda}_{\rm ACC}$. This result means that the uncertainty of  $\widehat{\bLambda}_{\rm ACC}$ may be ignored in the variance estimation of $\widehat{\btheta}_{\rm ACC; \bLambda}$. 

%
\emph{When  $\btheta$ and $\bkappa$ are unknown and we use $\bLambda_{\rm ACC}$:} We use a similar approach as that used when only $\btheta$ is unknown. This time, we choose a $\bLambda_{\rm ACC}$ that makes the efficiency condition in equation \eqref{eqn:efficiency-condition} true for any choice of $\bPsi_{\rm ACC}(y, \bz; \btheta)$.
We will compare the influence functions of $\widehat{\btheta}_{\rm CC}$ and $\widehat{\btheta}_{\rm ACC; \bLambda, \bkappa}$, and find  $\bLambda_{\rm ACC; \bkappa}$ that makes the  efficiency condition in equation \eqref{eqn:efficiency-condition} true. Let $\bUpsilon_\bkappa(Y,\bZ) = -E\{\partial\bS_{\bkappa}(Y,\bZ;\bkappa_0)/\partial\bkappa^T\}^{-1}\bS_{\bkappa}(Y,\bZ;\bkappa_0)$ be the influence function of $\bkappa$, and $\bA^*_{\rm ACC;\bkappa} = E\{\partial\bPhi_{\rm ACC; \bLambda}(\bO;\btheta_0,\bkappa_0)/\partial\bkappa^T\}$ as derived when establishing the asymptotic normality of the ACC estimator. Then, the efficiency condition is equivalent to 
\bse
&& \cov \left(\Delta \bS_\btheta^F(Y,W,\bZ;\btheta) + \bLambda_{\rm ACC; \bkappa} \left[ \left\{ \Delta - \pi_{Y,\bZ}(Y,\bZ; \bkappa_0) \right\} \bPsi_{\rm ACC}(Y,\bZ;\btheta) + \bA^*_{\rm ACC; \bkappa} \bUpsilon_{\bkappa}(Y,\bZ) \right] \right) \\
&&= \cov( \Delta \bS_\btheta^F(Y,W,\bZ;\btheta), \\
 && \quad \quad \quad  \Delta \bS_\btheta^F(Y,W,\bZ;\btheta) + \bLambda_{\rm ACC; \bkappa} \left[ \left\{ \Delta - \pi_{Y,\bZ}(Y,\bZ; \bkappa_0) \right\} \bPsi_{\rm ACC}(Y,\bZ;\btheta) + \bA^*_{\rm ACC; \bkappa} \bUpsilon_{\bkappa}(Y,\bZ) \right] ).
\ese
Solving for $\bLambda_{\rm ACC; \bkappa}$ leads to
\bse
\bLambda_{\rm ACC; \bkappa} && = -  E \biggr( \left[ \left\{ \Delta - \pi_{Y,\bZ}(Y, \bZ; \bkappa_0) \right\} \bPsi_{\rm ACC}(Y,\bZ; \btheta) + \bA^*_{\rm ACC;\bkappa} \bUpsilon_{\bkappa}(Y,\bZ) \right] \Delta \bS_\btheta^F(Y,W,\bZ;\btheta)^T  \biggr) \\
&& \times E \left(  \left[ \left\{ \Delta - \pi_{Y,\bZ}(Y, \bZ; \bkappa_0) \right\} \bPsi_{\rm ACC}(Y,\bZ; \btheta) + \bA^*_{\rm ACC;\bkappa} \bUpsilon_{\bkappa}(Y,\bZ) \right]^{\otimes 2} \right)^{-T}. 
\ese 
For any choice of $\bPsi_{\rm ACC}(y,\bz; \btheta)$, the above $\bLambda_{\rm ACC; \bkappa}$ guarantees higher efficiency over the CC estimator. The sample mean $\wh\bLambda_{\rm ACC; \bkappa}$ is used in the estimation of $\btheta$. That sample mean is
\bse
\widehat{\bLambda}_{\rm ACC; \bkappa} &=&  \sumi   \biggr( \left[ \left\{ \delta_i - \pi_{Y,\bZ}(y_i, \bz_i; \wh\bkappa) \right\} \bPsi_{\rm ACC}(y_i,\bz_i; \wh\btheta) + \bA^*_{\rm ACC;\bkappa} \bUpsilon_{\bkappa}(y_i,\bz_i) \right] \delta_i \bS_\btheta^F(y_i,w_i,\bz_i;\wh\btheta)^T  \biggr) \\
&& \quad \times \left(  \sumi   \left[ \left\{ \delta_i - \pi_{Y,\bZ}(y_i, \bz_i; \wh\bkappa) \right\} \bPsi_{\rm ACC}(y_i,\bz_i; \wh\btheta) + \bA^*_{\rm ACC;\bkappa} \bUpsilon_{\bkappa}(y_i,\bz_i) \right]^{\otimes 2}  \right)^{-T},
\ese
where $(\wh\btheta, \wh\bkappa)$ are obtained from consistent estimators (e.g., CC estimator and logistic regression). Since $\bLambda_{\rm ACC; \bkappa}$ only affects the variance of $\btheta$ and not its consistency, the influence function $\bUpsilon_{\rm ACC; \bLambda, \bkappa}$ may be used in the estimation of $\btheta$ when we include $\widehat{\bLambda}_{\rm ACC; \bkappa}$\parencite[Theorem 6.2]{newey1994large}. 
The uncertainty of  $\widehat{\bLambda}_{\rm ACC; \bkappa}$ may thus be ignored in the variance estimation of $\widehat{\btheta}_{\rm ACC; \bLambda, \bkappa}$.

\item {\bf MACC estimator:}

\emph{When only $\btheta$ is unknown and we do not use $\bLambda_{\rm MACC}$:}
We follow a similar process as that used with the ACC estimator, except this time, we compare the influence functions of $\btheta_{\rm CC}$ and $\btheta_{\rm MACC}$, and we find the form of $\bPsi_{\rm MACC}(y, \bz; \btheta_0)$ that makes the efficiency condition in equation \eqref{eqn:efficiency-condition} true. Satisfying the efficiency condition requires
\bse
\bzero &=& E_{W,\Delta|Y,\bZ} \biggr(  \left\{ 1 - \frac{\Delta}{\pi_{X,\bZ}(W, \bz;\boldeta_0)} \right\} \\
&& \quad \times \biggr[ \Delta \bS_\btheta^F(y,W,\bz;\btheta_0) + \left\{ 1 - \frac{\Delta}{\pi_{X,\bZ}(W, \bz;\boldeta_0)} \right\} \bPsi_{\rm MACC}(y,\bz; \btheta_0) \biggr] ^T \biggr) \\
&=&  E_{W,\Delta|Y,\bZ} \left[  \left\{ 1 - \frac{\Delta}{\pi_{X,\bZ}(W, \bz;\boldeta_0)} \right\} \Delta \bS_\btheta^F(y,W,\bz;\btheta_0)^T  \right] \\
&& \quad + E_{W,\Delta|Y,\bZ} \left[ \left\{ 1 - \frac{\Delta}{\pi_{X,\bZ}(W, \bz;\boldeta_0)} \right\}^2 \right]  \bPsi_{\rm MACC}(y,\bz; \btheta_0)^T.
\ese
Solving for $\bPsi_{\rm MACC}(y,\bz;\btheta_0)$ results in:
\bse
&& \bPsi_{\rm MACC, eff}(y,\bz; \btheta_0) \\
&& \quad = - E_{W,\Delta|Y,\bZ} \left[  \left\{ 1 - \frac{\Delta}{\pi_{X,\bZ}(W, \bz;\boldeta_0)} \right\} \Delta \bS_\btheta^F(y,W,\bz;\btheta_0)  \right] E_{W,\Delta|Y,\bZ} \left[ \left\{ 1 - \frac{\Delta}{\pi_{X,\bZ}(W, \bz;\boldeta_0)} \right\}^2 \right]^{-T}  \\
&& \quad = \frac{E_{X|Y,\bZ} \left[  \left\{ \pi_{X,\bZ}(X, \bz;\boldeta_0) - 1 \right\}  \bS_\btheta^F(y,X,\bz;\btheta_0)  \right]} {E_{X|Y,\bZ} \left\{ 1 - 1/\pi_{X,\bZ}(X, \bz;\boldeta_0) \right\} }. 
\ese 
With this choice of $\bPsi_{\rm MACC}(y,\bz; \btheta_0)$, the MACC estimator is more efficient than the CC estimator.

\emph{When  $\btheta$ and $\boldeta$ are unknown and we do not use $\bLambda_{\rm MACC}$:} Unlike the ACC estimator, the MACC estimator may require a different formulation of $\bPhi_{\rm MACC}(y,\bz;\btheta_0)$ to guarantee efficiency gains over the CC estimator. This difference means that the current choice of $\bPsi(y,\bz; \btheta_0)$ may not guarantee the efficiency condition in equation (\ref{eqn:efficiency-condition}). Finding a  $\bPhi_{\rm MACC}(y,\bz;\btheta_0)$ that guarantees efficiency over the CC estimator when $\balpha$ is estimated using the strategy discussed in Section \ref{sec:est_nuisance_distributions} is challenging. The difficulty is that we cannot easily solve for  $\bPhi_{\rm MACC}(y,\bz;\btheta_0)$ because $\bA^*_{\rm MACC; \balpha}$ is an expectation of a function on $\bPhi_{\rm MACC}(y,\bz;\btheta_0)$. Therefore, we recommend using the MACC estimator with $\bLambda_{\rm MACC}$ over the regular MACC estimator, as using $\bLambda_{\rm MACC}$ will guarantee higher efficiency than the CC estimator regardless of the choice of $\bPhi_{\rm MACC}(y,\bz;\btheta_0)$.  

\emph{When only $\btheta$ is unknown and we use $\bLambda_{\rm MACC}$:}
We compare the influence functions of $\widehat{\btheta}_{\rm CC}$ and $\widehat{\btheta}_{\rm MACC;\bLambda}$ to find a fixed matrix $\bLambda_{\rm MACC}$ that will make the MACC estimator more efficient than the CC estimator for any choice of $\bPsi_{\rm MACC}(y, \bz; \btheta_0)$. To achieve that goal, we strive to find $\bLambda_{\rm MACC}$ that makes the following efficiency condition in equation  (\ref{eqn:efficiency-condition}) true.
Equivalently,
\bse
&& \cov \left[\Delta \bS_\btheta^F(Y,W,\bZ;\btheta_0) + \biggr\{1- \frac{\Delta}{\pi_{X,\bZ}(W, \bZ; \boldeta_0)} \biggr\} \bLambda_{\rm MACC} \bPsi_{\rm MACC}(Y,\bZ; \btheta_0)  \right] \\
&&= \cov\biggr( \Delta \bS_\btheta^F(Y,W,\bZ;\btheta_0), \Delta \bS_\btheta^F(Y,W,\bZ;\btheta_0) + \biggr\{1- \frac{\Delta}{\pi_{X,\bZ}(W, \bZ; \boldeta_0)} \biggr\} \bLambda_{\rm MACC} \bPsi_{\rm MACC}(Y,\bZ; \btheta_0) \biggr] \biggr).
\ese
Solving for $\bLambda_{\rm MACC}$ leads to
\bse
\bLambda_{\rm MACC} && = -  E \biggr[ \biggr\{1- \frac{\Delta}{\pi_{X,\bZ}(W, \bZ; \boldeta_0)} \biggr\} \bPsi_{\rm MACC}(Y,\bZ; \btheta_0) \Delta \bS_\btheta^F(Y,W,\bZ;\btheta_0)^T \biggr] \nonumber \\
&&  \quad \quad \times E \biggr(  \biggr[ \biggr\{1- \frac{\Delta}{\pi_{X,\bZ}(W, \bZ; \boldeta_0)} \bPsi_{\rm MACC}(Y,\bZ; \btheta_0) \biggr]^{\otimes 2} \biggr)^{-T}.
\ese
This choice of $\bLambda_{\rm MACC}$ ensures that the MACC estimator is more efficient than the CC estimator for any choice of $\bPsi_{\rm MACC}(y,\bz; \btheta_0)$. In practice, we estimate $\bLambda_{\rm MACC}$ by its sample mean, $\wh\bLambda_{\rm MACC}$, which is
\bse
\widehat{\bLambda}_{\rm MACC} &=&  \sumi   \biggr( \left[ \biggr\{1- \frac{\delta}{\pi_{X,\bZ}(w, \bz; \boldeta_0)} \biggr\} \bPsi_{\rm MACC}(y_i,\bz_i; \wh{\btheta}) \right] \delta_i \bS_\btheta^F(y_i,w_i,\bz_i;\wh{\btheta})^T  \biggr) \\
&& \quad \times \left(  \sumi   \left[\biggr\{1- \frac{\delta}{\pi_{X,\bZ}(w, \bz; \boldeta_0)} \biggr\} \bPsi_{\rm MACC}(y_i,\bz_i; \wh{\btheta})  \right]^{\otimes 2}  \right)^{-T},
\ese
where the estimates $\wh\btheta$ are from a consistent estimator (e.g., the CC estimator).
As we saw with the ACC estimator, $\bLambda_{\rm MACC}$ affects the variance, but not the consistency of $\btheta$. When using $\widehat{\bLambda}_{\rm MACC}$, we may thus use the influence function $\bUpsilon_{\rm MACC; \bLambda}$ when estimating   $\btheta$. Consequently, the uncertainty of  $\widehat{\bLambda}_{\rm MACC}$ may be ignored in the variance estimation of $\widehat{\btheta}_{\rm MACC; \bLambda}$.


\emph{When  $\btheta$ and $\boldeta$ are unknown and we use $\bLambda_{\rm MACC}$:} 
Similar to when only $\btheta$ is unknown, we strive to find a fixed matrix $\bLambda_{\rm MACC; \balpha}$ that makes the MACC estimator more efficient than the CC estimator for any choice of $\bPsi_{\rm MACC}(y, \bz; \btheta_0)$. Let $\bUpsilon_{\balpha}(\bO^*) = -E\{\partial\bPhi_{\balpha}(\bO^*; \balpha)/\partial\balpha^T\}^{-1}\bPhi_{\balpha}(\bO^*; \balpha)$. Then, to satisfy the efficiency condition, we find that using 
\bse
&& \bLambda_{\rm MACC; \balpha} \\
&& = -  E \biggr( \biggr[ \biggr\{1- \frac{\Delta}{\pi_{X,\bZ}(W, \bZ; \boldeta_0)} \biggr\} \bPsi_{\rm MACC}(Y,\bZ; \btheta_0) + \bA^*_{\rm MACC; \balpha} \bUpsilon_{\balpha}(\bO^*) \biggr]  \delta \bS_\btheta^F(Y,W,\bZ;\btheta_0)^T \biggr)^T \nonumber \\
&& \quad \quad \times E \biggr(  \biggr[ \biggr\{1- \frac{\Delta}{\pi_{X,\bZ}(W, \bZ; \boldeta_0)} \biggr\} \bPsi_{\rm MACC}(Y,\bZ; \btheta_0) + \bA^*_{\rm MACC; \balpha} \bUpsilon_{\balpha}(\bO^*) \biggr]^{\otimes 2} \biggr)^{-T}
\ese 
will guarantee the MACC estimator is more efficient than the CC estimator for any choice of $\bPsi_{\rm MACC}(y,\bz; \btheta_0)$.  When estimating $\btheta$, we use the sample mean estimate of $\bLambda_{\rm MACC;\balpha}$, denoted by $\wh\bLambda_{\rm MACC;\balpha}$. Like we did when computing $\wh\bLambda_{\rm MACC}$, we obtain  $\wh\bLambda_{\rm MACC;\balpha}$ by replacing $(\btheta, \boldeta)$  with estimates $(\wh\btheta, \wh\boldeta)$ obtained from a consistent estimator (e.g., the CC estimator and the estimator from Section \ref{sec:est_nuisance_distributions}) and replacing expectations with summands.
Similar to our argument when only $\btheta$ is unknown, the uncertainty of  $\widehat{\bLambda}_{\rm MACC; \balpha}$ may be ignored in the variance estimation of $\widehat{\btheta}_{\rm MACC; \bLambda, \balpha}$.

\item {\bf AIPW estimator:}

\emph{When only $\btheta$ is unknown and we do not use $\bLambda_{\rm AIPW}$:} By comparing the influence functions of $\widehat{\btheta}_{\rm IPW}$ and $\widehat{\btheta}_{\rm AIPW}$, we find  the form of $\bPsi_{\rm AIPW}(y, \bz; \btheta_0)$ that makes the AIPW estimator more efficient than the IPW estimator. This gain in efficiency is satisfied when $\bzero = \cov(\bUpsilon_{\rm AIPW}  - \bUpsilon_{\rm IPW}, \bUpsilon_{\rm AIPW})$. Applying a similar argument as that for the MACC estimator, we find that using
\bse
&& \bPsi_{\rm AIPW, eff}(y,\bz; \btheta_0) \\
&& \quad = - E_{W,\Delta|Y,\bZ} \left[  \left\{ 1 - \frac{\Delta}{\pi_{X,\bZ}(W, \bz;\boldeta_0)} \right\} \frac{\Delta \bS_\btheta^F(y,W,\bz;\btheta_0)}{\pi_{X,\bZ}(W, \bz;\boldeta_0)}  \right] E_{W,\Delta|Y,\bZ} \left[ \left\{ 1 - \frac{\Delta}{\pi_{X,\bZ}(W, \bz;\boldeta_0)} \right\}^2 \right]^{-T}  \\
&& \quad = \frac{E_{X|Y,\bZ} \left[  \left\{ 1-  1/\pi_{X,\bZ}(X, \bz;\boldeta_0) \right\}  \bS_\btheta^F(y,X,\bz;\btheta_0)  \right]} {E_{X|Y,\bZ} \left\{ 1 - 1/\pi_{X,\bZ}(X, \bz;\boldeta_0) \right\} }
\ese 
will ensure the AIPW estimator is more efficient than the IPW estimator.

\emph{When $\btheta$ and $\boldeta$ are unknown and we do not use $\bLambda_{\rm AIPW}$:} Similar to the MACC estimator for unknown $f_{C|\bZ}$, the same $\bPsi_{\rm AIPW}(y, \bz; \btheta_0)$ does not guarantee the efficiency condition in equation (\ref{eqn:efficiency-condition}). Finding the form of $\bPsi_{\rm AIPW}(y, \bz; \btheta)$ that satisfies the efficiency condition is non-trivial since $\bPsi_{\rm AIPW}(y, \bz; \btheta)$ is embedded in an expectation, i.e., $\bA^*_{\rm AIPW; \balpha}$. Therefore, we recommend using the AIPW estimator with $\bLambda_{\rm AIPW}$, as that estimator will have higher efficiency than the IPW estimator for any choice of $\bPsi_{\rm AIPW}(y, \bz; \btheta_0)$.

\emph{When only $\btheta$ is unknown and we  use $\bLambda_{\rm AIPW}$:}
Instead of finding the form of $\bPsi_{\rm AIPW}(y, \bz; \btheta_0)$ that will make the AIPW estimator more efficient than the IPW estimator, we find a fixed matrix $\bLambda_{\rm AIPW}$ that makes the AIPW estimator more efficient than the IPW estimator for any choice of $\bPsi_{\rm AIPW}(y, \bz; \btheta_0)$. As before, we compare the influence functions of $\widehat{\btheta}_{\rm IPW}$ and $\widehat{\btheta}_{\rm AIPW}$ and find the form of $\bLambda_{\rm AIPW}$ that satisfies the efficiency condition in equation \ref{eqn:efficiency-condition}. Following similar calculations to those in ACC and MACC, we find this efficiency condition in equation \ref{eqn:efficiency-condition} is satisfied when
\bse
\bLambda_{\rm AIPW} && = -  E \biggr[ \biggr\{1- \frac{\Delta}{\pi_{X,\bZ}(W, \bZ; \boldeta_0)} \biggr\} \bPsi_{\rm AIPW}(Y,\bZ; \btheta_0)  \frac{\Delta \bS_\btheta^F(Y,W,\bZ;\btheta_0)^T}{\pi_{X,\bZ}(W, \bZ; \boldeta_0)} \biggr] \nonumber \\
&&  \quad \quad \times E \biggr(  \biggr[ \biggr\{1- \frac{\Delta}{\pi_{X,\bZ}(W, \bZ; \boldeta_0)} \biggr\} \bPsi_{\rm AIPW}(Y,\bZ; \btheta_0) \biggr]^{\otimes 2} \biggr)^{-T}.
\ese 
We use the sample mean of $\bLambda_{\rm AIPW}$, denoted by $\wh\bLambda_{\rm AIPW}$, when estimating $\btheta$.  That sample mean is
\bse
\widehat{\bLambda}_{\rm AIPW} &=&  \left[\sumi    \biggr\{1- \frac{\delta_i}{\pi_{X,\bZ}(w_i, \bz_i; \boldeta_0)} \biggr\} \bPsi_{\rm AIPW}(y_i,\bz_i; \wh{\btheta})  \frac{\delta_i \bS_\btheta^F(y_i,w_i,\bz_i;\wh{\btheta})^T}{\pi_{X,\bZ}(w_i, \bz_i; \boldeta_0)} \right] \\
&& \quad \times \left(  \sumi   \left[\biggr\{1- \frac{\delta_i}{\pi_{X,\bZ}(w_i, \bz_i; \boldeta_0)} \biggr\} \bPsi_{\rm AIPW}(y_i,\bz_i; \wh{\btheta}) \right]^{\otimes 2}  \right)^{-T},
\ese
where $\wh\btheta$ is from a consistent estimator (e.g., the IPW estimator). Following similar arguments to those used for the ACC and MACC estimators, the uncertainty of  $\widehat{\bLambda}_{\rm AIPW}$ may be ignored in the variance estimation of $\widehat{\btheta}_{\rm AIPW; \bLambda}$. 

\emph{When $\btheta$ and $\boldeta$ are unknown and we  use $\bLambda_{\rm AIPW}$:} We follow a similar argument as that when only $\btheta$ is unknown, except that now we find the form of $\bLambda_{\rm AIPW; \balpha}$ that makes the  efficiency condition in equation (\ref{eqn:efficiency-condition}) true. Equivalently,
\bse
\bzero &&= \cov \biggr[\bPhi_{\rm AIPW; \bLambda}(\bO;\btheta_0,\balpha_0) + (\bA^*_{\rm IPW, \balpha} + \bLambda_{\rm AIPW; \balpha} \bA^*_{\rm AIPW; \balpha}) \bUpsilon_{\balpha}(\bO^*) \\
&& \quad \quad \quad  \quad - \biggr\{ \bPhi_{\rm IPW}(\bO;\btheta_0,\balpha_0) + \bA^*_{\rm IPW, \balpha} \bUpsilon_{\balpha}(\bO^*) \biggr\}, \\
&& \quad \quad \quad  \quad \bPhi_{\rm AIPW; \bLambda}(\bO;\btheta_0,\balpha_0) + (\bA^*_{\rm IPW, \balpha} + \bLambda_{\rm AIPW; \balpha} \bA^*_{\rm AIPW; \balpha}) \bUpsilon_{\balpha}(\bO^*) \biggr]\\
&&= \cov \biggr( \bLambda_{\rm AIPW; \balpha} \biggr[ \biggr\{1- \frac{\Delta}{\pi_{X,\bZ}(W, \bZ; \balpha_0)} \biggr\} \bPsi_{\rm AIPW}(Y,\bZ; \btheta_0)  + \bA^*_{\rm AIPW; \balpha} \bUpsilon_{\balpha}(\bO^*) \biggr], \\
&& \quad \quad \quad  \quad \bPhi_{\rm AIPW; \bLambda}(\bO;\btheta_0,\balpha_0) + (\bA_{\rm IPW, \balpha} + \bLambda_{\rm AIPW; \balpha} \bA^*_{\rm AIPW; \balpha}) \bUpsilon_{\balpha}(\bO^*) \biggr).
\ese
Solving for $\bLambda_{\rm AIPW; \balpha}$ leads to
\bse
&& \bLambda_{\rm AIPW; \balpha} \\
&& = -\biggr\{ E \biggr( \left[ \biggr\{1- \frac{\Delta}{\pi_{X,\bZ}(W, \bZ; \balpha_0)} \biggr\} \bPsi_{\rm AIPW}(Y,\bZ; \btheta_0) + \bA^*_{\rm AIPW; \balpha} \bUpsilon_{\balpha}(\bO^*) \right] \\
&& \sextant \sextant \times \left\{ \frac{\Delta \bS_{\btheta}^F(Y,W,\bZ;\btheta)}{\pi_{X,\bZ}(W, \bZ; \balpha_0)   } + \bA^*_{\rm IPW; \balpha} \bUpsilon_{\balpha}(\bO^*) \right\} \biggr) \biggr\}^T \\
&& \times E \biggr( \biggr[ \biggr\{1- \frac{\Delta}{\pi_{X,\bZ}(W, \bZ; \balpha_0)} \biggr\} \bPsi_{\rm AIPW}(Y,\bZ; \btheta_0)  + \bA^*_{\rm AIPW; \balpha} \bUpsilon_{\balpha}(\bO^*) \biggr]^{\otimes 2} \biggr) ^{-T}.
\ese 
For any choice of $\bPsi_{\rm AIPW}(y,\bz; \btheta_0)$, the above $\bLambda_{\rm AIPW; \balpha}$ guarantees higher efficiency over the IPW estimator.  The sample mean of $\bLambda_{\rm AIPW; \balpha}$ is used in the estimation of $\btheta$, which is computed by replacing expectations with summands in $\bLambda_{\rm AIPW; \balpha}$ and replacing $(\btheta, \boldeta)$ with estimates obtained from a consistent estimator for each (e.g.,the CC estimator and the estimator described in Section \ref{sec:est_nuisance_distributions}).
Following arguments similar to those for the ACC and MACC estimators, the uncertainty of  $\widehat{\bLambda}_{\rm AIPW; \balpha}$ may be ignored in the variance estimation of $\widehat{\btheta}_{\rm AIPW; \bLambda, \balpha}$.


\end{enumerate}

\subsection{Changes to accommodate dependent right-censoring} 

Up to now, we assumed independent covariate right-censoring. Now we show the changes to the estimators when we assume dependent covariate right-censoring.

\begin{enumerate}
    \item {\bf Nuisance distribution estimators:}

As explained in Section \ref{sec:implement}, $\balpha$ is non-identifiable without additional assumptions beyond those made for independent covariate right-censoring. For $\balpha$ to be identifiable under dependent covariate right-censoring, we require that the true conditional covariance between $X$ and $C$ given $\bZ$ is known and that $f_{C,X|\bZ}$ is sufficiently smooth around $\balpha_0$. Under this assumption, a parametric distribution can be imposed for the log-likelihood and the decomposition can be reduced to
\bse
&& \log f_{W,\Delta| \bZ}(w_i,\delta_i, \bz_i; \balpha) \\
&& = \ \delta_i \log\biggr\{ \int_{w_i\leq C} f_{C|X,\bZ}(c,w_i,\bz_i;\boldeta^*) f_{X|\bZ}(w_i,\bz_i;\bgamma^*) dc \biggr\} + \\
&& \quad \quad (1-\delta_i) \log\biggr\{ \int_{w_i\leq X} f_{C|X, \bZ}(w_i,x,\bz_i\boldeta^*) f_{X|\bZ}(x, \bz_i;\bgamma^*) dx \biggr\}. 
\ese
The main difference in this decomposition compared to that when covariate right-censoring is \emph{independent} is in the conditional decomposition of $ f_{C,X|\bZ}(c,w_i,\bz_i;\balpha)$. Since the covariance between $X$ and $C$ given $\bZ$ is assumed to be known (not zero), the vector $\balpha$ that needs to be estimated remains the same. Therefore, the proof of consistency and asymptotic normality remains the same under dependent covariate right-censoring.

\item {\bf CC estimator:}

The estimating equation for the CC estimator is only a function of $f_{Y|X,\bZ}$, therefore, the CC estimator is independent of $f_{C, X|\bZ}$. As a result, the CC estimator does not change when covariate right-censoring is dependent. Therefore, the proof for consistency and asymptotic normality, as well as the proof for robustness remain the same

\item {\bf IPW estimator:}

Under dependent covariate right-censoring, the IPW estimating equations use the probabilities defined by $\pi^{\rm cens, dep}_{X,\bZ}(w,\bz) = \int_{w \leq C} f_{C|X,\bZ} (w,\bz) dc $ as opposed to the probabilities defined by $\pi^{\rm cens, ind}_{X,\bZ}(w,\bz)= \int_{w \leq C} f_{C|\bZ} (w,\bz) dc$ under independent covariate right-censoring. Both probabilities remain a function of $(X,\bZ)$, therefore, by interchanging the roles of $\pi^{\rm cens, dep}_{X,\bZ}(w,\bz)$ and $\pi^{\rm cens, ind}_{X,\bZ}(w,\bz)$, the proof for consistency and asymptotic normality, as well as the proof for robustness remain the same. The consistency and asymptotic normality of $\balpha$ requires knowing the true conditional covariance of $(X,C)$ given $\bZ$. 

\item {\bf MLE:}

Under dependent covariate right-censoring, the MLE estimating equations use the density $f_{X|C,\bZ}$ as opposed to $f_{X|\bZ}$. By interchanging the roles of $f_{X|C,\bZ}$ and $f_{X|\bZ}$, the proof for consistency and asymptotic normality, as well as the proof for robustness remain the same. The consistency and asymptotic normality of $\balpha$ requires knowing the true conditional covariance of $(X,C)$ given $\bZ$. 

\item {\bf ACC estimator:}

The ACC estimator uses the probability $\pi_{Y,\bZ}$ which may be specified/estimated without involving $f_{X,C|\bZ}$. Therefore, the proof for consistency and asymptotic normality, as well as the proof for robustness remain the same under dependent covariate right-censoring. These results hold true even when we use $\bLambda_{\rm ACC}$ as this is a non-random matrix.

\item {\bf MACC and AIPW estimators:}

Under dependent covariate right-censoring, the estimating equation for the MACC and AIPW estimators use the density $f_{C|X,\bZ}$ as opposed to $f_{C|\bZ}$ to define the probability $\pi_{X,\bZ}(w,\bz; \balpha)$. By interchanging the roles of $f_{C|X,\bZ}$ and $f_{C|\bZ}$, the proof for consistency and asymptotic normality, as well as the proof for robustness remain the same. The consistency and asymptotic normality of $\balpha$ requires knowing the true conditional covariance of $(X,C)$ given $\bZ$.  These results hold true even when we use $\bLambda_{\rm MACC}$ and $\bLambda_{\rm AIPW}$, as these are non-random matrices.

\end{enumerate}

\section{Proof of Theorem \ref{thm:all-robust-consistent} for the missing covariate problem}
\label{sec:thm1-proof-part2}

\subsection{Specification or estimation of nuisance distributions}

Similar to the right-censored covariate problem, estimation of $\btheta$ under the presence of a missing covariate requires the specification or estimation of nuisance parameters as discussed in Section \ref{sec:implement}. For example, under independent covariate missingness: 

\begin{itemize}
    \item IPW estimator: Requires the density $f_{R|\bZ}$ to compute the probability $\pi_{\bZ}(\bz)$;
    \item MLE: Requires the density $f_{X|\bZ}(x,\bz)$;
    \item ACC, MACC, and AIPW estimators: Requires the density $f_{R|\bZ}$ to compute the probability $\pi_{\bZ}(\bz)$ and $f_{X|\bZ}$ to compute $\bPsi(y,\bz;\btheta)$.  
\end{itemize}

Throughout, we consider the parametric models $f_{X|\bZ}(x,\bz; \biota)$ and $f_{R|\bZ}(r,\bz; \btau)$, where $(\biota^T, \btau^T)^T$ is a set of finite-dimensional parameters.  Under a \emph{mechanistic} approach, the parameter space is assumed known, but in practice, these parameters need to be estimated. We point out when a consistent estimator of these nuisance distribution is required. In what follows, we focus on proving the consistency and asymptotic properties of the estimators, as well as defining their robustness properties.

\subsection{Consistency}

\begin{enumerate}
    \item {\bf CC estimator:}

    Observe that
    \bse
    E\left\{ R \bS_{\btheta}^F (Y,X,\bZ; \btheta_0)\right\} &=& E_{Y,\bZ} [ E_{XR,R|Y,\bZ} \left\{R \bS_{\btheta}^F (Y,X,\bZ; \btheta_0)\right\} ] \\
    &=& E_{Y,\bZ} [ E_{X|Y,\bZ} \left\{\pi_{\bZ}(\bZ; \btau_0) \bS_{\btheta}^F (Y,X,\bZ; \btheta_0)\right\} ] \\
    &=&  E_{X,\bZ} [ \pi_{\bZ}(\bZ; \btau_0) E_{Y|X,\bZ} \{  \bS_{\btheta}^F (Y,X,\bZ; \btheta_0) \} ]  = \bzero.
    \ese

    Therefore, the CC estimating equation is unbiased, and the resulting estimator is consistent.
    

\item \textbf{IPW estimator:}  

 Observe that 
\bse
E\left\{\frac{R \bS_{\btheta}^F (Y,X,\bZ; \btheta_0)}{ \pi_{\bZ}(\bz; \btau_0)}\right\} &=& E_{Y,\bZ} \biggr[ E_{XR,R|Y,\bZ} \left\{\frac{R \bS_{\btheta}^F (Y,X,\bZ; \btheta_0)}{ \pi_{\bZ}(\bZ; \btau_0)}\right\} \biggr] \\
&=& E_{Y,\bZ} \biggr[ E_{X|Y,\bZ} \left\{\frac{ \pi_{\bZ}(\bZ; \btau_0) }{ \pi_{\bZ}(\bZ; \btau_0)} \times \bS_{\btheta}^F (Y,X,\bZ; \btheta_0)\right\} + \0 \biggr] \\
&=& E_{Y,X,\bZ} \left\{ \bS_{\btheta}^F (Y,X,\bZ; \btheta_0)\right\}  = \bzero.
\ese
Therefore, the IPW estimating equation is unbiased, and the resulting estimator is consistent. 

\item 
{\bf MLE:}

The nuisance distribution for MLE in both the right-censored and missing covariate problems is $f_{X|\bZ}$. Therefore, the proof of consistency remains the same as that shown for the right-censored covariate problem. The only difference is that the domain of integration covers the entire domain of $X$ in the missing covariate problem.

\item \textbf{ACC estimator:} 

We previously proved the CC estimating equation is unbiased. It thus suffices to show  that the expectation of the augmentation term is equal to $\bzero$. It follows that
\bse
&& E_{Y,XR,R,\bZ} [ \{R- \pi_{\bZ}(\bZ; \btau_0) \}\bPsi_{\rm ACC}(Y, \bZ; \btheta_0)] \\
&& = E_{Y,\bZ}[ E_{XR,R|Y,\bZ}\{R - \pi_{\bZ}(\bZ; \btau_0) \}\bPsi_{\rm ACC}(Y, \bZ; \btheta_0)].
\ese 
Consistency is achieved if $E_{XR,R|Y,\bZ}\{R - \pi_{\bZ}(\bZ; \btau_0) \} = \0$. Using conditional expectation rules it follows that
\bse
E_{XR,R|Y,\bZ} \{R- \pi_{\bZ}(\bZ; \btau_0) \} &=& E_{XR,R|Y,\bZ}(R) - \pi_{\bZ}(\bZ; \btau_0) \\
&=& \pi_{Y,\bZ}(Y,\bZ) - \pi_{\bZ}(\bZ; \btau_0)\\
&=& \pi_{\bZ}(\bZ; \btau_0) - \pi_{\bZ}(\bz; \btau_0) = 0.
\ese 
The last line follows since $\pi_{Y,\bZ}(Y,\bZ)$ is reduced to $\pi_{\bZ}(\bZ; \btau_0)$. Since $E_{XR,R|Y,\bZ}\{R - \pi_{\bZ}(\bZ; \btau_0) \} = 0$ then it follows that $E_{Y,XR,R,\bZ} \{ \bPhi_{\rm ACC}(\bO; \balpha_0, \btau_0) \} = \bzero$. Since $\bLambda_{\rm ACC}$ is a non-random matrix, the proof of robustness stays the same when modifying the ACC to include $\bLambda_{\rm ACC}$.

\item {\bf MACC estimator:}

As noted in the main manuscript, the estimating equations of the ACC and MACC estimators are identical for the choice of $\bPsi_{\rm MACC,eff}(y,\bz;\btheta)$. Therefore, the proof of consistency is the same as that for the ACC estimator. Since $\bLambda_{\rm MACC}$ is a non-random matrix, the proof of consistency stays the same when modifying the MACC estimator to include $\bLambda_{\rm MACC}$.

\item \textbf{AIPW estimator:}

The AIPW estimating equations use the IPW estimating equation and an augmentation part. We previously showed that the IPW estimating equations have mean zero. Therefore, it remains to be shown that the augmentation part of the AIPW estimator has mean $\bzero$ to prove consistency. The augmentation term $\bPsi_{\rm AIPW}(y, \bz; \btheta)$ is any function of $(Y,\bZ;\btheta)$, therefore by interchanging the roles of $\bPsi_{\rm AIPW}(y, \bz; \btheta)$ and $\bPsi_{\rm MACC}(y, \bz; \btheta)$, the proof of consistency remains the same as for the ACC and MACC estimators. Since $\bLambda_{\rm AIPW}$ is a non-random matrix, the proof of robustness stays the same when modifying the AIPW to include $\bLambda_{\rm AIPW}$.

\end{enumerate}

\subsection{Robustness}
\label{sec:thm1-proof-part2-robustness}

\begin{enumerate}
    \item \textbf{CC estimator:}

As noted in previous works, such as \textcite{Little1992} and \textcite{AshnerGarcia2023}, the CC estimator is valid as long as the missingness is independent (or conditionally independent) of the outcome. In our paper we assume that $Y \independent (C, R) | X,\bZ$, meaning that the right-censoring and missingness are conditionally independent of the outcome given $(X,\bZ)$. Therefore, the same robustness and asymptotic properties apply when interchanging the roles of $\Delta$ and $R$. That is, the CC estimator is robust to misspecification of $f_{R|\bZ}$, $f_{X|\bZ}$, and $f_{\bZ}$.

\item \textbf{IPW estimator:}

Consider the case that $f_{R|\bZ}$ is incorrectly specified as $f_{R|\bZ}^*$. Let  $\pi_{\bZ}^*(\bz; \btau_0) = f_{R|\bZ}^*(r=1, \bz; \btau_0)$. Then,
\bse
E\left\{\frac{R \bS_{\btheta}^F (Y,X,\bZ; \btheta)}{ \pi_{\bZ}(\bZ; \btau_0)}\right\}
&=& E_{Y,\bZ} \biggr[ E_{X|Y,\bZ} \left\{\frac{ \pi_{\bZ}^*(\bZ; \btau_0)}{ \pi_{\bZ}(\bZ; \btau_0)} \times \bS_{\btheta}^F (Y,X,\bZ; \btheta)\right\}\biggr] \\
&=& E_{X,\bZ} \biggr[ \frac{\pi_{\bZ}^*(\bZ; \btau_0)}{ \pi_{\bZ}(\bZ; \btau_0)} \times  E_{Y|X,\bZ} \left\{ \bS_{\btheta}^F (Y,X,\bZ; \btheta)\right\} \biggr] \\
&=& E_{X,\bZ} \biggr[ \frac{ \pi_{\bZ}^*(\bZ; \btau_0)}{ \pi_{\bZ}(\bZ; \btau_0)} \times  \bzero \biggr] = \bzero.
\ese
The last line follows since $E_{Y|X,\bZ} \left\{ \bS_{\btheta}^F (Y,x,\bz; \btheta)\right\} = \bzero$. Therefore, the IPW estimator is robust to the misspecification of $f_{R|\bZ}$, $f_{X|\bZ}$, and $f_{\bZ}$.
\vskip 0.1in

\item {\bf MLE:}

The only nuisance distribution in the MLE is $f_{X|\bZ}$. Therefore, the proof of robustness remains the same when changing the domain of integration for the missing covariate setting to the entire domain of $X$ as opposed that used for the right-censored covariate problem, i.e., $X<C$. We conclude that the the MLE is robust to the misspecficiation of $f_{R|\bZ}$ and $f_{\bZ}$, but not $f_{X|\bZ}$.

\item \textbf{ACC estimator:} 

To show that the ACC is doubly robust under independent missingness, it suffices to show that the estimator is consistent when the probabilities are correctly specified or the augmentation component is correctly specified. We let $\bPsi_{\rm ACC}(y,\bz;\btheta_0) = -E_{X|Y,\bZ}\{\bS_{\btheta}^F(y,X,\bz;\btheta_0)\}$. Later, we show that this choice makes the ACC estimator more efficient than the CC estimator. Consider the case that $\pi_{\bZ}(\bz; \btau_0)$ is incorrectly specified as $\pi_{\bZ}^*(\bz; \btau_0)$ but $\bPhi_{\rm ACC}(y,\bz;\btheta_0)$ is correctly specified:
\bse
&& E_{Y,XR,R,\bZ}[R \bS_\btheta^F (Y,X,\bZ; \btheta_0) - \{R - \pi_{\bZ}^*(\bZ; \btau_0) \} E_{X|Y,\bZ}\{\bS_{\btheta}^F(Y,X,\bZ;\btheta_0)\} ] \\
&& = -E_{Y,XR,R,\bZ}[\{R - \pi_{\bZ}^*(\bZ; \btau_0) \} E_{X|Y,\bZ}\{\bS_{\btheta}^F(Y,X,\bZ;\btheta_0)\} ] \\
&& = -E_{Y,X,\bZ}[\{\pi_{\bZ}(\bZ; \btau_0) - \pi_{\bZ}^*(\bZ; \btau_0) \} E_{X|Y,\bZ}\{\bS_{\btheta}^F(Y,X,\bZ;\btheta_0)\} ] \\
&& = -E_{Y,X,\bZ}[\{\pi_{\bZ}(\bZ; \btau_0) - \pi_{\bZ}^*(\bZ; \btau_0) \} \bS_{\btheta}^F(Y,X,\bZ;\btheta_0) ] \\
&& = -E_{X,\bZ}[\{\pi_{\bZ}(\bZ; \btau_0) - \pi_{\bZ}^*(\bZ; \btau_0) \} E_{Y|X,\bZ} \{\bS_{\btheta}^F(Y,X,\bZ;\btheta_0) \}] \\
&& = -E_{X,\bZ}[\{\pi_{\bZ}(\bZ; \btau_0) - \pi_{\bZ}^*(\bZ; \btau_0) \} \times \bzero \}] = \bzero.
\ese 
The first line follows since the CC estimating equations are unbiased. The second line follows by the law of conditional expectation. 
The last two lines follow from the law of conditional expectation and since  $E_{Y|X,\bZ} \{\bS_{\btheta}^F(Y,x,\bz;\btheta_0) \} = \bzero$. Now consider the case when the probability $\pi_{\bZ}(\bz; \btau)$ is correctly specified but $\bPhi_{\rm ACC}(y,\bz;\btheta_0)$ is incorrectly specified as $\bPhi_{\rm ACC}^*(y,\bz;\btheta_0)$. It follows that 
\bse
&& E_{Y,XR,R,\bZ}[R \bS_\btheta^F (Y,X,\bZ; \btheta_0) + \{R - \pi_{\bZ}(\bZ; \btau_0) \} \bPhi_{\rm ACC}^*(Y,\bZ;\btheta_0)\} ] \\
&& = E_{Y,XR,R,\bZ}[\{R - \pi_{\bZ}(\bZ; \btau_0) \} \bPhi_{\rm ACC}^*(Y,\bZ;\btheta_0)\} ] \\
&& = E_{Y,X,\bZ}[\{\pi_{\bZ}(\bZ; \btau_0) - \pi_{\bZ}(\bZ; \btau_0) \} \bPhi_{\rm ACC}^*(Y,\bZ;\btheta_0) ] \\
&& = E_{Y,X,\bZ}\{ 0 \times  \bPhi_{\rm ACC}^*(Y,\bZ;\btheta_0) \} = \bzero. 
\ese 
Therefore we have proven that the ACC estimator only requires correct specification of either the probabilities or the augmentation component. Consequently, the ACC estimator is doubly robust under independent covariate missingness. Since $\bLambda_{\rm ACC}$ is a non-random matrix, the same robustness properties apply as those of the  ACC estimator without $\bLambda_{\rm ACC}$.

\item {\bf MACC estimator:}

The estimating equations of the ACC and MACC estimators are identical for the choice of $\bPsi_{\rm MACC}(y,\bz;\btheta)$. Therefore, the same robustness properties apply for the independent missingness case. Since $\bLambda_{\rm MACC}$ is a non-random matrix, the  robustness properties with and without using  $\bLambda_{\rm MACC}$ are the same.

\item  \textbf{AIPW estimator:} 

To show that the AIPW estimator is doubly robust under independent missingness, it suffices to show that the estimator is consistent when the probabilities are correctly specified or the augmentation component is correctly specified. We let $\bPsi_{\rm AIPW}(y,\bz;\btheta_0) = E_{X|Y,\bZ}\{\bS_{\btheta}^F(y,x,\bz;\btheta_0)\}$. We will later show that this choice makes the AIPW estimator more efficient than the IPW estimator. Since $\bPsi_{\rm AIPW}(y,\bz;\btheta_0) = -\bPsi_{\rm ACC}(y,\bz;\btheta_0)$, then we can show that the AIPW estimator is doubly robust using a similar argument as that which we used for the ACC estimator. The robustness properties do not change when using $\bLambda_{\rm AIPW}$ since $\bLambda_{\rm AIPW}$ is a non-random matrix.

\end{enumerate}

\subsection{Asymptotic normality}
\label{sec:thm1-proof-part2-normality}
%
\begin{enumerate}
    \item {\bf CC estimator}: 

    By interchanging the roles of $(W,\Delta)$ with $(XR,R)$, the proof of asymptotic normality remains the same. 

\item \textbf{IPW estimator:}

By interchanging $\bPhi_{\rm IPW}(\bO_i; \btheta_0, \btau_0)$ from the missing covariate problem with $\bPhi_{\rm IPW}(\bO_i; \btheta_0, \boldeta_0)$ from the right-censored covariate problem, the proof of asymptotic normality remains the same when $f_{R|\bZ}$ is known. When $f_{R|\bZ}$ needs to be estimated, the corresponding consistent and asymptotically normal estimator $\wh\btau$ needs to be specified. Then, the influence function of $\wh\btau$ is used in the construction of the asymptotic normality of $\wh\btheta_{\rm IPW; \btau}$. 

\item {\bf MLE:}

By interchanging $\bPhi_{\rm MLE}(\bO_i; \btheta_0, \biota_0)$ from the missing covariate setting with $\bPhi_{\rm MLE}(\bO_i; \btheta_0, \bgamma_0)$ from the right-censored covariate problem, the proof of asymptotic normality remains the same when $f_{X|\bZ}$ is known. When, $f_{X|\bZ}$ needs to be estimated, the corresponding consistent and asymptotically normal estimator $\wh\biota$ needs to be specified. Then, the influence function of $\wh\biota$ is used in the construction of the asymptotic normality of $\wh\btheta_{\rm MLE; \biota}$. 

\item \textbf{ACC estimator:} 

By interchanging $\pi_{\bZ}(\bz;\btau_0)$ from the missing covariate with $\pi_{Y,\bZ}(y,\bz;\bkappa_0)$ from the right-censored covariate problem, as well as the augmented components $\bPsi_{\rm ACC}(y,\bz; \btheta)$ for both problems, the proof of asymptotic normality remains the same when $f_{R|\bZ}$ is known. When $f_{R|\bZ}$ needs to be estimated, the corresponding consistent and asymptotically normal estimator $\wh\btau$ needs to be specified. Then, the influence function of $\wh\btau$ is used in the construction of the asymptotic normality of $\wh\btheta_{\rm ACC; \btau}$ when using and when not using $\bLambda_{\rm ACC}$.

\item {\bf MACC estimator:}

As noted in the main manuscript, the estimating equations of the ACC and MACC estimator are identical for the choice of $\bPsi_{\rm MACC, eff}(y,\bz;\btheta)$. Therefore, the same asymptotic properties apply for the independent missingness case, and likewise, when we use $\bLambda_{\rm MACC}$.

\item \textbf{AIPW estimator:}

By interchanging $\pi_{\bZ}(\bz;\btau_0)$ from the missing covariate problem with $\pi_{X,\bZ}(w,\bz;\boldeta_0)$ from the right-censored covariate problem, as well as the augmented components $\bPsi_{\rm AIPW}(y,\bz; \btheta)$ for both problems, the proof of asymptotic normality remains the same when $f_{R|\bZ}$ is known. When $f_{R|\bZ}$ needs to be estimated, the corresponding consistent and asymptotically normal estimator $\wh\btau$ needs to be specified. Then, the influence function of $\wh\btau$ is used in the construction of the asymptotic normality of $\wh\btheta_{\rm AIPW; \btau}$ when using and when not using $\bLambda_{\rm AIPW}$.

\end{enumerate}

\subsection{Efficiency}
\label{sec:thm1-proof-part2-efficiency}

For brevity, in this section we show what the forms of the augmented components are for the missing covariate problem after applying a similar argument as that for the right-censored covariate problem. After finding the form of $\bLambda$ that guarantees efficiency over the original estimators, one can use the sample mean in the estimation of $\btheta$.

\begin{enumerate}
    \item {\bf ACC estimator:}

\emph{When $\btheta$ is the only unknown parameter and we do not use $\bLambda_{\rm ACC}$:} After applying a similar argument as that for the right-censored covariate case, we find 
\bse
&& \bPsi_{\rm ACC, eff}(y,\bz; \btheta_0) \\
&& \quad = - E_{XR,R|Y,\bZ} \left[  \left\{R - \pi_{\bZ}(\bz;\btau_0) \right\} R \bS_\btheta^F(y,X,\bz;\btheta_0)  \right] E_{XR,R|Y,\bZ} \left[ \left\{R - \pi_{\bZ}( \bz;\btau_0) \right\}^2 \right]^{-T}  \\
&& \quad = - \frac{\pi_{\bZ}( \bz;\btau_0)\{1-\pi_{\bZ}( \bz;\btau_0)\} E_{X|Y,\bZ} \left\{   \bS_\btheta^F(y,X,\bz;\btheta_0)  \right\}} {\pi_{\bZ}( \bz;\btau_0)\{1-\pi_{\bZ}( \bz;\btau_0)\} }  \\
&& \quad = - E_{X|Y,\bZ} \left\{   \bS_\btheta^F(y,X,\bz;\btheta_0)  \right\}. 
\ese 

\emph{When $\btheta$ and $\btau$ are  unknown parameters and we do not use $\bLambda_{\rm ACC}$:} Using a similar argument as that of \textcite{Bartlettetal2014}, it follows that the same $\bPhi_{\rm ACC}(y,\bz;\btheta)$ can be used to guarantee efficiency gains over the CC estimator when $\btau_0$ is estimated using logistic regression. 

\emph{When only  $\btheta$ is unknown  and we use $\bLambda_{\rm ACC}$:} After applying a similar argument as that for the right-censored covariate case, we find 
\bse
\bLambda_{\rm ACC} && = -  E \left[ \left\{ R - \pi_{\bZ}(\bZ; \btau_0) \right\} \bPsi_{\rm ACC}(Y,\bZ; \btheta_0) R \bS_\btheta^F(Y,X,\bZ;\btheta_0)^T  \right] \\
&& \quad \quad \times \left(  E \left[ \left\{ R - \pi_{\bZ}(\bZ; \btau_0) \right\} \bPsi_{\rm ACC}(Y,\bZ; \btheta_0) \right]^{\otimes 2} \right)^{-T} .
\ese 

\emph{When   $\btheta$ and $\btau$ are unknown and we use $\bLambda_{\rm ACC}$:} Before introducing the form of $\bLambda_{\rm ACC;\btau}$ that guarantees higher efficiency over the CC estimator, we let
\bse
\bA^*_{\rm ACC;\btau} &=& E \left\{ \partial \bPhi_{\rm ACC} (\bO; \btheta_0, \btau_0) / \partial \btau^T \right\}; \\
\bUpsilon_{\btau}(\bz) &=& - E\left\{ \partial \bS^F_{\btau} (\bZ; \btau_0) / \partial \btau^T \right\}^{-1} \bS^F_{\btau} (\bz; \btau_0),
\ese
where $\bS^F_{\btau} (\bz; \btau)$ is the score equation of the log-likelihood of $f_{R|\bZ}$. Then it follows that,
\bse
\bLambda_{\rm ACC; \btau} && = -  E \biggr( \left[ \left\{ R - \pi_{\bZ}(\bZ; \btau_0) \right\} \bPsi_{\rm ACC}(Y,\bZ; \btheta) + \bA^*_{\rm ACC;\btau} \bUpsilon_{\btau} (\bZ) \right] R \bS_\btheta^F(Y,X,\bZ;\btheta)^T  \biggr) \\
&& \times \left( E  \left[ \left\{ R - \pi_{\bZ}(\bZ; \btau_0) \right\} \bPsi_{\rm ACC}(Y,\bZ; \btheta) + \bA^*_{\rm ACC;\btau} \bUpsilon_{\btau} (\bZ) \right]^{\otimes 2} \right)^{-T}. 
\ese 

\item {\bf MACC estimator:}

\emph{When $\btheta$ is the only unknown parameter and we do not use $\bLambda_{\rm MACC}$:} After applying a similar argument as that for the right-censored covariate case, we find 
\bse
&& \bPsi_{\rm MACC, eff}(y,\bz; \btheta_0) \\
&& \quad = - E_{XR,R|Y,\bZ} \left[  \left\{ 1- R/\pi_{\bZ}(\bZ;\btau_0) \right\} R \bS_\btheta^F(y,X,\bz;\btheta_0)  \right] E_{XR,R|Y,\bZ} \left[ \left\{1 - R/\pi_{\bZ}( \bz;\btau_0) \right\}^2 \right]^{-T}  \\
&& \quad = - \frac{\{\pi_{\bZ}( \bz;\btau_0) - 1 \} E_{X|Y,\bZ} \left\{   \bS_\btheta^F(y,X,\bz;\btheta_0)  \right\}} {-1+1/\pi_{\bZ}( \bz;\btau_0) }  \\
&& \quad = - \frac{\{\pi_{\bZ}( \bz;\btau_0) - 1 \} E_{X|Y,\bZ} \left\{   \bS_\btheta^F(y,X,\bz;\btheta_0)  \right\}} {-1+1/\pi_{\bZ}( \bz;\btau_0) }  \\
&& \quad = E_{X|Y,\bZ} \left\{   \bS_\btheta^F(y,X,\bz;\btheta_0) \right\} \pi_{\bZ}( \bz;\btau_0). 
\ese 

\emph{When $\btheta$ and $\btau$ are  unknown parameters and we do not use $\bLambda_{\rm MACC}$:} Using a similar argument as that for the ACC estimator, and assuming that $\btheta_0$ and $\btau_0$ are independent, it follows that the same $\bPhi_{\rm MACC}(y,\bz;\btheta)$ can be used to guarantee efficiency gains over the CC estimator.

\emph{When only  $\btheta$ is unknown  and we  use $\bLambda_{\rm MACC}$:}
After applying a similar argument as that for the ACC estimator, we find 
\bse
\bLambda_{\rm MACC} && = -  E \left[ \left\{1- R/\pi_{\bZ}(\bZ; \btau_0) \right\} \bPsi_{\rm MACC}(Y,\bZ; \btheta_0) R \bS_\btheta^F(Y,X,\bZ;\btheta_0)^T  \right] \\
&& \quad \quad \times \left( E \left[ \left\{1- R/\pi_{\bZ}(\bZ; \btau_0) \right\} \bPsi_{\rm MACC}(Y,\bZ; \btheta_0) \right]^{\otimes 2} \right)^{-1} .
\ese 

\emph{When   $\btheta$ and $\btau$ are unknown and we use $\bLambda_{\rm MACC}$:}
Before introducing the form of $\bLambda_{\rm ACC;\btau}$ that guarantees higher efficiency over the CC estimator, we let
\bse
\bA^*_{\rm MACC;\btau} &=& E \left\{ \partial \bPhi_{\rm MACC} (\bO; \btheta_0, \btau_0) / \partial \btau^T \right\}.
\ese
Then it follows that
\bse
\bLambda_{\rm MACC; \btau} && = -  E \biggr( \left[ \left\{ 1-R/\pi_{\bZ}(\bZ; \btau_0) \right\} \bPsi_{\rm MACC}(Y,\bZ; \btheta) + \bA^*_{\rm MACC;\btau} \bUpsilon_{\btau}(\bZ) \right] R \bS_\btheta^F(Y,X,\bZ;\btheta_0)^T  \biggr) \\
&& \times \left( E  \left[ \left\{ 1 -R/\pi_{\bZ}(\bZ; \btau_0) \right\} \bPsi_{\rm MACC}(Y,\bZ; \btheta) + \bA^*_{\rm MACC;\btau} \bUpsilon_{\btau}(\bZ) \right]^{\otimes 2} \right)^{-T}. 
\ese 

\item {\bf AIPW estimator:}

\emph{When only  $\btheta$ is unknown  and we do not use $\bLambda_{\rm MACC}$:} We investigate a choice of $\bPsi_{\rm AIPW}(y, \bz; \btheta_0)$ to make the AIPW estimator more efficient than the IPW estimator. To do so, we compare the influence functions of $\widehat{\btheta}_{\rm AIPW}$ and $\widehat{\btheta}_{\rm IPW}$, and find a choice of $\bPsi_{\rm AIPW}(y, \bz; \btheta_0)$ such that $\bzero = \cov(\bUpsilon_{\rm AIPW}  - \bUpsilon_{\rm IPW}, \bUpsilon_{\rm AIPW})$. Applying a similar argument as before, we find 
\bse
&& \bPsi_{\rm AIPW, eff}(y,\bz; \btheta_0) \\
&& \quad = - E_{XR,R|Y,\bZ} \left[  \left\{ 1 - R/ \pi_{\bZ}(\bz;\btau_0) \right\} R \bS_\btheta^F(y,X,\bz;\btheta_0)/\pi_{\bZ}(\bz;\btau_0)  \right] E_{XR,R|Y,\bZ} \left[ \left\{1-r/\pi_{\bZ}( \bz;\btau_0) \right\}^2 \right]^{-T}  \\
&& \quad = - \frac{\{1-1/\pi_{\bZ}(\bz;\btau_0)\} E_{X|Y,\bZ} \left\{   \bS_\btheta^F(y,X,\bz;\btheta_0)  \right\}} {\{1/\pi_{\bZ}(w, \bz;\btau_0) - 1\} }  \\
&& \quad = E_{X|Y,\bZ} \left\{   \bS_\btheta^F(y,X,\bz;\btheta_0)  \right\}. 
\ese

\emph{When $\btheta$ and $\btau$ are  unknown parameters and we do not use $\bLambda_{\rm AIPW}$:} Using a similar argument as that for the ACC estimator, and assuming that $\btheta_0$ and $\btau_0$ are independent, it follows that the same $\bPhi_{\rm AIPW}(y,\bz;\btheta)$ can be used to guarantee efficiency gains over the IPW estimator.
\vskip 0.1 in

\emph{When only  $\btheta$ is unknown  and we  use $\bLambda_{\rm AIPW}$:} By interchanging the roles of $\bPhi_{\rm MACC}(\bO; \btheta_0, \btau_0)$ with $\bPhi_{\rm AIPW}(\bO; \btheta_0, \btau_0)$, the same argument applies.  We find that
\bse
\bLambda_{\rm AIPW} && = -  E \left[ \left\{1- R/\pi_{\bZ}(\bZ; \btau_0) \right\} \bPsi_{\rm AIPW}(Y,\bZ; \btheta_0) R \bS_\btheta^F(Y,X,\bZ;\btheta_0)^T/\pi_{\bZ}(\bZ; \btau_0)  \right] \\
&& \quad \quad \times \left( E  \left[ \left\{1- R/\pi_{\bZ}(\bZ; \btau_0) \right\} \bPsi_{\rm AIPW}(Y,\bZ; \btheta_0) \right]^{\otimes 2} \right)^{-T}. 
\ese 

\emph{When   $\btheta$ and $\btau$ are unknown and we use $\bLambda_{\rm MACC}$:} Before introducing the form of $\bLambda_{\rm AIPW;\btau}$ that guarantees higher efficiency over the IPW estimator, we let
\bse
\bA^*_{\rm IPW;\btau} &=& E \left\{ \partial \bPhi_{\rm IPW} (\bO; \btheta_0, \btau_0) / \partial \btau^T \right\}; \\
\bA^*_{\rm AIPW;\btau} &=& E \left[\frac{\partial}{\partial \btau^T}  \left\{ 1 -\frac{R}{\pi_{\bZ}(\bZ;\btau)} \right\} \bPsi_{\rm AIPW} (Y,\bZ; \btheta_0) \right].
\ese
We find that
\bse
\bLambda_{\rm AIPW; \btau} &=& -  E \biggr( \left[ \left\{ 1-R/\pi_{\bZ}(\bZ; \btau_0) \right\} \bPsi_{\rm AIPW}(Y,\bZ; \btheta) + \bA^*_{\rm AIPW;\btau} \bUpsilon_{\btau} (\bZ) \right] \\
&&  \quad \quad \quad \times \{ \bPhi_{\rm IPW}(\bO; \btheta_0, \btau_0) + \bA^*_{\rm IPW;\btau} \bUpsilon_{\btau} (\bZ) \} \biggr)^T \\
&& \times  E \left( \left[ \left\{ 1 -R/\pi_{\bZ}(\bz; \btau_0) \right\} \bPsi_{\rm AIPW}(Y,\bZ; \btheta_0) + \bA^*_{\rm AIPW;\btau} \bUpsilon_{\btau}(\bZ) \right]^{\otimes 2} \right)^{-1}. 
\ese 

\end{enumerate}

\subsection{Changes to accommodate dependent covariate missingness} 

\begin{enumerate}
    \item {\bf CC estimator:}
    
    No changes occur.

    \item {\bf IPW estimator:} 
    
    Under dependent covariate missingness, the IPW estimating equations use the probabilities defined by $\pi^{\rm miss, dep}_{X,\bZ}(x,\bz)$ as opposed to $\pi^{\rm miss, ind}_{\bZ}(\bz)$. The probabilities in the dependent covariate right-censoring and missingness settings are now a function of $(X,\bZ)$. Therefore, the proofs for consistency and robustness remain the same when interchanging $\pi^{\rm miss, dep}_{X,\bZ}(x,\bz)$ from the missing covariate problem with $\pi^{\rm cens, dep}_{X,\bZ}(x,\bz)$ from the right-censored covariate problem, and $R$ with $\Delta$. Let the probabilities $\pi^{\rm miss, dep}_{X,\bZ}(x,\bz;\bxi) = f_{R|X,\bZ}(r=1, x,\bz; \bxi)$ be indexed by parameter $\bxi$. When $f_{R|X,\bZ}$ is assumed known, the asymptotic normality of $\wh\btheta_{\rm IPW}$ remains the same as for independent covariate missingness, but instead of probabilities defined by $f_{R|\bZ}$, we use $f_{R|X,\bZ}$. When $f_{R|X,\bZ}$ is unknown, a consistent and asymptotically normal estimator needs to be obtained first. As described in Section \ref{sec:implement}, obtaining this estimator is particularly difficult for the missing covariate problem, as that requires more unverifiable assumptions to satisfy identifiability than are needed for the dependent right-censored covariate problem. If identifiability is established, then we can follow a similar argument as in the right-censored covariate problem to derive the asymptotic distribution of $\wh\btheta_{\rm IPW}$ by interchanging the influence functions of the nuisance parameters with those from the independent missingness case.

    \item {\bf MLE:}
     
    Under dependent covariate missingness, the MLE estimating equations use $f_{X|R,\bZ}$ as opposed to $f_{X|\bZ}$, and the domain of integration is $\D(x)$. By (i) interchanging the roles of $f_{X|R,\bZ}$ with $f_{X|C,\bZ}$ from the right-censored covariate problem, and (ii) defining the domain of integration of $X$ (i.e., $\D(x)$), the proofs of consistency and robustness remain the same as those in the right-censored covariate problem. That is, the MLE is robust to $f_{R|\bZ}$ and $f_{\bZ}$, but not $f_{X|R,\bZ}$. When assuming that $f_{X|R,\bZ}$ is known, the proof of asymptotic normality distribution of $\wh\btheta_{\rm MLE}$ remains the same as that for independent covariate missingness using $f_{X|R,\bZ}$ instead of $f_{X|\bZ}$. When $f_{X|R,\bZ}$ is unknown, we run into a similar issue as we did for the IPW estimator. That is, a consistent and asymptotically normal estimator for the nuisance distribution needs to be obtained first, but doing so will involve  more unverifiable assumptions than those needed for dependent covariate right-censoring. If identifiability is established, then  a similar argument as that for independent covariate missingness is made to define the asymptotic distribution of $\wh\btheta_{\rm MLE}$.

    \item {\bf ACC estimator:}
    %
    
    The proofs of robustness and asymptotic normality of the ACC estimator under dependent covariate missingness can be found in the supplementary material of \textcite{Bartlettetal2014}. In that paper, the authors use semi-parametric theory to define orthogonal spaces for the the parameters governing the distributions of the the mean form of the restricted mean model (i.e., $\bbeta$ in our case) and those from the probability of observing $X$ given $(Y,\bZ)$. By interchanging the roles of our score equation in our paper with the restricted mean model in their paper, the proof remains the same. 
    \item {\bf MACC estimator:}

    Under dependent covariate missingness, the MACC estimator uses the probabilities $\pi_{X,\bZ}$ as opposed to $\pi_{\bZ}$. Let the probabilities $\pi_{X,\bZ}(x,\bz;\bxi) = f_{R|X,\bZ}(r=1,x,\bz;\bxi)$ be indexed by parameter $\bxi$. By interchanging the roles of $(W,\Delta)$ with $(XR,R)$, and  $\pi_{X,\bZ}^{\rm miss,dep}(x,\bz;\bxi)$ with $\pi_{X,\bZ}^{\rm cens,dep}(x,\bz;\boldeta)$, the proofs of consistency and robustness remain the same as those for the right-censored covariate problem. Additionally, the asymptotic normality of $\wh\btheta_{\rm MACC}$ when $f_{R|X,\bZ}$ is assumed known remains the same as that for the right-censored covariate problem when $f_{C|X,\bZ}$ is assumed known. When $f_{R|X,\bZ}$ is unknown, a consistent and asymptotically normal estimator needs to be obtained first. Identifiability requires more unverifiable assumptions (Section \ref{sec:implement}), but if identifiability is established, then  a similar argument to the ACC estimator is made to define the asymptotic distribution of $\wh\btheta_{\rm MACC;\bxi}$, i.e., by interchanging the influence functions of the nuisance parameters. When the MACC estimator uses $\bLambda_{\rm MACC}$, the same considerations apply.


    \item {\bf AIPW estimator:}
    
    The proofs of robustness and asymptotic normality of the AIPW estimator under dependent covariate missingness can be found in \textcite{rotnitzky1997analysis}. In that paper, the authors use semi-parametric theory to define orthogonal spaces for the parameters governing the distributions of the the mean form of the restricted mean model (i.e., $\bbeta$ in our case) and those from the probability of observing $X$ given the observed data (i.e., $\bxi$ in our case). By interchanging the roles of our score equation with their restricted mean model, the proof remains the same. When the MACC uses $\bLambda_{\rm MACC}$, the same considerations apply.
    

\end{enumerate}
\section{Proof of Theorem \ref{thm:all-efficiency}}
\label{sec:thm2-proof}

By definition, the MLE achieves optimal efficiency when correctly specified, making it the most efficient estimator when compared to the IPW, ACC, MACC, and AIPW estimators. The ACC and MACC estimators are more statistically efficient than the CC estimator as they were constructed to satisfy the efficiency condition in equation (\ref{eqn:efficiency-condition}).  Similarly, the AIPW estimator is more statistically efficient than the IPW estimator by construction.

In the subsequent sections, we first present  preliminary results in the form of corollaries, then we use these to show that the CC estimator is more efficient than the IPW (Section \ref{sec:cc-ipw-efficiency}) and AIPW (Section \ref{sec:cc-aipw-efficiency}) estimators. We show this result by comparing the diagonal elements of the covariance matrix for the CC, IPW, and AIPW estimators, which represent the variances of each individual component of $\btheta_0$. We show that the individual variance components of the CC estimator are equal to or smaller than those of the IPW and AIPW estimators. While we show that the diagonal elements of the CC estimator are equally or smaller than those provided by the IPW and AIPW estimators, it may be the case that the conditional mean of the outcome $Y$ given the set of covariate values $(X=x, \bZ=\bz)$ using the IPW and AIPW estimators may be smaller than that of the CC estimator since in our proof of higher efficiency of the CC estimator, we do not establish the relationship between the off-diagonal elements of the covariance matrix.

\subsection{Supporting corollaries}

\begin{enumerate}[label=(C\arabic*),ref=(C\arabic*)]
    \item\label{sec:a_cc=-b_cc} Proof $\bA_{\rm CC} = - \bB_{\rm CC}$

    To show that $\bA_{\rm CC} = - \bB_{\rm CC}$, we must show that $E \{ \Delta \partial \bS_\btheta^F (Y,W,\bZ; \btheta_0)/ \partial \btheta\} = $ \\ $ - E \{ \Delta \bS_\btheta^F (Y,W,\bZ; \btheta_0) ^{\otimes 2} \}$. Given that $E \{ \Delta \bS_\btheta^F (Y,W,\bZ; \btheta_0)\} =\bzero$  then it follows that
\bse
\bzero &=& \frac{\partial}{\partial \btheta}\int \delta \bS_{\btheta}^F(\by,w,\bz;\btheta_0) f_{Y,W,\Delta, \bZ} (y,w,\delta, \bz; \btheta_0) dy dw d\delta d\bz \\
&=&  \int \delta \bS_{\btheta}^F(\by,w,\bz;\btheta_0) \biggr\{ \frac{\partial}{\partial \btheta} f_{Y,W,\Delta, \bZ} (y,w,\delta, \bz; \btheta_0) \biggr\} + \\
&& \quad \quad \biggr\{ \frac{\partial}{\partial \btheta}   \delta \bS_{\btheta}^F(\by,w,\bz;\btheta_0) \biggr\} f_{Y,W,\Delta, \bZ} (y,w,\delta, \bz; \btheta_0) \biggr] dy dw d\delta d\bz  \\
&=&  \int \delta \bS_{\btheta}^F(\by,w,\bz;\btheta_0) \biggr\{ \frac{\partial}{\partial \btheta} f_{Y,W,\Delta, \bZ} (y,w,\delta, \bz; \btheta_0) \biggr\} dy dw d\delta d\bz  + E \{ \Delta \partial \bS_\btheta^F (Y,W,\bZ; \btheta_0)/ \partial \btheta\}.
\ese
The partial derivative of the integral is equal to
\bse
 &&  \frac{\partial}{\partial \btheta} f_{Y,W,\Delta, \bZ} (y,w,\delta, \bz; \btheta_0) \\
 && = \frac{\partial}{\partial \btheta} \biggr\{ \int_{x\leq C} f_{Y,X,C,\bZ}(y,w,c,\bz; \btheta_0) dc \biggr\} ^{\delta} \biggr\{ \int_{c<X}  f_{Y,X,C,\bZ}(y,x,w,\bz; \btheta_0) dx \biggr\} ^{1-\delta} \\
 && = \frac{\partial}{\partial \btheta} \biggr\{ f_{Y|X,\bZ}(y,w,\bz;\btheta_0) \int_{x\leq C} f_{X,C,\bZ}(w,c,\bz) dc \biggr\} ^{\delta} \biggr\{ \int_{c<X}  f_{Y,X,C,\bZ}(y,x,w,\bz; \btheta_0) dx \biggr\} ^{1-\delta} \\
 && = \biggr\{ f_{Y|X,\bZ}(y,w,\bz;\btheta_0) \int_{x\leq C} f_{X,C,\bZ}(w,c,\bz) dc \biggr\} ^{\delta} (1-\delta) \biggr\{ \int_{c<X}  f_{Y,X,C,\bZ}(y,x,w,\bz; \btheta_0) dx \biggr\}^{-\delta} \\
 && \times \biggr\{ \frac{\partial}{\partial \btheta} \int_{c<X}  f_{Y,X,C,\bZ}(y,x,w,\bz; \btheta_0) dx \biggr\} + \biggr\{ \int_{c<X}  f_{Y,X,C,\bZ}(y,x,w,\bz; \btheta_0) dx \biggr\} ^{1-\delta} \\
 && \times \delta \biggr\{ f_{Y|X,\bZ}(y,w,\bz;\btheta_0) \int_{x\leq C} f_{X,C,\bZ}(w,c,\bz) dc \biggr\} ^{\delta-1} \biggr\{ \frac{\partial}{\partial \btheta} f_{Y|X,\bZ}(y,w,\bz;\btheta_0) \int_{x\leq C} f_{X,C,\bZ}(w,c,\bz) dc \biggr\}. 
\ese
The integral with respect to $\delta=0$ is equal to $\bzero$ and the integral with respect to $\delta=1$ leads to
\bse
&&\int \delta \bS_{\btheta}^F(\by,w,\bz;\btheta_0) \biggr\{ \frac{\partial}{\partial \btheta} f_{Y,W,\Delta, \bZ} (y,w,\delta, \bz; \btheta_0) \biggr\} dy dw d\delta d\bz \\
&& = \int \bS_{\btheta}^F(\by,x,\bz;\btheta_0) \frac{\partial}{\partial \btheta} f_{Y|X,\bZ}(y,x,\bz;\btheta_0) \int_{x\leq C} f_{X,C,\bZ}(w,c,\bz) dc dy dx d\bz \\
&& = \int I(X\leq C) \bS_{\btheta}^F(\by,x,\bz;\btheta_0)^{\otimes 2} f_{Y|X,\bZ}(y,x,\bz;\btheta_0)  f_{X,C,\bZ}(w,c,\bz) dc dy dx d\bz  \\
&& = E \{ \Delta \bS_\btheta^F (Y,W,\bZ; \btheta_0) ^{\otimes 2} \}
\ese
where the second-to-last line follows since $\partial f_{Y|X,\bZ}(y,x,\bz;\btheta_0)/\partial \btheta = \bS_{\btheta}^F(\by,x,\bz;\btheta_0) f_{Y|X,\bZ}(y,x,\bz;\btheta_0)$. It follows that
\bse
\bzero &=& \frac{\partial}{\partial \btheta}\int \delta \bS_{\btheta}^F(\by,w,\bz;\btheta_0) f_{Y,W,\Delta, \bZ} (y,w,\delta, \bz; \btheta_0) dy dw d\delta d\bz \\
&=& E \{ \Delta \bS_\btheta^F (Y,W,\bZ; \btheta_0) ^{\otimes 2} \} + E \{ \Delta \partial \bS_\btheta^F (Y,W,\bZ; \btheta_0)/ \partial \btheta\}.
\ese
Therefore, $E \{ \Delta \partial \bS_\btheta^F (Y,W,\bZ; \btheta_0)/ \partial \btheta\} = - E \{ \Delta \bS_\btheta^F (Y,W,\bZ; \btheta_0) ^{\otimes 2} \}$, and $\bA_{\rm CC} = - \bB_{\rm CC}$.

\item\label{sec:-a_ipw} Proof $E \{ \bPhi_{\rm IPW}(\bO; \btheta_0, \balpha_0) \bPhi_{\rm CC}(\bO; \btheta_0)^T \} = -\bA_{\rm IPW}$

To prove consistency, we showed  $E \{ \bPhi_{\rm IPW}(\bO; \btheta_0, \balpha_0) \} = E\{\delta \bS_{\btheta}^F (Y,W,\bZ; \btheta_0) / \pi_{X,\bZ}(W,\bZ; \boldeta_0)\} = \bzero$. Using a similar argument as in Corollary \ref{sec:a_cc=-b_cc}, it follows  that
\bse
\bzero &=&  \frac{\partial}{\partial \btheta^T} E \{ \bPhi_{\rm IPW}(\bO; \btheta_0, \balpha_0) \} \\
 &=& \int \frac{\partial}{\partial \btheta^T} \frac{\delta \bS_{\btheta}^F(\by,w,\bz;\btheta_0)}{\pi_{X,\bZ}(w,\bz; \boldeta_0)} f_{Y,W,\Delta, \bZ} (y,w,\delta, \bz) dy dw d\delta d\bz \\
&=& \int I(X\leq C) \frac{ \bS_{\btheta}^F(\by,x,\bz;\btheta_0)}{\pi_{X,\bZ}(x,\bz; \boldeta_0)}  I(X\leq C) \bS_{\btheta}^F(\by,x,\bz;\btheta_0)^T f_{Y|X,\bZ}(y,x,\bz;\btheta_0) f_{X,C,\bZ}(w,c,\bz) dc dy dx d\bz \\
&& + E \biggr\{ \frac{\partial}{\partial \btheta^T} \frac{\Delta \bS_{\btheta}^F(Y,W,\bZ;\btheta_0)}{\pi_{X,\bZ}(W,\bZ; \boldeta_0)} \biggr\} \\
&=& E \biggr\{ \frac{\Delta \bS_{\btheta}^F(Y,W,\bZ;\btheta_0)}{\pi_{X,\bZ}(W,\bZ; \boldeta_0)} \delta \bS_{\btheta}^F(Y,W,\bZ;\btheta_0)^T \biggr\} + E \biggr\{ \frac{\partial}{\partial \btheta^T} \frac{\Delta \bS_{\btheta}^F(Y,W,\bZ;\btheta_0)}{\pi_{X,\bZ}(W,\bZ; \boldeta_0)} \biggr\} \\
&=& E \{ \bPhi_{\rm IPW}(\bO; \btheta, \balpha_0) \bPhi_{\rm CC}(\bO; \btheta_0)^T \} + E \biggr\{ \frac{\partial}{\partial \btheta^T} \frac{\Delta \bS_{\btheta}^F(Y,W,\bZ;\btheta_0)}{\pi_{X,\bZ}(W,\bZ; \boldeta_0)} \biggr\}.
\ese
Since $\bA_{\rm IPW} =E \left\{  \partial/ \partial \btheta^T \Delta \bS_{\btheta}^F(Y,W,\bZ;\btheta_0) /\pi_{X,\bZ}(W,\bZ; \boldeta_0)  \right\}$, it follows that
\bse
E \{ \bPhi_{\rm IPW}(\bO; \btheta_0, \balpha_0) \bPhi_{\rm CC}(\bO; \btheta_0)^T \} = -\bA_{\rm IPW}.
\ese

\item \label{sec:cc_upsilon_alpha} Proof $ E\{ \bUpsilon^*_{\balpha} (\bO^*) \bPhi_{\rm CC}(\bO; \btheta_0)^T \} = \bzero$

Let $\bUpsilon^*_{\balpha}(\bO^*) = \bA^*_{\rm IPW; \balpha} \bUpsilon_{\balpha}(\bO^*)$ where $\bUpsilon_{\balpha}(\bO^*) = - [E\{\partial \bPhi_{\balpha}(\bO^*; \balpha_0)/\partial\balpha^T\}]^{-1}\bPhi_{\balpha}(\bO^*; \balpha_0)$ and $\bA^*_{\rm IPW; \balpha} = E \{ \partial \bPhi_{\rm IPW}(\bO_i; \btheta_0, \balpha_0) / \partial \balpha^T \}$. Then, 
\bse
&& E \{ \bUpsilon^*_{\balpha} (\bO^*) \bPhi_{\rm CC}(\bO; \btheta_0)^T \} \\
&& = E_{Y,W,\Delta, \bZ} \{ \bA^*_{\rm IPW; \balpha} \bUpsilon_{\balpha} (W,\Delta,\bZ) \Delta \bS_{\btheta}^F(Y,W,\bZ;\btheta_0)^T \} \\
&& = \bA^*_{\rm IPW; \balpha}E_{Y,X,\bZ} \{ \bUpsilon_{\balpha}(X,\Delta=1,\bZ) \pi_{X,\bZ}(X,\bZ; \boldeta_0) \bS_{\btheta}^F(Y,X,\bZ;\btheta_0)^T \} \\
&& = \bA^*_{\rm IPW; \balpha} E_{X,\bZ} [\bUpsilon_{\balpha}(X,\Delta=1,\bZ) \pi_{X,\bZ}(X,\bZ; \boldeta_0)  E_{Y|X,\bZ}\{\bS_{\btheta}^F(Y,X,\bZ;\btheta_0)^T \}].
\ese

The second line follows since $\bA^*_{\rm IPW; \balpha}$ is a fixed matrix, and the last line follows since $E_{Y|X,\bZ}\{\bS_{\btheta}^F(y,x,\bz;\btheta_0)^T \} = \bzero$. In the case of the consistent and asymptotically normal estimator of $\balpha$ discussed in Section \ref{sec:est_nuisance_distributions}, it follows that
\bse
&&E \{ \bUpsilon^*_{\balpha} (\bO^*) \bPhi_{\rm CC}(\bO; \btheta_0)^T \} \\
&&= - \bA^*_{\rm IPW; \balpha} E_{Y,W,\Delta, \bZ} \{ \bA^{-1}_{\balpha} \bS_{\balpha}(W,\Delta,\bZ;\balpha) \Delta \bS_{\btheta}^F (Y,W,\bZ; \btheta_0)^T \} \\
&& = - \bA^*_{\rm IPW; \balpha} \bA^{-1}_{\balpha} \ E_{X,\bZ} \biggr\{ \frac{\int_{x \leq C} \{ \frac{\partial}{\partial \balpha^T} \log f_{C,X|\bZ}(c,x,\bz; \balpha)\}  f_{C,X|\bZ}(c,x,\bz; \balpha) dc}{\int_{x\leq C} f_{C,X|\bZ}(c,x,\bz; \balpha) dc} \\
&& \quad \quad \times E_{Y| X,\bZ} \{ \pi_{X,\bZ}(X,\bZ) \bS_\btheta^F(Y,X,\bZ;\btheta_0) \} ^T \biggr\} \\
&& = -\bA^*_{\rm IPW; \balpha} \bA^{-1}_{\balpha} E_{X,\bZ} \left[ \bS_{\balpha}(X,\Delta=1,\bZ)  E_{Y| X,\bZ} \{ \pi_{X,\bZ}(X,\bZ) \bS_\btheta^F(Y,X,\bZ;\btheta_0) \} ^T \right] \\
&& = - \bA^*_{\rm IPW; \balpha} \bA^{-1}_{\balpha} \ E_{X,\bZ} \left[ \bS_{\balpha}(X,\Delta=1,\bZ)   \pi_{X,\bZ}(X,\bZ) \times \bzero^T \right] = \bzero.
\ese

\item \label{sec:cc_aipw_psi} Proof that the diagonal elements of $E [ \{ 1 - \Delta/\pi_{X,\bZ}(W,\bZ; \boldeta_0)\} \bPsi_{\rm AIPW, eff}(Y,\bZ; \btheta_0) \bPhi_{\rm CC}(\bO; \btheta_0)^T ]$ are non-negative.

It follows that
\bse
&& E_{Y,W,\Delta,\bZ} [ \{ 1 - \Delta/\pi_{X,\bZ}(W,\bZ; \boldeta_0)\} \bPsi_{\rm AIPW, eff}(Y,\bZ; \btheta_0) \bPhi_{\rm CC}(\bO; \btheta_0)^T ]\\
&&= E_{Y,W,\Delta, \bZ} [ \{ 1 - \delta/\pi_{X,\bZ}(w,\bz)\} \bPsi_{\rm AIPW, eff}(y,\bz; \btheta) \delta \bS_{\btheta}^F (y,w,\bz)^T ] \\
&& =  E_{Y,X, \bZ} [ \{ \pi_{X,\bZ}(X,\bZ; \boldeta_0)  -1 \} \bPsi_{\rm AIPW, eff}(Y,\bZ; \btheta_0) \bS_\btheta^F(Y,X,\bZ;\btheta_0)^T ].
\ese
The resulting expression is equal to
 \bse
&& E_{Y,X, \bZ} [ \{ \pi_{X,\bZ}(X,\bZ; \boldeta_0)  -1 \} \bPsi_{\rm AIPW, eff}(Y,\bZ; \btheta_0) \bS_\btheta^F(Y,X,\bZ;\btheta_0)^T ] \\
&& = E_{Y,\bZ} 
\biggr( \frac{1}{ E_{X|Y,\bZ}\{1/\pi_{X,\bZ}(X,\bZ; \boldeta_0)-1\} } \\
&& \quad \quad \quad \quad \times E_{X|Y,\bZ}[\{1/\pi_{X,\bZ}(X,\bZ; \boldeta_0) - 1 \} \bS^F_{\btheta} (Y,X,\bZ; \btheta_0)] \\
&& \quad \quad \quad \quad \times  E_{X|Y,\bZ}[\{1-\pi_{X,\bZ}(X,\bZ; \boldeta_0)\} \bS^F_{\btheta} (Y,X,\bZ; \btheta_0)^T]   \biggr) 
 \ese
Assuming that $0 < \pi_{X,\bZ}(x,\bz; \boldeta_0) < 1$, then $E_{X|Y,\bZ}\{1/\pi_{X,\bZ}(X,\bz; \boldeta_0)\} > 1$ because the expectation is with respect to a function whose range  is strictly greater than 1. Therefore, we conclude that the denominator will lead to a positive number. We prove that the diagonal elements of the numerator are non-negative by contradiction by assuming that the diagonal elements of the numerator are strictly less than zero. We achieve this result if the two expectation terms are opposite signs. Suppose that $E_{X|Y,\bZ}[\{1/\pi_{X,\bZ}(X,\bz; \boldeta_0) - 1 \} \bS^F_{\btheta} (y,X,\bz; \btheta_0)]$ is a positive and non-zero vector, and $E_{X|Y,\bZ}[\{1-\pi_{X,\bZ}(X,\bz; \boldeta_0)\} \bS^F_{\btheta} (y,X,\bz; \btheta_0)]$ is a negative and non-zero vector. This assumption implies that
\bse
\bzero &<& E_{X|Y,\bZ}[\{1/\pi_{X,\bZ}(X,\bz; \boldeta_0) - 1 \} \bS^F_{\btheta} (y,X,\bz; \btheta_0)] \\
&=& E_{X|Y,\bZ}\{\bS^F_{\btheta} (y,X,\bz; \btheta_0)/\pi_{X,\bZ}(X,\bz; \boldeta_0) \}  - E_{X|Y,\bZ}\{\bS^F_{\btheta} (y,X,\bz; \btheta_0) \}; \\
\bzero &>& E_{X|Y,\bZ}[\{1-\pi_{X,\bZ}(X,\bz; \boldeta_0)\} \bS^F_{\btheta} (y,X,\bz; \btheta_0)] \\
&=& E_{X|Y,\bZ}\{\bS^F_{\btheta} (y,X,\bz; \btheta_0) \}-E_{X|Y,\bZ}\{ \pi_{X,\bZ}(X,\bz; \boldeta_0) \bS^F_{\btheta} (y,X,\bz; \btheta_0)\}. 
\ese
These two results imply that for all $\btheta_0$,
\bse
\begin{array}{c}
E_{X|Y,\bZ}\{\bS^F_{\btheta} (y,X,\bz; \btheta_0)/\pi_{X,\bZ}(X,\bz; \boldeta_0) \}\\
E_{X|Y,\bZ}\{ \pi_{X,\bZ}(X,\bz; \boldeta_0) \bS^F_{\btheta} (y,X,\bz; \btheta_0)\}.
\end{array}
&>& E_{X|Y,\bZ}\{\bS^F_{\btheta} (y,X,\bz; \btheta_0) \}
\ese
This statement is a contradiction because only one statement can be correct. Either (i) $E_{X|Y,\bZ}\{\bS^F_{\btheta} (y,X,\bz; \btheta_0)/\pi_{X,\bZ}(X,\bz; \boldeta_0) \} > E_{X|Y,\bZ}\{\bS^F_{\btheta} (y,X,\bz; \btheta_0) \}$ \\ or (ii)
$E_{X|Y,\bZ}\{ \pi_{X,\bZ}(X,\bz; \boldeta_0) \bS^F_{\btheta} (y,X,\bz; \btheta_0)\} > E_{X|Y,\bZ}\{\bS^F_{\btheta} (y,X,\bz; \btheta_0) \}$, but not both. Therefore, we conclude that the product of $E_{X|Y,\bZ}[\{1/\pi_{X,\bZ}(X,\bz; \boldeta_0) - 1 \} \bS^F_{\btheta} (y,X,\bz; \boldeta_0)]$ and $E_{X|Y,\bZ}[\{1-\pi_{X,\bZ}(X,\bz; \boldeta_0)\} \bS^F_{\btheta} (y,X,\bz; \btheta_0)]$ is a non-negative vector. By proving that this product is in the non-negative domain, we prove that the diagonal elements of $E [ \{ 1 - \Delta/\pi_{X,\bZ}(W,\bZ; \boldeta_0)\} \bPsi_{\rm AIPW, eff}(Y,\bZ; \btheta_0) \bPhi_{\rm CC}(\bO; \btheta_0)^T ]$ are non-negative.

\vskip 1em

\item \label{sec:cc_ipw_positive} Proof that the diagonal elements of $\bA^{-1}_{\rm CC}$ and $\bA^{-1}_{\rm IPW}$ are of equivalent signs. 

In Corollary \ref{sec:a_cc=-b_cc}, we showed that $\bA_{\rm CC} = -\bB_{\rm CC}$, therefore, $\bA^{-1}_{\rm CC} = -\bB^{-1}_{\rm CC}$. It follows that the diagonal elements of $\bA^{-1}_{\rm CC}$ are negative since (i) the diagonal elements of $\bB_{\rm CC}$ are strictly positive due to the property of the sandwich variance estimator that $\diag \{ \bA^{-1}_{\rm CC} \bB_{\rm CC} \bA^{-T}_{\rm CC} \} > \bzero$ and (ii) inverting the $\bB_{\rm CC}$  matrix does not change the sign of the diagonal elements. 

Now, it remains to be shown that the diagonal elements of $\bA^{-1}_{\rm IPW}$ are non-negative. We will begin by showing the diagonal elements of  $\bA_{\rm IPW}$ are non-negative. In Corollary \ref{sec:-a_ipw},  we showed that $\bA_{\rm IPW}= -E \{ \bPhi_{\rm IPW}(\bO; \btheta_0; \boldeta_0) \bPhi_{\rm CC}(\bO; \btheta_0)^T \}$. It follows that

\bse
\bA_{\rm IPW} &=& -E \{ \bPhi_{\rm IPW}(\bO; \btheta_0, \boldeta_0) \bPhi_{\rm CC}(\bO; \btheta_0)^T \} \\
&=&  - E \left\{  \frac{\Delta \bS_{\btheta}^F (Y,W,\bZ; \btheta_0)}{\pi_{X,\bZ}(W,\bZ ;\boldeta_0)} \bS_{\btheta}^F (Y,W,\bZ; \btheta_0)^T \right\} \\
&=&  - E_{Y,X, \bZ} \left\{  \frac{ \pi_{X,\bZ}(X,\bZ ;\boldeta_0) \bS_{\btheta}^F (Y,X,\bZ; \btheta_0)}{\pi_{X,\bZ}(X,\bZ ;\boldeta_0)} \bS_{\btheta}^F (Y,X,\bZ; \btheta_0)^T \right\} \\
&=&  - E_{Y,X,\bZ} \left\{  \bS_{\btheta}^F (Y,X,\bZ; \btheta_0)^{\otimes 2} \right\}.
\ese

Since $E_{Y,X,\bZ} \left\{  \bS_{\btheta}^F (Y,X,\bZ; \btheta_0)^{\otimes 2} \right\}$ is the expectation of a matrix with non-negative diagonal values, then the resulting expectation  in non-negative and we conclude that the diagonal values of $\bA_{\rm IPW}$ are less than zero. Finally, we conclude that the diagonal values of $\bA^{-1}_{\rm IPW}$ are strictly negative since inverting the $\bA_{\rm IPW}$  matrix does not change the sign of the diagonal elements.

\end{enumerate}

\subsection{Efficiency comparison of the CC and IPW estimators}
\label{sec:cc-ipw-efficiency}

We show that the CC estimator is as or more efficient than the IPW estimator whether the probabilities are known or estimated. 

\begin{enumerate}

\item {\bf Known probabilities:} \\
To prove that the CC estimator is as or more efficient than the IPW estimator it is sufficient to show that $\diag\{ {\rm Var}({\btheta}_{\rm IPW}) \} \geq \diag \{ {\rm Var}({\btheta}_{\rm CC}) \}$. The variances of the CC and IPW estimators are equal to
\bse
{\rm Var}({\btheta}_{\rm CC}) &=& {\rm Var}\{\bUpsilon_{\rm CC}(\bO)\},  {\rm\ where}\ \bUpsilon_{\rm CC}(\bO) = -\bA_{\rm CC} \bPhi_{\rm CC}(\bO;  \btheta_0) \\ 
&=& \bA_{\rm CC}^{-1} \bB_{\rm CC} \bA_{\rm CC}^{-T},  {\rm\ where}\ \bB_{\rm CC} = E\{ \bPhi_{\rm CC}(\bO;  \btheta_0) ^{\otimes 2}\}; \\ 
{\rm Var}({\btheta}_{\rm IPW}) &=& {\rm Var}\{ \bUpsilon_{\rm IPW}(\bO)\},  {\rm\ where}\ \bUpsilon_{\rm IPW}(\bO) = -\bA_{\rm IPW} \bPhi_{\rm IPW}(\bO;  \btheta_0, \boldeta_0) \\
&=& \bA_{\rm IPW}^{-1} \bB_{\rm IPW} \bA_{\rm IPW}^{-T}; {\rm\ where}\ \bB_{\rm IPW} = E\{ \bPhi_{\rm IPW}(\bO;  \btheta_0, \boldeta_0) ^{\otimes 2}\}.
\ese
Using the property that the expectation of a non-negative function is non-negative, it follows that 
$ \diag\left( E [\{\bUpsilon_{\rm CC}(\bO) - \bUpsilon_{\rm IPW}(\bO)\}^{\otimes 2}] \right) \geq \bzero$. This inequality indicates that each individual cell component of the matrix is equal to or greater than zero. We have that
\bse
\diag \left( E [\{\bUpsilon_{\rm CC}(\bO) - \bUpsilon_{\rm IPW}(\bO)\}^{\otimes 2}] \right)  &=&  \diag \biggr[  \bA_{\rm CC}^{-1} \bB_{\rm CC} \bA_{\rm CC}^{-T} + \bA_{\rm IPW}^{-1} \bB_{\rm IPW} \bA_{\rm IPW}^{-T}  \\
&& - \bA_{\rm CC}^{-1} E \{ \bPhi_{\rm CC}(\bO; \btheta_0) \bPhi_{\rm IPW}(\bO; \btheta, \boldeta_0)^T \} \bA_{\rm IPW}^{-T} \\
&& - \bA_{\rm IPW}^{-1}E \{ \bPhi_{\rm IPW}(\bO; \btheta, \boldeta_0) \bPhi_{\rm CC}(\bO; \btheta_0)^T \} \bA_{\rm CC}^{-T} \biggr].
\ese
In Corollaries \ref{sec:a_cc=-b_cc} and \ref{sec:-a_ipw}, respectively, we show that $\bA_{\rm CC} = -\bB_{\rm CC}$ and  \\ $E \{ \bPhi_{\rm IPW}(\bO; \btheta, \boldeta_0) \bPhi_{\rm CC}(\bO; \btheta_0)^T \} = -\bA_{\rm IPW}$. From these results, it follows that
\bse
\diag \left[ E\{ (\bUpsilon_{\rm CC} - \bUpsilon_{\rm IPW})^{\otimes 2} \} \right] &=& \diag \left( -\bA^{-T}_{\rm CC} +\bA_{\rm IPW}^{-1} \bB_{\rm IPW} \bA_{\rm IPW}^{-T} + \bA_{\rm CC}^{-1} + \bA^{-T}_{\rm CC} \right) \\
&=& \diag \left( \bA_{\rm IPW}^{-1} \bB_{\rm IPW} \bA_{\rm IPW}^{-T} + \bA_{\rm CC}^{-1} \right) \\
&=& \diag \left(\bA_{\rm IPW}^{-1} \bB_{\rm IPW} \bA_{\rm IPW}^{-T} - \bA_{\rm CC}^{-1} \bB_{\rm CC} \bA_{\rm CC}^{-T} \right),
\ese
where the second-to-last line follows since $\bA_{\rm CC} = -\bB_{\rm CC}$ and $\bA_{\rm CC}^{-1} = - \bA_{\rm CC}^{-1} \bB_{\rm CC} \bA_{\rm CC}^{-T}$. Lastly,
since the expression is greater than or equal to  $\0$, we have that 
\bse
\diag\left( \bA_{\rm IPW}^{-1} \bB_{\rm IPW} \bA_{\rm IPW}^{-T} \right) &\geq& \diag \left( \bA_{\rm CC}^{-1} \bB_{\rm CC} \bA_{\rm CC}^{-T} \right), 
\ese
which means $\diag\left\{ {\rm Var}({\btheta}_{\rm IPW}) \right\} \geq \diag\left\{ {\rm Var}({\btheta}_{\rm CC}) \right\}$. We conclude that the CC estimator is  as or more efficient than the IPW estimator when $f_{C|X,\bZ}$ is known. By interchanging the roles of $f_{C|X,\bZ}$ with $f_{C|\bZ}$, the same result follows under independent  covariate right-censoring. This result continues to hold for independent and dependent covariate missingness.  

\item {\bf Unknown probabilities:} \\
When the probabilities are unknown, we need to estimate the set of parameters $\boldeta$ that index the $f_{C|X,\bZ}$ distribution. In this case, we consider the influence function
\bse
&& \bUpsilon_{\rm IPW;\balpha} (\bO) \\
&& = -\bA_{\rm IPW}^{-1} \{(\bPhi_{\rm IPW}(\bO;\btheta_0,\balpha_0) \\
&& \quad - E\{\partial\bPhi_{\rm IPW}(\bO;\btheta_0,\balpha_0)/\partial\balpha^T\}[E\{\partial \bPhi_{\balpha}(\bO^*; \balpha_0)/\partial\balpha^T\}]^{-1}\bPhi_{\balpha}(\bO^*; \balpha_0)) \}\\
&& = -\bA_{\rm IPW}^{-1} \{ \bPhi_{\rm IPW}(\bO;\btheta_0,\balpha_0) + \bA^*_{\rm IPW; \balpha} \bUpsilon_{\balpha}(\bO^*) \},
\ese
where $\bA^*_{\rm IPW; \balpha} = E \{ \partial \bPhi_{\rm IPW}(\bO_i; \btheta_0, \balpha_0) / \partial \balpha^T \}$ is a fixed matrix. Without loss of generality,  let $\bUpsilon^*_{\balpha}(\bO^*) = \bA^*_{\rm IPW; \balpha} \bUpsilon_{\balpha}(\bO^*) $. It follows that
\bse
{\rm Var}({\btheta}_{\rm IPW;\balpha}) &=& {\rm Var} (\bUpsilon_{\rm IPW;\balpha} )  = \bA_{\rm IPW}^{-1} \bB_{\rm IPW; \balpha} \bA_{\rm IPW}^{-T}, 
\ese
where $\bB_{\rm IPW; \balpha}  = E [ \{\bPhi_{\rm IPW}(\bO; \btheta_0, \balpha_0) + \bUpsilon^*_{\balpha} (\bO^*) \} ^{\otimes 2} ]$. Applying a similar argument as that when $f_{C|X,\bZ}$ was known,
\bse
&& \diag \left( E[ \{\bUpsilon_{\rm CC}(\bO) - \bUpsilon_{\rm IPW;\balpha} (\bO)\}^{\otimes 2} \} ] \right) \\
&& = \diag \biggr[ \bA_{\rm CC}^{-1} \bB_{\rm CC} \bA_{\rm CC}^{-T} + \bA_{\rm IPW}^{-1} \bB_{\rm IPW} \bA_{\rm IPW}^{-T}   \\
&& - \bA_{\rm CC}^{-1} E \{ \bPhi_{\rm CC}(\bO; \btheta_0) \bPhi_{\rm IPW}(\bO; \btheta_0, \balpha_0)^T \} \bA_{\rm IPW}^{-T}  - \bA_{\rm IPW}^{-1}E \{ \bPhi_{\rm IPW}(\bO; \btheta_0, \balpha_0) \bPhi_{\rm CC}(\bO; \btheta_0)^T \} \bA_{\rm CC}^{-T} \\
&& - \bA_{\rm CC}^{-1} E \{ \bPhi_{\rm CC}(\bO; \btheta_0)  \bUpsilon^*_{\balpha} (\bO^*)^T \} \bA_{\rm IPW}^{-T}  - \bA_{\rm IPW}^{-1} E \{ \bUpsilon^*_{\balpha} (\bO^*) \bPhi_{\rm CC}(\bO; \btheta_0)^T \} \bA_{\rm CC}^{-T} \biggr].
\ese
As shown in Corollary \ref{sec:cc_upsilon_alpha}, $ E \{ \bUpsilon^*_{\balpha} (\bO^*) \bPhi_{\rm CC}(\bO; \btheta_0)^T \} = \bzero$, so the expression may be reduced  to
\bse
&& \diag \left( E[ \{\bUpsilon_{\rm CC}(\bO) - \bUpsilon_{\rm IPW;\balpha} (\bO)\}^{\otimes 2} \} ] \right) \\
&& = \diag \biggr[  \bA_{\rm CC}^{-1} \bB_{\rm CC} \bA_{\rm CC}^{-T} + \bA_{\rm IPW}^{-1} \bB_{\rm IPW} \bA_{\rm IPW}^{-T}  - \bA_{\rm CC}^{-1} E \{ \bPhi_{\rm CC}(\bO; \btheta_0) \bPhi_{\rm IPW}(\bO; \btheta_0, \balpha_0)^T \} \bA_{\rm IPW}^{-T} \\
&& \quad - \bA_{\rm IPW}^{-1}E \{ \bPhi_{\rm IPW}(\bO; \btheta_0, \balpha_0) \bPhi_{\rm CC}(\bO; \btheta_0)^T \} \bA_{\rm CC}^{-T} \biggr].
\ese
The remainder of the proof remains the same as that used when the probabilities were known. We conclude that the CC estimator is as or more efficient than the IPW estimator when the probabilities are unknown. By interchanging the roles of $f_{C|X,\bZ}$  with $f_{C|\bZ}$, as well as their corresponding influence functions, the same result follows under independent covariate right-censoring. A similar argument shows that this result holds for independent and dependent covariate missingness.

\end{enumerate}

\subsection{Efficiency comparison of the CC and AIPW estimators}
\label{sec:cc-aipw-efficiency}

Here we show that the CC estimator is as or more efficient than the AIPW estimator, regardless of whether the probabilities are known or estimated. We will compare the CC estimator to the AIPW estimator with the augmentation term $\bPsi_{\rm AIPW, eff}(y,\bz; \btheta_0)$. That is, 
\bse
\bPsi_{\rm AIPW, eff}(y,\bz; \btheta_0) &=& 
 -\frac{E_{X|Y,\bZ}[\{1-1/\pi_{X,\bZ}(X,\bz;\boldeta_0)\} \bS^F_{\btheta} (y,X,\bz; \btheta_0)]}{ E_{X|Y,\bZ}[\{1/\pi_{X,\bZ}(X,\bz;\boldeta_0)-1\}]}. 
\ese

\begin{enumerate}

\item {\bf Known probabilities:} \\
To prove that the CC estimator is as or more efficient than the AIPW estimator it is sufficient to show that $\diag \left\{ {\rm Var}({\btheta}_{\rm AIPW}) \right\} \geq  \left\{ {\rm Var}({\btheta}_{\rm CC}) \right\}$. The variance of the CC and AIPW estimators is equal to
\bse
{\rm Var}({\btheta}_{\rm CC}) &=& {\rm Var}\{\bUpsilon_{\rm CC}(\bO)\},  {\rm\ where}\ \bUpsilon_{\rm CC}(\bO) = -\bA_{\rm CC} \bPhi_{\rm CC}(\bO;  \btheta_0) \\ 
&=& \bA_{\rm CC}^{-1} \bB_{\rm CC} \bA_{\rm CC}^{-T},  {\rm\ where}\ \bB_{\rm CC} = E\{ \bPhi_{\rm CC}(\bO;  \btheta_0) ^{\otimes 2}\}; \\ 
{\rm Var}({\btheta}_{\rm AIPW}) &=& {\rm Var}\{ \bUpsilon_{\rm AIPW}(\bO)\},  {\rm\ where}\ \bUpsilon_{\rm AIPW}(\bO) = -\bA_{\rm AIPW} \bPhi_{\rm AIPW}(\bO;  \btheta_0, \boldeta_0) \\
&=& \bA_{\rm AIPW}^{-1} \bB_{\rm AIPW} \bA_{\rm AIPW}^{-T}; {\rm\ where}\ \bB_{\rm AIPW} = E\{ \bPhi_{\rm AIPW}(\bO;  \btheta_0, \boldeta_0) ^{\otimes 2}\}.
\ese
From the proofs of consistency and asymptotic normality, we know that $\bA_{\rm AIPW} = \bA_{\rm IPW}$. Similar to the proof for the CC and IPW estimators, we use the fact that $ \diag\left( E [\{\bUpsilon_{\rm CC}(\bO) - \bUpsilon_{\rm AIPW}(\bO)\}^{\otimes 2}] \right) \geq \bzero$. It follows that
\bse
 && \diag\left( E[\{\bUpsilon_{\rm CC}(\bO) - \bUpsilon_{\rm AIPW}(\bO)\}^{\otimes 2}] \right) \\
&& = \diag\bigr[  \bA_{\rm CC}^{-1} \bB_{\rm CC} \bA_{\rm CC}^{-T} + \bA_{\rm AIPW}^{-1} \bB_{\rm AIPW} \bA_{\rm AIPW}^{-T}  - \bA_{\rm CC}^{-1} E \{ \bPhi_{\rm CC}(\bO; \btheta_0) \bPhi_{\rm AIPW}(\bO; \btheta_0, \boldeta_0)^T \} \bA_{\rm AIPW}^{-T} \\
&& \quad - \bA_{\rm AIPW}^{-1}E \{ \bPhi_{\rm AIPW}(\bO; \btheta_0, \boldeta_0) \bPhi_{\rm CC}(\bO; \btheta_0)^T \} \bA_{\rm CC}^{-T} \bigr] \\
&& =  \diag\bigr( -{\rm Var}({\btheta}_{\rm CC}) + {\rm Var}({\btheta}_{\rm AIPW})  \\
&& \quad - \bA_{\rm CC}^{-1} E[ \bPhi_{\rm CC}(\bO; \btheta_0)\{ 1 - \Delta/\pi_{X,\bZ}(W,\bZ; \boldeta_0)\} \bPsi_{\rm AIPW}(Y,\bZ; \btheta_0)^T ] \bA_{\rm IPW}^{-T} \\
&& \quad - \bA_{\rm IPW}^{-1}E [ \{ 1 - \Delta/\pi_{X,\bZ}(W,\bZ; \boldeta_0)\} \bPsi_{\rm AIPW}(Y,\bZ; \btheta_0) \bPhi_{\rm CC}(\bO; \btheta_0)^T ] \bA_{\rm CC}^{-T} \bigr).
\ese
In \ref{sec:cc_aipw_psi}, it is shown that $ \diag\left(E [ \{ 1 - \Delta/\pi_{X,\bZ}(W,\bZ; \boldeta_0)\} \bPsi_{\rm AIPW}(Y,\bZ; \btheta_0) \bPhi_{\rm CC}(\bO; \btheta_0)^T ] \right) \geq \bzero$, and in \ref{sec:cc_ipw_positive}, it is shown that $\diag(\bA_{\rm CC}^{-1})$ and $\diag(\bA_{\rm IPW}^{-1})$ are of equivalent sign. Therefore, it follows that
\bse
&& \diag\left\{ {\rm Var}({\btheta}_{\rm AIPW}) \right\} \\
&& \geq \diag \bigr( {\rm Var}({\btheta}_{\rm CC}) + \bA_{\rm CC}^{-1} E[ \bPhi_{\rm CC}(\bO, \btheta_0)\{ 1 - \Delta/\pi_{X,\bZ}(W,\bZ; \boldeta_0)\} \bPsi_{\rm AIPW}(Y,\bZ; \btheta_0)^T ] \bA_{\rm IPW}^{-T} \\
&& \quad + \bA_{\rm IPW}^{-1}E [ \{ 1 - \Delta/\pi_{X,\bZ}(W,\bZ; \boldeta_0)\} \bPsi_{\rm AIPW}(Y,\bZ; \btheta_0) \bPhi_{\rm CC}(\bO, \btheta_0)^T ] \bA_{\rm CC}^{-T}  \bigr) \\
&&\geq \diag \left\{ {\rm Var}({\btheta}_{\rm CC}) \right\}.
\ese
We conclude that the CC estimator is as or more efficient than the AIPW estimator when $f_{C|X,\bZ}$ is known. By interchanging the roles of $f_{C|X,Z}$ with $f_{C|Z}$, the same result follows under independent covariate right-censoring. A similar argument can be applied as that for the missing covariate problem. 

\item {\bf Unknown weights:} In our derivations for the AIPW estimator, we did not provide the form of $\bPsi_{\rm AIPW, eff}(y,\bz; \btheta_0)$ for unknown weights. Therefore, the proof holds only for known weights.  
    
\end{enumerate}

\section{More details of the simulation study}

\subsection{Data generation}
\label{sec:sims_data_paper_additional}

In all $N=3,000$ simulations, we generated data with a sample size of $n = 1,000$ and a censoring rate of $50\%$. Simulations were repeated using a smaller sample size of $n=100$ to validate results. Data were generated from the regression model in Equation \eqref{eqn:equation_paper2}, where  $m(X,Z;\btheta) = \beta_0 + \beta_{AX}(A-X) + \beta_ZZ$,  $\epsilon\sim\Normal(0,1)$, and $A\sim\Normal(0,1)$.  The true parameter values were $\beta_0=1$, $\beta_{AX}=3$, and $\beta_Z=2$. We used $(A-X)$ as opposed to $X$
alone to mimic the case when time to diagnosis is of interest; here, $A$ is the current age and $X$ is the age of diagnosis, so $A-X$ is the time to diagnosis. We generated data $(X,C,Z)$ from a normal distribution as follows:
\bse
\label{eqn:mvn}
(X,C,Z)^{\trans} \sim \Normal \left\{ \bmu = \begin{bmatrix}
0 \\
0 \\
0
\end{bmatrix}, \bSigma = \begin{bmatrix}
\sigma_X^2 = 1 & - & - \\
\sigma_{XC} & \sigma_C^2 = 4 & - \\
\sigma_{XZ} = 0.50 & \sigma_{CZ} = 0.50 & \sigma_Z^2 = 1
\end{bmatrix}
\right\}.
\ese
To guarantee independent covariate right-censoring, we set $\sigma_{XC} = 0.25$, so that the covariance of $(X,C)$ conditional on $Z$ equals 0. To guarantee dependent covariate right-censoring, we set $\sigma_{XC} = 0.60$, so that the covariance of $(X,C)$ conditional on $Z$ does not equal 0. We generated $N=1,000$ datasets following the above procedure and then estimated $(\beta_0,\beta_{AX},\beta_Z)$. Next, we explain what the nuisance parameters of the distribution $f_{C,X|Z}$ are based on our trivariate normal distribution, and how these are used to calculate the probabilities $\pi_{Y,\bZ}(y,z)$ and $\pi_{X,\bZ}(x,\bz)$ under independent and dependent covariate right-censoring.

\subsubsection{Independent covariate right-censoring}
The distribution of $C$  conditional on $Z$ is given by
\bse
\mu_{C|Z} &=& \mu_C + (\sigma_{CZ}/\sigma_Z^2)*(z_i-\mu_Z) \\
&=& \alpha_{10} + \alpha_{11} z_i; \\
\sigma_{C|Z}^2 &=& \sigma_C^2-\sigma_{CZ}^2/\sigma_Z^2.
\ese
Similarly, the distribution of $X$  conditional on $Z$ is given by
\bse
\mu_{X|Z} &=& \mu_X + (\sigma_{XZ}/\sigma_Z^2)*(z_i-\mu_Z) \\
&=& \alpha_{20} + \alpha_{21} z_i; \\
\sigma_{X|Z}^2 &=& \sigma_X^2-\sigma_{XZ}^2/\sigma_Z^2.
\ese
Under our simulation settings, $\balpha_1 = (\alpha_{10}, \alpha_{11}, \sigma_{C|Z}^2)^T = (0, 0.5, 3.75)^T$ and $\balpha_2 = (\alpha_{20}, \alpha_{21}, \sigma_{X|Z}^2)^T = (0, 0.5, 0.75)^T$. The probability of  observing $X$ is defined by
\bse
\pi_{X,Z} (x_i,z_i) &\equiv& \pr(x_i \leq C| Z = z_i) =  \int_{x_i<c} f_{C|Z}(c,z) dc \\
\pi_{Y,Z} (y_i,z_i) &\equiv& \int \pi_{X,Z} (x,z_i) f_{X|Y,Z}(x,y,z) dx \\
&=& \frac{\int \pi_{X,Z} (x,z_i) f_{Y|X,Z}(y_i,x,z_i;\btheta) f_{X|Z}(x,z_i) dx}{\int  f_{Y|X,Z}(y_i,x,z_i;\btheta) f_{X|Z}(x,z_i) dx}.
\ese
The probability $\pi_{X,Z} (x_i,z_i)$ corresponds to the upper-tail of the cumulative distribution function (CDF) of a conditional normal distribution governed by parameters $\mu_{C|Z}$ and $\sigma_{C|Z}^2$, and evaluated at $z_i$. The probability $\pi_{Y,Z} (y_i,z_i)$ is calculated by computing the ratio of integrals. Since all variables are normally distributed, this calculation narrows down to a straightforward numeral integration.  

\subsubsection{Dependent {covariate} right-censoring}

Under dependent {covariate right-censoring}, the distribution of $C$ conditional on $X$ and $Z$ is 
also governed by a mean and variance. The mean is equal to
\bse
\mu_{C|X,Z} &=& \mu_C 
            + \frac{(\sigma_{XC}\sigma_Z^2-\sigma_{CZ}\sigma_{XZ})(x_i - \mu_X) +
            (\sigma_{CZ}\sigma_X^2-\sigma_{XC}*\sigma_{XZ})(z_i - \mu_Z)}{\sigma_X^2\sigma_Z^2 - \sigma_{XZ}^2} \\
            &=& \alpha_{30} + \alpha_{31} x_i + \alpha_{32} z_i
\ese
such that 
\bse
\alpha_{30} &=& \mu_C - \frac{(\sigma_{XC}\sigma_Z^2-\sigma_{CZ}\sigma_{XZ})\mu_X}{\sigma_X^2\sigma_Z^2 - \sigma_{XZ}^2} - \frac{
            (\sigma_{CZ}\sigma_X^2-\sigma_{XC}*\sigma_{XZ})\mu_Z}{\sigma_X^2\sigma_Z^2 - \sigma_{XZ}^2}; \\
\alpha_{31} &=& \frac{(\sigma_{XC}\sigma_Z^2-\sigma_{CZ}\sigma_{XZ})}{\sigma_X^2\sigma_Z^2 - \sigma_{XZ}^2} ; \alpha_{32} = \frac{
            (\sigma_{CZ}\sigma_X^2-\sigma_{XC}\sigma_{XZ})}{\sigma_X^2\sigma_Z^2 - \sigma_{XZ}^2}.
\ese
The variance is equal to
\bse
\sigma_{C|X,Z}^2 = \sigma_C^2 - \frac
                              {(\sigma_{XC}\sigma_Z^2-\sigma_{CZ}\sigma_{XZ})\sigma_{XC} +
                                 (\sigma_{CZ}\sigma_X^2-\sigma_{XC}\sigma_{XZ})\sigma_{CZ}}{\sigma_X^2\sigma_Z^2 - \sigma_{XZ}^2}.
\ese
Under our simulation settings, the parameters are equal to $\balpha_1 = (\alpha_{30}, \alpha_{31}, \alpha_{32}, \sigma_{C|X,Z}^2) = (0, 0.35/0.75, 0.2/0.75,  299/75)$. The distribution of $X$  conditional on $Z$ is the same as for independent covariate censoring (i.e., the same $\balpha_2$). The probability of observing $X$ is now defined as
\bse
\pi_{X,Z} (x_i,z_i) &\equiv& \pr(x_i \leq C| X = x_i, Z = z_i) =  \int_{x<c} f_{C|X,Z}(c,x,\bz) dc \\
\pi_{Y,Z} (y_i,z_i) &\equiv& \int \pi_{X,Z} (x,z_i) f_{X|Y,Z}(x,y,z) dx; \\
&=& \frac{\int \pi_{X,Z} (x,z_i) f_{Y|X,Z}(y_i,x,z_i;\btheta) f_{X|Z}(x,z_i) dx}{\int  f_{Y|X,Z}(y_i,x,z_i;\btheta) f_{X|Z}(x,z_i) dx}.
\ese
These probabilities are then computed similarly as for independent covariate right-censoring.

We have outlined the form of the nuisance parameters governing $f_{X,C|\bZ}$. Now, we describe how these nuisance parameters were specified in our simulation study. We employed two approaches: a mechanistic approach, where the nuisance parameters were fully specified using the true values, and an empirical approach, where the nuisance parameters were estimated from the data.
 
\subsection{Nuisance parameters are specified}

When  the nuisance parameters were specified, the estimators were implemented as follows: 

\vspace{-0.5em}
\begin{enumerate}

    \item \textbf{IPW estimator.} Two variations for $\pi_{X,\bZ}(w,\bz)$ were considered: (i) correctly specified and generated from the multivariate normal distribution; and (ii) incorrectly specified and generated from uniform(0.1, 0.9). We also considered the correct specification of $\pi_{Y,\bZ}$ to show that this choice will lead to an inconsistent estimator.

     \item \textbf{MLE.} Two variations for $f_{X|\bZ}$ ($f_{X|C,\bZ}$ under dependent censoring) were considered: (i)  correctly specified from the trivariate normal distribution; and (ii)  incorrectly specified as a normal distribution with mean $-2$ and variance 1.

    \item \textbf{ACC, MACC, and AIPW estimators.} Four cases were considered: (i) $\pi_{Y,\bZ}(y,\bz)$ or $\pi_{X,\bZ}(w,\bz)$  and $\bPsi(y,\bz;\btheta)$ correctly specified; (ii) only $\bPsi(y,\bz;\btheta)$ correctly specified; (iii) only $\pi_{Y,\bZ}(y,\bz)$ or $\pi_{X,\bZ}(w,\bz)$ correctly specified; and (iv) $\pi_{Y,\bZ}(y,\bz)$ or $\pi_{X,\bZ}(w,\bz)$  and $\bPsi(y,\bz;\btheta)$ incorrectly specified. Incorrectly specified $\pi_{Y,\bZ}(y,\bz)$ or $\pi_{X,\bZ}(w,\bz)$  was generated from uniform(0.1, 0.9), and the incorrect $\bPsi(y,\bz;\btheta)$ was chosen to be $\bPsi_{\rm close}(y,\bz; \btheta)$. Correctly specified $\pi_{Y,\bZ}(y,\bz)$ or $\pi_{X,\bZ}(w,\bz)$ was calculated from the trivariate normal distribution, and correct $\bPsi(y,\bz;\btheta)$ was calculated as in Tables \ref{tab:estimators} and \ref{tab:estimators-dependent}. 
    
    \item \textbf{ACC, MACC, and AIPW with $\bLambda$ estimators.} The augmented component was chosen to be $\bPsi_{\rm updated}(y,\bz; \btheta) = \bLambda \bPsi_{\rm close}(y,\bz; \btheta)$. Two variations for the $\pi_{Y,\bZ}(y,\bz)$ or $\pi_{X,\bZ}(w,\bz)$ were considered: (i) correctly specified and generated from the trivariate normal distribution; and (ii) incorrectly specified and generated from uniform(0.1, 0.9).
   
\end{enumerate}

We also implemented the oracle, naive, and complete case estimators. The oracle estimator estimates $\btheta$ using the full but unobserved data $(Y,X,\bZ)$. Constructing the oracle estimator is not possible in practice since $X$ is right-censored, but it gives us a gold standard to assess performance. In contrast, the naive estimator, which uses all observations and naively treats $W$ as $X$ in the estimation of $\btheta$, was used to showcase the worst-case scenario. Finally, the complete case estimator was used to showcase the performance of an estimator that only includes complete observations.

\subsection{Nuisance distributions are estimated}

When the nuisance parameters were estimated, we used a conditional bivariate normal distribution and the parameters governing this distribution $\balpha^T = (\mu_{X|Z}, \mu_{C|Z}, \sigma_{X|Z}, \sigma_{C|Z})$ were estimated. All estimators were considered for the independent {covariate right-censoring} case and only correctly specified scenarios were evaluated: the weights $\pi_{X,\bZ}(w,\bz)$ were calculated using $f_{C|\bZ}(c,\bz; \wh\balpha)$, and $f_{X|\bZ}(x,\bz; \wh\balpha)$ was used for the augmented estimators and the MLE. For dependent censoring, $f_{X,C|\bZ}$ is not always identifiable; therefore only estimators that do not require  this distribution were evaluated: oracle, naive, complete case, and ACC. To showcase that the ACC estimator in the dependent missing problem can be used for the right-censored covariate problem, we modified the simulation study from  \textcite{Bartlettetal2014} by letting $C = X - beta\{\exp(Z), 1\}$ when $X$ was not observed. 
 
We computed all simulations using the \textit{geex} package in R \parencite{saul2020calculus}. Then, we evaluated all estimators by computing the mean estimate of $\btheta$ (i.e., $N^{-1}\sum_{i=1}^N\wh{\btheta}_i$)  and its percent bias (i.e.,  $N^{-1}\sum_{i=1}^N (\wh{\btheta}_i-\btheta_0)/\btheta_0$); the empirical standard deviation of $\wh\btheta$ across all simulations; and the empirical mean of the estimated standard errors (i.e.,  $N^{-1}\sum_{i=1}^{N} \wh{\rm SE}_i$).  Estimated standard errors were computed using the asymptotic variances derived in our theorems, with all expectations replaced by empirical averages.  Lastly, we calculated the empirical coverage of the estimated 95\% confidence intervals. All R code used for the simulation studies and can be found at \url{https://github.com/jesusepfvazquez/right-censored-covariates}.

\subsection{Simulation results}

\label{sec:additional_simulations}

\subsubsection{Independent covariate right-censoring, mechanistic approach}

\begin{landscape}
\begin{table}[hb]
\caption{\label{tab:independent_known_1000} \textbf{Independent {covariate right-}censoring with known $f_{C,X|\bZ}(c,x,\bz; \balpha)$.} Simulation results of the mean estimate, mean estimated standard errors (SE) and empirical standard deviations (SD) scaled by 100, and confidence interval coverage  (95\% Cov) for each estimator. Results are based on 1,000 simulated datasets, each with a sample size of 1,000 and a censoring rate of 50\%.}
\centering
\resizebox{1\linewidth}{!}{
\begin{tabular}[t]{llrrrrrrrrrrrrrrrrrrrr}
\toprule
\textbf{Estimator} & \textbf{Specification} & \textbf{Estimate} & \textbf{Bias} & \textbf{SE} & \textbf{SD} & \textbf{95\% Cov} & \textbf{Estimate} & \textbf{Bias} & \textbf{SE} & \textbf{SD} & \textbf{95\% Cov} & \textbf{Estimate} & \textbf{Bias} & \textbf{SE} & \textbf{SD} & \textbf{95\% Cov} \\
\midrule
\addlinespace[0.3em]
&&\multicolumn{5}{c}{\textbf{Intercept: $\beta_0 = 1$}} &\multicolumn{5}{c}{\textbf{$A-X$: $\beta_{AX} = 3$}} &\multicolumn{5}{c}{\textbf{$Z$: $\beta_{Z} = 2$}}\\
\hspace{1em}Oracle & &  1.00 & 0.00 & 3.16 & 3.19 & 94.10 & 3.00 & 0.00 & 2.38 & 2.37 & 94.35 & 2.00 & 0.00 & 3.37 & 3.31 & 95.37 \\
\addlinespace
\hspace{1em}Naive &  & -0.38 & -1.38 & 9.10 & 9.07 & 0.00  & 1.63 & -1.37 & 7.77 & 7.81 & 0.00 & 1.31 & -0.69 & 10.73 & 10.57 & 0.00 \\
\addlinespace
\hspace{1em}CC & &    1.00 & -0.00 & 4.57 & 4.57 & 94.96 & 3.00 & 0.00 & 3.45 & 3.38 & 95.27 & 2.00 & 0.00 & 4.78 & 4.77 & 95.07  \\
\addlinespace
 & correct $\pi_{X,\bZ}(w,\bz)$ &  1.00 & -0.00 & 4.77 & 4.83 & 94.61 & 3.00 & 0.00 & 3.77 & 3.81 & 95.12 & 2.00 & 0.00 & 5.11 & 5.19 & 94.86\\
\multirow[t]{-2}{*}{\raggedright\arraybackslash \hspace{1em}IPW} 
& incorrect $\pi_{X,\bZ}(w,\bz)$ & 1.00 & -0.00 & 5.50 & 5.54 & 95.07 & 3.00 & 0.00 & 4.13 & 4.11 & 94.86 & 2.00 & 0.00 & 5.74 & 5.72 & 95.07\\
& correct $\pi_{Y,\bZ}(y,\bz)$ & 0.88 & -0.12 & 4.87 & 4.94 & 29.91 & 3.02 & 0.02 & 3.80 & 3.85 & 92.62  & 2.01 & 0.01 & 5.17 & 5.24 & 94.30 \\
\addlinespace
 & correct $f_{X|\bZ}$ & 1.00 & -0.00 & 4.42 & 4.14 & 94.81 & 3.00 & 0.00 & 3.48 & 3.17 & 94.76 & 2.00 & 0.00 & 4.67 & 4.36 & 95.22  \\
\multirow[t]{-2}{*}{\raggedright\arraybackslash \hspace{1em}MLE} 
& incorrect $f_{X|\bZ}$  &  0.06 & -0.94 & 8.94 & 8.97 & 0.10 & 2.80 & -0.20 & 7.39 & 7.39 & 18.57 & 1.71 & -0.29 & 8.40 & 8.29 & 5.49  \\
\addlinespace
&\multicolumn{5}{l}{\underline{without using $\bLambda$}}\\
\multirow[t]{-2}{*}{\raggedright\arraybackslash \hspace{1em}ACC} 
 & correct $\pi_{X,\bZ}(w,\bz)$ and correct $\bPsi_{\rm ACC}(y,\bz)$&  1.00 & -0.00 & 4.40 & 4.40 & 94.51 & 3.00 & 0.00 & 3.31 & 3.24 & 95.57 & 2.00 & 0.00 & 4.62 & 4.62 & 94.71 \\
& incorrect $\pi_{X,\bZ}(w,\bz)$ and correct $\bPsi_{\rm ACC}(y,\bz)$ & 0.84 & -0.16 & 5.10 & 5.11 & 10.27 & 3.00 & -0.00 & 4.24 & 4.24 & 95.22 & 2.00 & 0.00 & 5.52 & 5.69 & 94.00  \\
 & correct $\pi_{X,\bZ}(w,\bz)$ and incorrect $\bPsi_{\rm ACC}(y,\bz)$  & 1.00 & -0.00 & 5.18 & 5.24 & 94.66 & 3.00 & 0.00 & 4.19 & 4.19 & 95.32 & 2.00 & 0.00 & 5.25 & 5.30 & 94.91  \\
& incorrect $\pi_{X,\bZ}(w,\bz)$ and incorrect $\bPsi_{\rm ACC}(y,\bz)$  &   0.67 & -0.33 & 6.08 & 6.24 & 0.00 & 3.02 & 0.02 & 5.20 & 5.34 & 91.40 & 2.01 & 0.01 & 6.63 & 6.84 & 93.74  \\
&\multicolumn{5}{l}{\underline{using $\bLambda$}}\\
 & correct $\pi_{X,\bZ}(w,\bz)$ &  1.00 & -0.00 & 4.40 & 4.44 & 94.76 & 3.00 & 0.00 & 3.30 & 3.25 & 95.78 & 2.00 & 0.00 & 4.62 & 4.64 & 94.61   \\
& incorrect $\pi_{X,\bZ}(w,\bz)$ & 0.91 & -0.09 & 4.43 & 4.58 & 47.20 & 3.00 & -0.00 & 3.30 & 3.45 & 94.40 & 2.00 & 0.00 & 4.63 & 4.89 & 93.54   \\
\addlinespace
&\multicolumn{5}{l}{\underline{without using $\bLambda$}}\\
\multirow[t]{-2}{*}{\raggedright\arraybackslash \hspace{1em}MACC} 
 & correct $\pi_{X,\bZ}(w,\bz)$ and correct $\bPsi_{\rm MACC}(y,\bz)$&  1.00 & -0.00 & 4.43 & 4.43 & 95.02 & 3.00 & 0.00 & 3.39 & 3.26 & 95.47 & 2.00 & 0.00 & 4.67 & 4.63 & 95.22 \\
& incorrect $\pi_{X,\bZ}(w,\bz)$ and correct $\bPsi_{\rm MACC}(y,\bz)$ &  0.72 & -0.29 & 5.63 & 5.77 & 0.25 & 2.94 & -0.06 & 5.14 & 5.10 & 80.93 & 1.97 & -0.03 & 6.30 & 6.46 & 92.78  \\
 & correct $\pi_{X,\bZ}(w,\bz)$ and incorrect $\bPsi_{\rm MACC}(y,\bz)$  & 1.00 & -0.00 & 5.42 & 5.52 & 94.40 & 3.00 & 0.00 & 4.38 & 4.42 & 95.37 & 2.00 & 0.00 & 5.45 & 5.52 & 94.96  \\
& incorrect $\pi_{X,\bZ}(w,\bz)$ and incorrect $\bPsi_{\rm MACC}(y,\bz)$  &  0.67 & -0.33 & 6.08 & 6.24 & 0.00 & 3.02 & 0.02 & 5.20 & 5.34 & 91.40 & 2.01 & 0.01 & 6.63 & 6.84 & 93.74  \\
&\multicolumn{5}{l}{\underline{using $\bLambda$}}\\
 & correct $\pi_{X,\bZ}(w,\bz)$ &  1.00 & -0.00 & 4.46 & 4.53 & 94.76 & 3.00 & 0.00 & 3.34 & 3.30 & 95.27 & 2.00 & 0.00 & 4.67 & 4.71 & 94.81  \\
& incorrect $\pi_{X,\bZ}(w,\bz)$ & 0.93 & -0.07 & 4.51 & 4.67 & 64.29 & 3.01 & 0.01 & 3.38 & 3.52 & 91.81 & 2.01 & 0.01 & 4.74 & 4.94 & 93.29   \\
\addlinespace
&\multicolumn{5}{l}{\underline{without using $\bLambda$}}\\
\multirow[t]{-2}{*}{\raggedright\arraybackslash \hspace{1em}AIPW} 
 & correct $\pi_{X,\bZ}(w,\bz)$ and correct $\bPsi_{\rm AIPW}(y,\bz)$ & 1.00 & -0.00 & 4.69 & 4.78 & 94.15 & 3.00 & 0.00 & 3.80 & 3.68 & 94.66 & 2.00 & 0.00 & 5.13 & 5.13 & 94.61   \\
& incorrect $\pi_{X,\bZ}(w,\bz)$ and correct $\bPsi_{\rm AIPW}(y,\bz)$ & 0.57 & -0.43 & 6.19 & 6.97 & 0.00 & 3.03 & 0.03 & 4.59 & 4.63 & 87.08 & 2.02 & 0.02 & 6.52 & 6.59 & 93.44   \\
 & correct $\pi_{X,\bZ}(w,\bz)$ and incorrect $\bPsi_{\rm AIPW}(y,\bz)$  & 1.00 & -0.00 & 4.95 & 5.07 & 94.20 & 3.00 & 0.00 & 3.81 & 3.87 & 94.66  & 2.00 & 0.00 & 5.28 & 5.42 & 94.20   \\
& incorrect $\pi_{X,\bZ}(w,\bz)$ and incorrect $\bPsi_{\rm AIPW}(y,\bz)$ &  0.73 & -0.27 & 5.90 & 6.15 & 0.41 & 3.05 & 0.05 & 4.40 & 4.51 & 77.92 & 2.03 & 0.03 & 6.20 & 6.26 & 92.07   \\
&\multicolumn{5}{l}{\underline{using $\bLambda$}}\\
 & correct $\pi_{X,\bZ}(w,\bz)$ & 1.00 & -0.00 & 4.60 & 4.76 & 93.95 & 3.00 & 0.00 & 3.50 & 3.64 & 94.05 & 2.00 & 0.00 & 4.88 & 5.06 & 94.00  \\
& incorrect $\pi_{X,\bZ}(w,\bz)$ &   0.88 & -0.12 & 5.35 & 5.97 & 37.44 & 3.02 & 0.02 & 3.93 & 4.53 & 86.67 & 2.02 & 0.02 & 5.56 & 6.06 & 92.01  \\
\addlinespace
\addlinespace[0.3em]
\bottomrule
\end{tabular}}
\end{table}
\end{landscape}

\begin{figure}[!ht]
\caption{Distribution of simulation results under independent {covariate right-}censoring with known $f_{C,X|\bZ}(c,x,\bz; \balpha)$.}
    \centering
    \label{fig:indepdent-boxplot}
\includegraphics[width=0.95\textwidth]{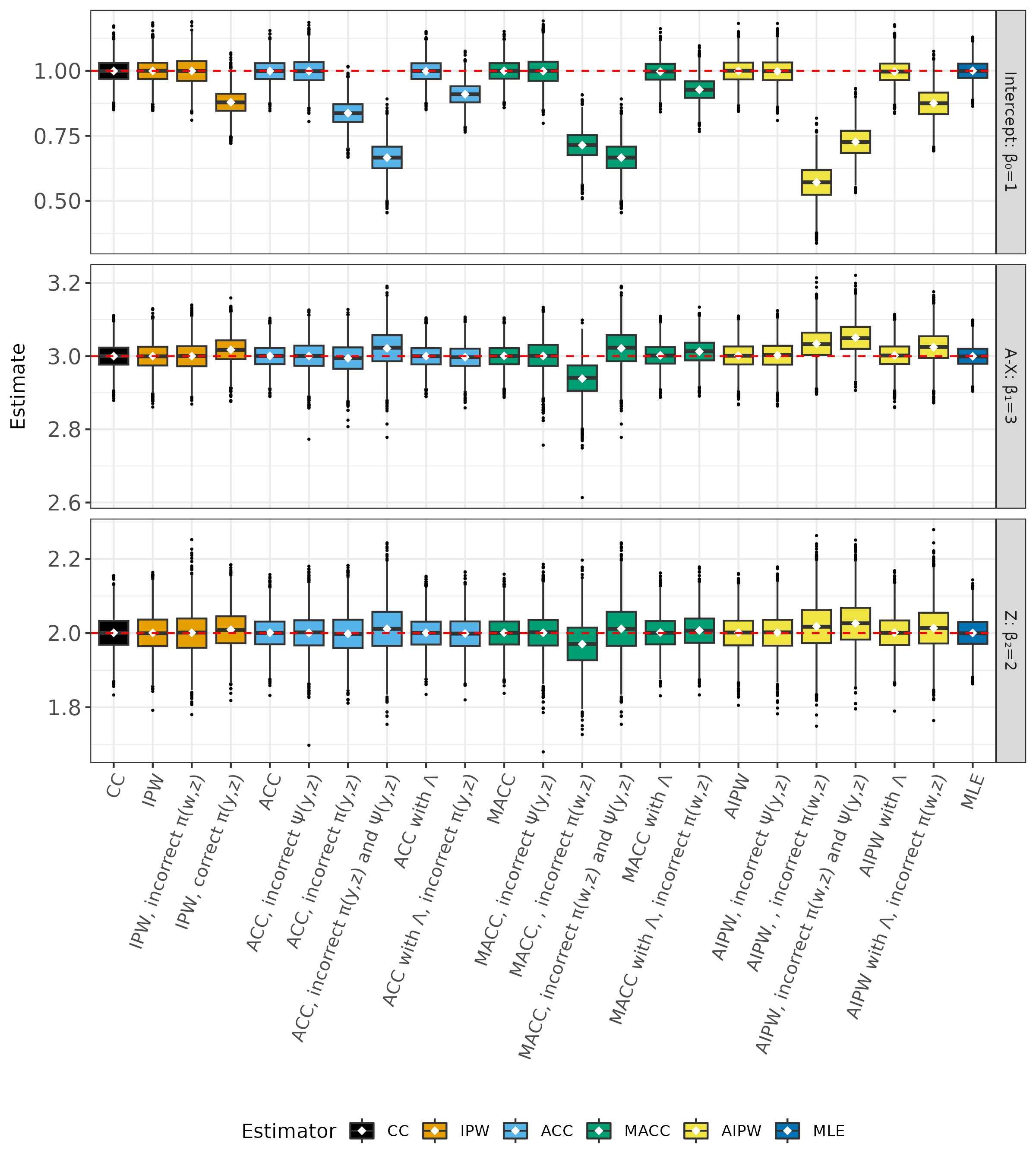}
    \caption*{ \footnotesize \textbf{Note:} Red-dashed line indicates oracle value.}
    \vskip -2em
\end{figure}

\begin{figure}[!ht]
\caption{Efficiency comparison of estimators against the CC estimator under independent covariate right-censoring with known $f_{C,X|\bZ}(c,x,\bz; \balpha)$.}
    \centering
    \label{fig:indepdent-efficiency}
\includegraphics[width=0.7\textwidth]{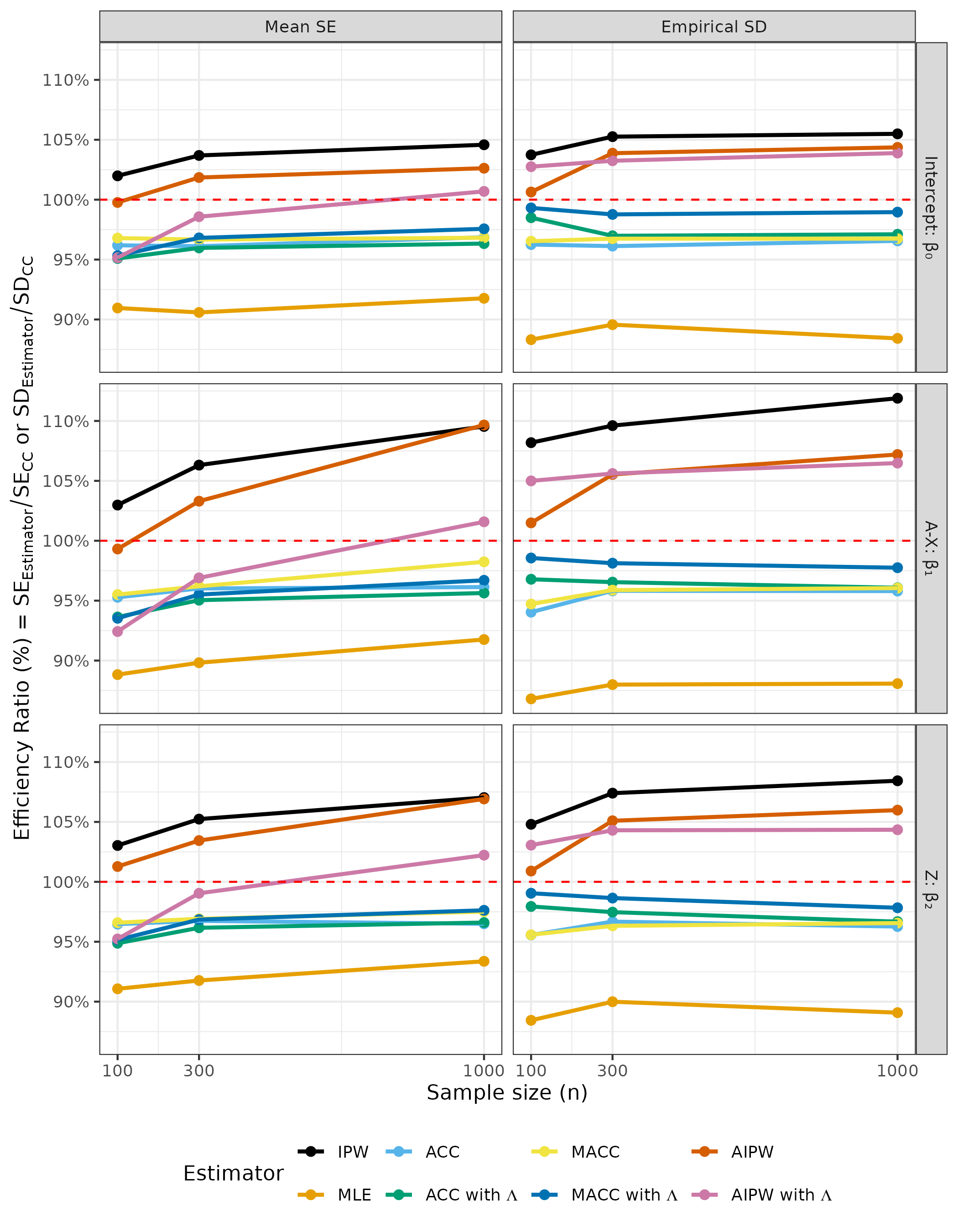}
    \caption*{ \footnotesize \textbf{Note:} The red-dashed line represents no gain in efficiency compared to the CC estimator. Values below this line ($<100$\%) indicate greater efficiency, while values above it ($>100$\%) indicate reduced efficiency.  The empirical mean of the estimated standard errors (Mean SE) and the empirical standard deviation (Empirical SD) of $\wh\btheta$ across all simulations are illustrated. The Mean SE were computed using the asymptotic variances derived in our theorems, with all expectations replaced by empirical averages.}
    \vskip -2em
\end{figure}

\begin{landscape}
\begin{table}[hb]
\caption{\label{tab:dependent_known_1000} \textbf{Dependent covariate right-censoring with known $f_{C,X|\bZ}(c,x,\bz; \balpha)$.} Simulation results of the mean estimate, mean estimated standard errors (SE) and empirical standard deviations (SD) scaled by 100, and confidence interval coverage  (95\% Cov) for each estimator. Results are based on 1,000 simulated datasets, each with a sample size of 1,000 and a censoring rate of 50\%.}
\centering
\resizebox{1\linewidth}{!}{
\begin{tabular}[t]{llrrrrrrrrrrrrrrrrrrrr}
\toprule
\textbf{Estimator} & \textbf{Specification} & \textbf{Estimate} & \textbf{Bias} & \textbf{SE} & \textbf{SD} & \textbf{95\% Cov} & \textbf{Estimate} & \textbf{Bias} & \textbf{SE} & \textbf{SD} & \textbf{95\% Cov} & \textbf{Estimate} & \textbf{Bias} & \textbf{SE} & \textbf{SD} & \textbf{95\% Cov} \\
\midrule
\addlinespace[0.3em]
&&\multicolumn{5}{c}{\textbf{Intercept: $\beta_0 = 1$}} &\multicolumn{5}{c}{\textbf{$A-X$: $\beta_{AX} = 3$}} &\multicolumn{5}{c}{\textbf{$Z$: $\beta_{Z} = 2$}}\\
\hspace{1em}Oracle & & 1.00 & 0.00 & 3.16 & 3.17 & 94.36 & 3.00 & 0.00 & 2.38 & 2.39 & 94.67 & 2.00 & -0.00 & 3.37 & 3.33 & 94.62  \\
\addlinespace
\hspace{1em}Naive &  & 1.00 & 0.00 & 4.50 & 4.50 & 94.36 & 3.00 & 0.00 & 3.40 & 3.44 & 94.36 & 2.00 & -0.00 & 4.77 & 4.76 & 95.14    \\
\addlinespace
\hspace{1em}CC & &   1.00 & 0.00 & 4.50 & 4.50 & 94.36 & 3.00 & 0.00 & 3.40 & 3.44 & 94.36 & 2.00 & -0.00 & 4.77 & 4.76 & 95.14   \\
\addlinespace
 & correct $\pi_{X,\bZ}(w,\bz)$ & 1.00 & 0.00 & 4.55 & 4.57 & 94.57 & 3.00 & 0.00 & 3.48 & 3.54 & 94.21 & 2.00 & -0.00 & 4.86 & 4.88 & 94.83   \\
\multirow[t]{-2}{*}{\raggedright\arraybackslash \hspace{1em}IPW} 
& incorrect $\pi_{X,\bZ}(w,\bz)$ & 1.00 & 0.00 & 5.43 & 5.42 & 94.93 & 3.00 & 0.00 & 4.08 & 4.14 & 94.21 & 2.00 & -0.00 & 5.73 & 5.79 & 94.83   \\
& correct $\pi_{Y,\bZ}(y,\bz)$ & 0.94 & -0.06 & 4.58 & 4.58 & 70.98 & 3.01 & 0.01 & 3.49 & 3.56 & 94.05 & 2.00 & 0.00 & 4.88 & 4.89 & 95.09  \\
\addlinespace
 & correct $f_{X|\bZ}$ & 1.00 & 0.00 & 4.12 & 4.08 & 94.46 & 3.00 & 0.00 & 3.17 & 3.00 & 94.46 & 2.00 & -0.00 & 4.46 & 4.32 & 94.57    \\
\multirow[t]{-2}{*}{\raggedright\arraybackslash \hspace{1em}MLE} 
& incorrect $f_{X|\bZ}$  & 0.14 & -0.86 & 8.16 & 8.24 & 0.05 & 2.74 & -0.25 & 6.74 & 6.61 & 1.91 & 1.70 & -0.30 & 7.66 & 7.64 & 2.22  \\
\addlinespace
&\multicolumn{5}{l}{\underline{without using $\bLambda$}}\\
\multirow[t]{-2}{*}{\raggedright\arraybackslash \hspace{1em}ACC} 
 & correct $\pi_{X,\bZ}(w,\bz)$ and correct $\bPsi_{\rm ACC}(y,\bz)$& 1.00 & 0.00 & 4.35 & 4.40 & 93.95 & 3.00 & 0.00 & 3.27 & 3.25 & 94.26 & 2.00 & -0.00 & 4.63 & 4.59 & 94.57  \\
& incorrect $\pi_{X,\bZ}(w,\bz)$ and correct $\bPsi_{\rm ACC}(y,\bz)$ &  0.91 & -0.09 & 4.86 & 4.86 & 52.77 & 3.00 & -0.00 & 3.85 & 3.81 & 95.71 & 2.00 & -0.00 & 5.23 & 5.15 & 94.88   \\
 & correct $\pi_{X,\bZ}(w,\bz)$ and incorrect $\bPsi_{\rm ACC}(y,\bz)$  & 1.00 & 0.00 & 4.62 & 4.75 & 94.98 & 3.00 & 0.00 & 3.56 & 3.52 & 94.31 & 2.00 & -0.00 & 4.88 & 4.89 & 94.77   \\
& incorrect $\pi_{X,\bZ}(w,\bz)$ and incorrect $\bPsi_{\rm ACC}(y,\bz)$  & 0.81 & -0.19 & 6.11 & 6.24 & 11.95 & 3.01 & 0.01 & 5.11 & 5.02 & 95.09 & 2.00 & 0.00 & 6.63 & 6.50 & 94.98   \\
&\multicolumn{5}{l}{\underline{using $\bLambda$}}\\
 & correct $\pi_{X,\bZ}(w,\bz)$ &  1.00 & 0.00 & 4.34 & 4.41 & 94.77 & 3.00 & 0.00 & 3.24 & 3.26 & 94.31  & 2.00 & -0.00 & 4.60 & 4.60 & 94.77  \\
& incorrect $\pi_{X,\bZ}(w,\bz)$ & 0.95 & -0.05 & 4.37 & 4.47 & 78.27 & 3.00 & -0.00 & 3.26 & 3.36 & 93.79 & 2.00 & -0.00 & 4.62 & 4.68 & 94.46  \\
\addlinespace
&\multicolumn{5}{l}{\underline{without using $\bLambda$}}\\
\multirow[t]{-2}{*}{\raggedright\arraybackslash \hspace{1em}MACC} 
 & correct $\pi_{X,\bZ}(w,\bz)$ and correct $\bPsi_{\rm MACC}(y,\bz)$&1.00 & 0.00 & 4.37 & 4.40 & 94.72 & 3.00 & 0.00 & 3.27 & 3.24 & 94.36 & 2.00 & -0.00 & 4.63 & 4.59 & 94.72  \\
& incorrect $\pi_{X,\bZ}(w,\bz)$ and correct $\bPsi_{\rm MACC}(y,\bz)$ & 0.85 & -0.15 & 5.01 & 5.03 & 15.62 & 2.98 & -0.02 & 4.03 & 3.98 & 93.95 & 1.99 & -0.01 & 5.42 & 5.31 & 94.72  \\
 & correct $\pi_{X,\bZ}(w,\bz)$ and incorrect $\bPsi_{\rm MACC}(y,\bz)$  &  1.00 & 0.00 & 4.68 & 4.82 & 94.98 & 3.00 & 0.00 & 3.61 & 3.58 & 94.26 & 2.00 & -0.00 & 4.93 & 4.95 & 94.62  \\
& incorrect $\pi_{X,\bZ}(w,\bz)$ and incorrect $\bPsi_{\rm MACC}(y,\bz)$  &  0.81 & -0.19 & 6.11 & 6.24 & 11.95 & 3.01 & 0.01 & 5.11 & 5.02 & 95.09 & 2.00 & 0.00 & 6.63 & 6.50 & 94.98  \\
&\multicolumn{5}{l}{\underline{using $\bLambda$}}\\
 & correct $\pi_{X,\bZ}(w,\bz)$ & 1.00 & 0.00 & 4.36 & 4.44 & 94.31 & 3.00 & 0.00 & 3.26 & 3.28 & 94.15 & 2.00 & -0.00 & 4.62 & 4.64 & 94.88  \\
& incorrect $\pi_{X,\bZ}(w,\bz)$ & 0.96 & -0.04 & 4.46 & 4.56 & 85.36 & 3.01 & 0.01 & 3.35 & 3.45 & 93.53 & 2.00 & 0.00 & 4.73 & 4.77 & 94.52   \\
\addlinespace
&\multicolumn{5}{l}{\underline{without using $\bLambda$}}\\
\multirow[t]{-2}{*}{\raggedright\arraybackslash \hspace{1em}AIPW} 
 & correct $\pi_{X,\bZ}(w,\bz)$ and correct $\bPsi_{\rm AIPW}(y,\bz)$ & 1.00 & 0.00 & 4.50 & 4.58 & 94.77 & 3.00 & 0.00 & 3.41 & 3.35 & 94.83 & 2.00 & -0.00 & 4.79 & 4.81 & 94.67  \\
& incorrect $\pi_{X,\bZ}(w,\bz)$ and correct $\bPsi_{\rm AIPW}(y,\bz)$ & 0.76 & -0.24 & 6.06 & 6.28 & 3.03 & 0.03 & 4.53 & 4.33 & 90.07 & 2.12 & 2.01 & 0.01 & 6.45 & 6.35 & 93.64  \\
 & correct $\pi_{X,\bZ}(w,\bz)$ and incorrect $\bPsi_{\rm AIPW}(y,\bz)$  & 1.00 & 0.00 & 4.55 & 4.68 & 94.46 & 3.00 & 0.00 & 3.43 & 3.41 & 94.26 & 2.00 & -0.00 & 4.85 & 4.89 & 94.46  \\
& incorrect $\pi_{X,\bZ}(w,\bz)$ and incorrect $\bPsi_{\rm AIPW}(y,\bz)$ &  0.84 & -0.16 & 5.92 & 6.03 & 25.25 & 3.03 & 0.03 & 4.39 & 4.32 & 88.46 & 2.01 & 0.01 & 6.26 & 6.26 & 94.10  \\
&\multicolumn{5}{l}{\underline{using $\bLambda$}}\\
 & correct $\pi_{X,\bZ}(w,\bz)$ & 1.00 & 0.00 & 4.40 & 4.50 & 94.72 & 3.00 & 0.00 & 3.30 & 3.35 & 94.10 & 2.00 & -0.00 & 4.68 & 4.73 & 94.72    \\
& incorrect $\pi_{X,\bZ}(w,\bz)$ &0.93 & -0.07 & 5.30 & 5.57 & 72.79 & 3.02 & 0.02 & 3.90 & 4.18 & 91.00 & 2.01 & 0.01 & 5.57 & 5.81 & 93.69  \\
\addlinespace
\addlinespace[0.3em]
\bottomrule
\end{tabular}}
\end{table}
\end{landscape}

\begin{figure}[!ht]
\caption{Distribution of simulation results under dependent {covariate right-}censoring with known $f_{C,X|\bZ}(c,x,\bz; \balpha)$.}
    \centering
    \label{fig:depdent-boxplot}
\includegraphics[width=0.95\textwidth]{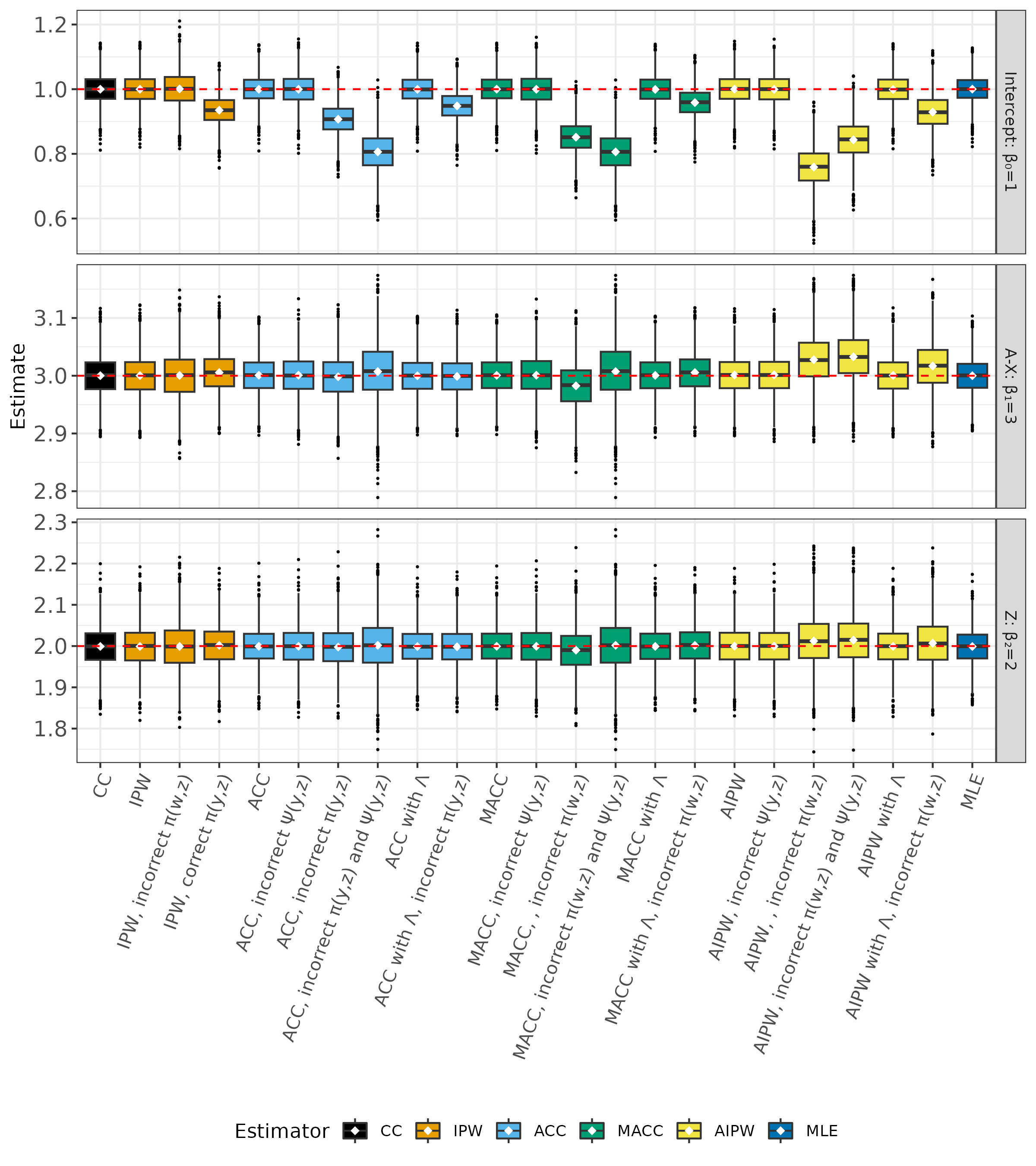}
    \caption*{ \footnotesize \textbf{Note:} Red-dashed line indicates oracle value.}
    \vskip -2em
\end{figure}

\begin{figure}[!ht]
\caption{Efficiency comparison of estimators against the CC estimator under dependent covariate right-censoring with known $f_{C,X|\bZ}(c,x,\bz; \balpha)$}
    \centering
    \label{fig:dependent-efficiency}
\includegraphics[width=0.85\textwidth]{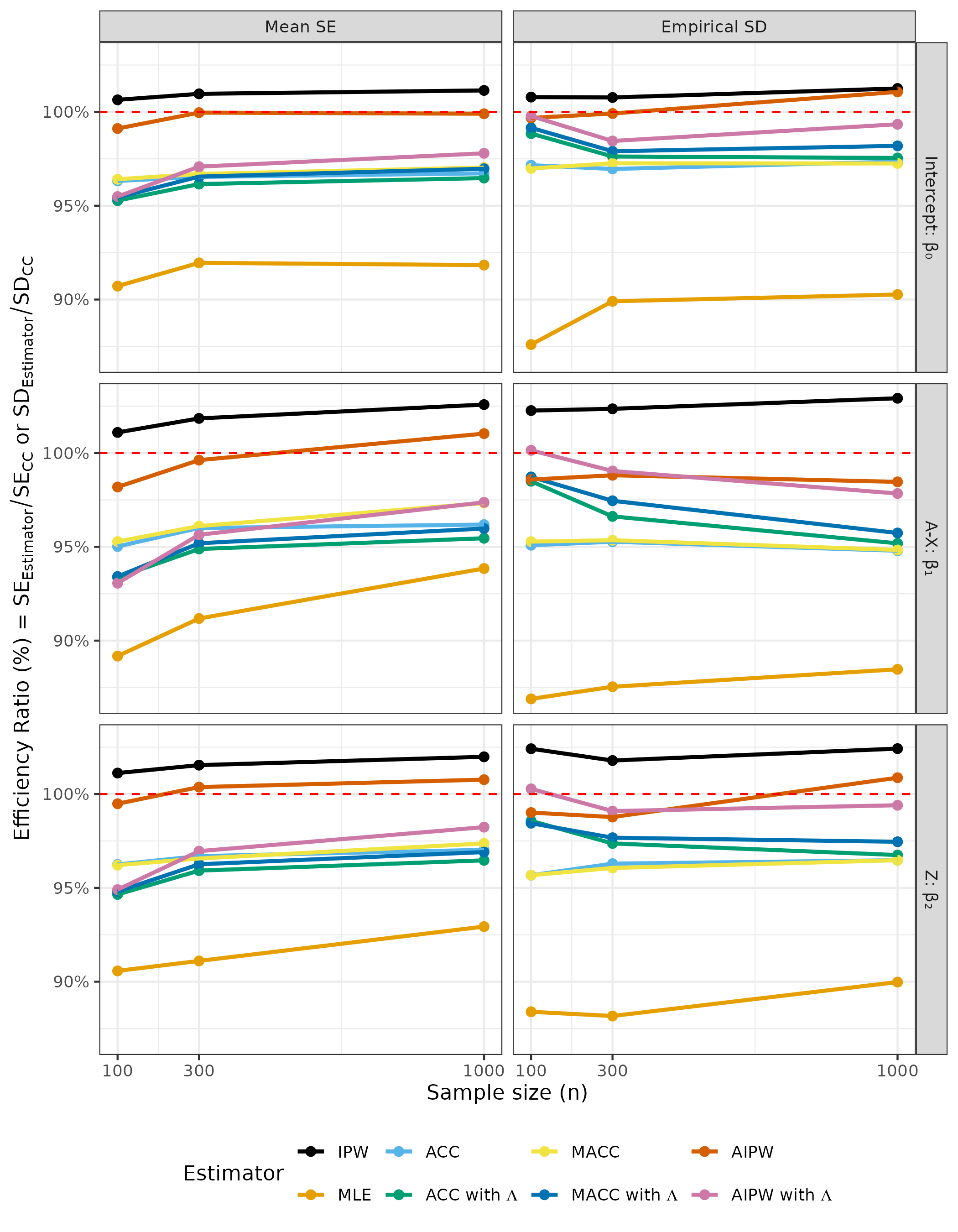}
    \caption*{ \footnotesize \textbf{Note:} The red-dashed line represents no gain in efficiency compared to the CC estimator. Values below this line ($<100$\%) indicate greater efficiency, while values above it ($>100$\%) indicate reduced efficiency. The empirical mean of the estimated standard errors (Mean SE) and the empirical standard deviation (Empirical SD) of $\wh\btheta$ across all simulations are illustrated. The Mean SE were computed using the asymptotic variances derived in our theorems, with all expectations replaced by empirical averages.}
    \vskip -2em
\end{figure}


\begin{table}[hbt!]
\caption{\label{tab:independent_estimated_1000} \textbf{Independent covariate right-censoring with unknown $f_{C,X|Z}(c,x,\bz; \balpha)$.} Simulation results of the mean estimate, mean estimated standard errors (SE) and empirical standard deviations (SD) scaled by 100, and confidence interval coverage  (95\% Cov) for each estimator. Results are based on 1,000 simulated datasets, each with a sample size of 1,000 and a censoring rate of 50\%.}
\centering
\resizebox{0.8\linewidth}{!}{
\begin{tabular}[t]{lllrrrr}
\toprule
\textbf{Estimator} & \textbf{Specification}  & \textbf{Estimate} & \textbf{Bias} & \textbf{SE} & \textbf{SD} & \textbf{95\% Cov}\\
\midrule
\addlinespace[0.3em]
& &\multicolumn{5}{c}{\textbf{Intercept: $\beta_0 = 1$}}\\
\hspace{1em}Oracle &  &1.00 & 0.00 & 3.16 & 3.15 & 95.29 \\
\hspace{1em}Naive &  & -0.37 & -1.37 & 8.35 & 8.60 & 0.00 \\
\hspace{1em}CC & & 1.00 & 0.00 & 4.50 & 4.58 & 94.38 \\
\hspace{1em}IPW & correct $\wh\pi_{X,\bZ}(w,\bz)$ &  1.00 & 0.00 & 4.73 & 4.83 & 94.28 \\
\hspace{1em}MACC & correct $\wh\pi_{X,\bZ}(w,\bz)$ and using $\bLambda$ &  1.00 & 0.00 & 4.48 & 4.56 & 94.58  \\
\hspace{1em}AIPW & correct $\wh\pi_{X,\bZ}(w,\bz)$ and using $\bLambda$ & 1.00 & 0.00 & 4.69 & 4.78 & 94.38  \\
\hspace{1em}MLE & correct $\wh f_{X|\bZ}$ & 1.00 & 0.00 & 4.17 & 4.24 & 94.79  \\
& &\multicolumn{5}{c}{\textbf{$A-X$: $\beta_{AX} = 3$}}\\
\hspace{1em}Oracle & &  3.00 & 0.00 & 2.38 & 2.43 & 95.29 \\
\hspace{1em}Naive & & 1.76 & -0.41 & 6.96 & 6.84 & 0.00 \\
\hspace{1em}CC & & 3.00 & 0.00 & 3.39 & 3.56 & 93.98 \\
\hspace{1em}IPW & correct $\wh\pi_{X,\bZ}(w,\bz)$ & 3.00 & 0.00 & 3.87 & 4.33 & 92.98 \\
\hspace{1em}MACC & correct $\wh\pi_{X,\bZ}(w,\bz)$ and using $\bLambda$ &   3.00 & 0.00 & 3.36 & 3.55 & 93.67  \\
\hspace{1em}AIPW & correct $\wh\pi_{X,\bZ}(w,\bz)$ and using $\bLambda$ & 3.00 & 0.00 & 3.70 & 4.01 & 92.78  \\
\hspace{1em}MLE & correct $\wh f_{X|\bZ}$ & 3.00 & 0.00 & 3.25 & 3.16 & 93.45  \\
&& \multicolumn{5}{c}{\textbf{$Z$: $\beta_{Z} = 2$}}\\
\hspace{1em}Oracle & & 2.00 & 0.00 & 3.38 & 3.39 & 94.48 \\
\hspace{1em}Naive & &  1.38 & -0.31 & 9.90 & 9.69 & 0.00 \\
\hspace{1em}CC & & 2.00 & 0.00 & 4.78 & 4.81 & 93.98 \\
\hspace{1em}IPW & & 2.00 & 0.00 & 5.35 & 5.58 & 94.38 \\
\hspace{1em}MACC & correct $\wh\pi_{X,\bZ}(w,\bz)$ and using $\bLambda$ &  2.00 & 0.00 & 4.76 & 4.80 & 94.48   \\
\hspace{1em}AIPW & correct $\wh\pi_{X,\bZ}(w,\bz)$ and using $\bLambda$ & 2.00 & 0.00 & 5.18 & 5.29 & 94.18  \\
\hspace{1em}MLE & correct $\wh f_{X|\bZ}$ & 2.00 & 0.00 & 4.44 & 4.47 & 93.78   \\
\addlinespace[0.3em]
\bottomrule
\end{tabular}}
\end{table}

\begin{table}[hbt!]
\caption{\label{tab:dependent_estimated_1000} \textbf{Dependent {covariate right-}censoring with unknown $f_{C,X|Z}(c,x,\bz; \balpha)$.} We report the mean estimate, mean estimated standard errors (SE) and empirical standard deviations (SD) scaled by 100, and confidence interval coverage  (95\% Cov) for each estimator. Results are based on 1,000 simulated datasets, each with a sample size of 1,000 and a censoring rate of 50\%.}
\centering
\resizebox{0.9\linewidth}{!}{
\begin{tabular}[t]{lllrrrr}
\toprule
\textbf{Estimator} & \textbf{Specification} & \textbf{Estimate} & \textbf{Bias}& \textbf{SE} & \textbf{SD} & \textbf{95\% Cov}\\
\midrule
\addlinespace[0.3em]
&&\multicolumn{5}{c}{\textbf{Intercept: $\beta_0 = 0$}}\\
\hspace{1em}Oracle  & & 0.00 & 0.00 & 3.40 &  3.47 & 93.40 \\
\hspace{1em}Naive & & 0.05 & 0.05 & 3.29 & 3.33 & 64.70 \\
\hspace{1em}CC & &  0.00 & 0.00 & 6.84 &  6.79 & 94.90 \\
\hspace{1em}ACC & correct $\wh\pi_{Y,\bZ}(y,\bz)$ and using $\bLambda$ & 0.00 & 0.00 
&6.40 & 6.40 & 94.71 \\
&&\multicolumn{5}{c}{\textbf{$X$: $\beta_X = 0.2$}}\\
\hspace{1em}Oracle  & &  0.20 & 0.00 & 2.92 & 2.95 & 94.10 \\
\hspace{1em}Naive & & 0.17 & 0.03 & 2.59 & 2.62 & 77.70 \\
\hspace{1em}CC & & 0.20 & 0.00 & 4.37 &  4.31 & 95.00  \\
\hspace{1em}ACC & correct $\wh\pi_{Y,\bZ}(y,\bz)$ and using $\bLambda$ & 0.19 & 0.01 & 4.32 & 4.28 & 95.10 \\
&&\multicolumn{5}{c}{\textbf{$Z$: $\beta_Z = 0.2$}}\\
\hspace{1em}Oracle  &  & 0.20 & 0.00 &2.92 & 2.97 & 94.10 \\
\hspace{1em}Naive  & & 0.21 & 0.01&  2.89 & 2.95 & 92.80 \\
\hspace{1em}CC &  & 0.20 &0.00&  4.37 &  4.34 & 94.60  \\
\hspace{1em}ACC & correct $\wh\pi_{Y,\bZ}(y,\bz)$ and using $\bLambda$ & 0.20 & 0.00 & 3.65 & 3.62 & 95.60 \\
\addlinespace[0.3em]
\bottomrule
\end{tabular}}
\end{table}

\clearpage\newpage


\defbibfilter{myfilter}{%
    keyword=bibonly      
    or keyword=mainbib
}
\printbibliography[filter=myfilter]

\end{document}